\documentclass{aa}  
\usepackage{graphicx}
\usepackage{txfonts}
\usepackage{natbib}
\usepackage{multirow}
\usepackage{txfonts}
\usepackage{tabularray}
\usepackage{subcaption}
\usepackage{footnotehyper}
\usepackage{array}
\usepackage{booktabs}
\usepackage{caption}
\usepackage{subcaption}
\usepackage{gensymb}
\usepackage{amssymb}
\usepackage{booktabs}
\usepackage{url}
\usepackage{multirow}
\usepackage{siunitx}
\usepackage{wrapfig}
\usepackage{lscape}
\usepackage{rotating}
\usepackage{longtable}
\usepackage{epstopdf}
\usepackage{caption}
\usepackage{txfonts}
\usepackage{natbib}
\usepackage{subcaption}
\usepackage{longtable}
\usepackage{array}
\usepackage{caption}
\usepackage{tablefootnote}
\usepackage{orcidlink}
\usepackage{threeparttablex} 
\newcommand{\solar}{$_{\odot}$}
\newcommand{\micron}{$\mu$m}
\newcommand{\standform}[1]{$\times$10$^{#1}$}
\newcommand{\angstrom}{$\AA$}
\newcommand{\up}[1]{$^{#1}$}
\newcommand{\down}[1]{$_{#1}$}
\newcolumntype{y}[1]{>{\centering\let\newline\\\arraybackslash\hspace{0pt}}p{#1}}
\bibpunct{(}{)}{;}{a}{}{,} 
\begin{document} 

    \title{Is the nitrogen-rich source PN K4-47 a true planetary nebula?}
    \author{T. Steinmetz\inst{1} \href{https://orcid.org/0000-0003-1592-3249}{\includegraphics[scale=0.75]{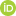}}
          \and
          T. Kami\'{n}ski\inst{1} \href{https://orcid.org/0000-0001-8541-8024}{\includegraphics[scale=0.75]{Images/ORCID-iD_icon_16x16.png}}
          \and
          D. Jones\inst{2,3} \href{https://orcid.org/0000-0003-3947-5946}{\includegraphics[scale=0.75]{Images/ORCID-iD_icon_16x16.png}}
          \and
          M. Hajduk\inst{4} \href{https://orcid.org/0000-0001-6028-9932}{\includegraphics[scale=0.75]{Images/ORCID-iD_icon_16x16.png}}
          \and
          D. R. Gon\c{c}alves\inst{5} \href{https://orcid.org/0000-0001-9388-7146}{\includegraphics[scale=0.75]{Images/ORCID-iD_icon_16x16.png}}
          \and
          S. Akras\inst{6} \href{https://orcid.org/0000-0003-1351-7204}{\includegraphics[scale=0.75]{Images/ORCID-iD_icon_16x16.png}}
          }

   \institute{Nicolaus Copernicus Astronomical Center, ul. Rabia\'{n}ska 8, 87-100 Toru\'{n}, Poland \email{thomas@ncac.torun.pl}
   \and 
   Instituto de Astrof\'{i}sica de Canarias, E-38205 La Laguna, Tenerife, Spain
   \and
   Departamento de Astrof\'{i}sica, Universidad de La Laguna, E-38206 La Laguna, Tenerife, Spain
   \and
   Department of Geodesy, Faculty of Geoengineering, University of Warmia and Mazury, ul. Oczapowskiego 2, 10-719 Olsztyn, Poland
   \and
   Observat\'{o}rio do Valongo, Universidade Federal do Rio de Janeiro, Ladeira Pedro Antonio 43, 20080-090 Rio de Janeiro, Brazil
   \and
   Institute for Astronomy, Astrophysics, Space Applications and Remote Sensing, National Observatory of Athens, GR 15236 Penteli, Greece
   }

   \date{Received XXX, accepted YYY}

\abstract
   {PN K4-47 is a young planetary nebula that exhibits low-ionisation structures in the form of two `lobes'. The unusual chemistry of the nebula has raised questions about whether K4-47 is a true planetary nebula, or if the origins are more exotic in nature.}
   {We aim to investigate the spatially-resolved structure of the nebula, including in the sub-millimetre for the first time. We want to examine the kinematics, chemistry, and mass-loss history of the nebula, and probe the stellar properties of the central system.}
   {We use a combination of optical imaging and spectroscopy and sub-millimetre interferometry, as well as archival radio interferometric data, to study the kinematics and morphology of the nebula and the differences between the atomic and molecular gas phases.}
   {We find extended CO (2--1) emission towards the northern optical lobe, which shows a clear velocity gradient along the same position angle as the optical bipolar nebula. Comparing the elemental and isotopic abundances to model predictions, we estimate an upper limit of 6 kpc for the distance to K4-47. The source hosts a fast ($\sim$350 km s\up{-1}) atomic jet and slower ($\sim$50--60 km s\up{-1}) molecular outflow that are spatially coincident. The outflow velocity infers an age of 336 $\pm$ 119 yr. Using atomic line diagnostics, we find that the core has an electron temperature and density of $\approx$20 kK and 2800 cm\up{-3}, respectively. We derive a Zanstra effective temperature of 81 $\pm$ 2 kK for the central star. We also see evidence of a significant circumstellar component in the line-of-sight extinction to the source. The progenitor mass of K4-47 is estimated to be 4--6 M\solar, based on the comparisons of measured abundances with model predictions. The central white dwarf of K4-47 may have an approximate mass of 0.9--1.1 M\solar, following the initial-final mass relation for white dwarfs. The released material in the nebula is dominated by neutral atomic gas.}
   {The resolved molecular environment indicates that the nebula was shaped by the asymmetric outflow now seen primarily in optical emission. The progenitor of K4-47 was likely an AGB star, possibly a J-type carbon star. We also see that shocks, possibly from the bipolar lobes passing through the circumstellar environment, may play a non-negligible role in the excitation of gas in the core, rather than just photoionisation. The analysis indicates that K4-47 is indeed a true, if peculiar, planetary nebula.}

\maketitle


\section{Introduction}\label{sect: intro}
Planetary Nebulae (PNe) are one of the last evolutionary stages for stars with masses M $\le$ 8 M\solar, occurring after the Asymptotic Giant Branch (AGB) phase. PNe show a wide variety of morphologies ranging from spherical to highly multipolar \citep{balickfrank2002}.  It is now thought that binarity plays a key role in the formation of the most strongly axisymmetric structures \citep{jones2017}, in particular via a common envelope evolution \citep{ivanova2013review}.

CEE can also lead to the formation of more extreme transients known as Luminous Red Novae (LRNe), which are believed to result from the merger of two non-compact stars \citep{soker2003}. These sources are characterised by intermediate luminosity eruptions, often with multiple outburst peaks \citep{metzger2017} and low effective temperatures \citep{tylenda2005}, which give rise to molecules and dust \citep{kaminski2018}. LRNe also often display disks/torii \citep{kaminski2010,kaminski2021v838mon} and bipolar structures \citep{kaminski2020,mobeen2024,steinmetz2024}.

K4-47 (IRAS 04166+5611) is a young \citep[$\sim$400--900 yr; ][]{corradi2000} PN that exhibits an abundance of carbon-bearing molecules \citep{edwards2014,schmidt2016hcnhco,schmidt2017hcnhnc,schmidt2017cch,schmidt2019carbonchem} and isotopic enrichment \citep{schmidt2018isotope}. Molecule-rich PNe are unusual, as the conditions surrounding the central stars of PNe do not support molecule survival. Indeed, few molecules other than CO and H\down{2} have commonly been detected in PNe \citep[see][and references within]{kimura2012}. Optical imaging of K4-47 reveals three distinct morphological components: a nebulous core, and two lobes that appear shock-dominated \citep{goncalves2004,akras2017} and feature low ionisation structures \citep[LISs;][]{goncalves2001}. The lobes are also apparent in infrared H$_2$ emission arising from shocked gas at an excitation temperature of 1400--1700 K \citep{lumsden2001nirspec,akras2017}.

\begin{figure}
    \centering
    \includegraphics[width=\linewidth]{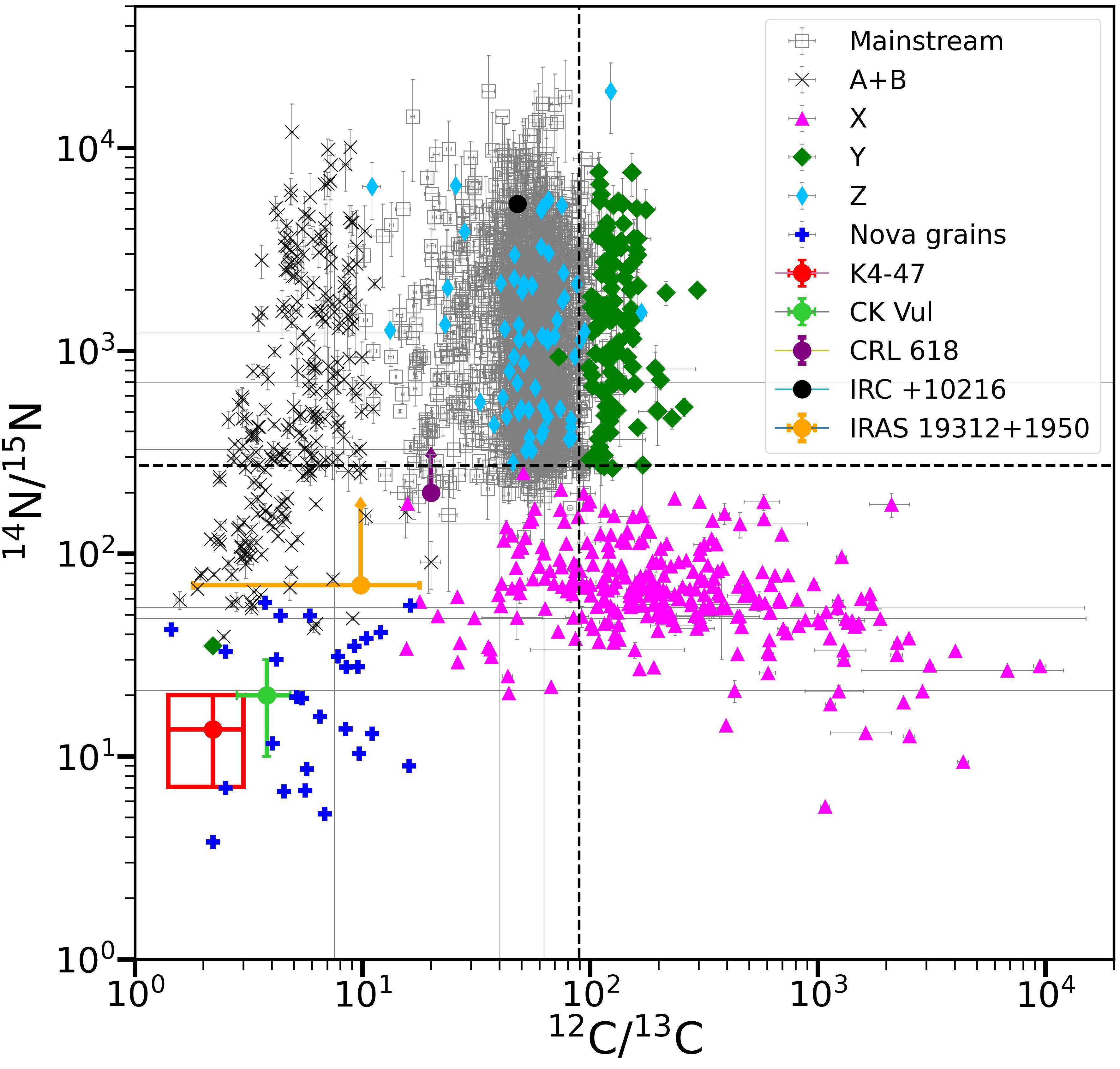}
    \caption{\up{12}C/\up{13}C vs \up{14}N/\up{15}N isotopic ratios for different types of presolar SiC grains \protect{\citep{stephan2024}}. Also plotted are measured isotopic ratios for K4-47 \protect{\citep{schmidt2018isotope}}, CK Vul \citep{kaminski2017}, CRL 618 \protect{\citep{wannier1991}}, IRC +10216 \citep{wannier1991}, and IRAS 19312+1950 \protect{\citep{qiu2023}}. The black dashed lines indicate solar system values of the isotopic ratios \protect{\citep{lodders2003}}.}
    \label{fig: isotopic ratios}
\end{figure}

The isotopic enrichment in K4-47 \citep{schmidt2018isotope} also bears a strong similarity to the oldest known Galactic LRN, CK Vul \citep{kato2003}. CK Vul is believed to be a merger between a red giant branch (RGB) star and a helium white dwarf (WD) \citep{tylenda2024}. It shows a bipolar structure in optical, infrared, and sub-millimetre emission \citep{shara1985,tylenda2019,kaminski2020}, as well as a faint extended hourglass nebula in H$\alpha$ which is reminiscent of many classical PNe \citep{hajduk2007}. The similarity in isotopic enrichment, partially illustrated in Fig. \ref{fig: isotopic ratios}, led \citet{schmidt2018isotope} to suggest that K4-47 may also be a stellar merger remnant.

In this paper, we examine this new potential origin of K4-47, and critically compare the properties of K4-47 and CK Vul in more detail using data across a wide range of wavelengths. Since only a few Galactic LRNe are known, identifying merger remnants at late epochs will help us to understand the evolution of these sources, and produce better statistics on their properties. Such comparisons are key to identify telltale signatures of red nova remnants years after the merger event, as well as avoid misidentifications of both planetary nebulae and red novae. Additionally, by this comparison we also aim to better understand the nature of K4-47.

The paper is organised in the following sections. Section \ref{sect: observations} describes the various observations in the optical, near-infrared, sub-millimetre, and radio. Section \ref{sect: results} describes the results, and Sect. \ref{sect: discussion} discusses the implications of these results, particularly for CK Vul. We summarise the paper in Sect. \ref{sect: summary}.

\section{Observations}\label{sect: observations}
\subsection{NOT} \label{sect: not obs}
Narrowband imaging in H$\alpha$, [\ion{O}{III}], [\ion{S}{II}], and [\ion{N}{II}] was obtained with the ALhambra Faint Object Spectrograph and Camera (ALFOSC) mounted on the 2.56m Nordic Optical Telescope (NOT) located at the Roque de los Muchachos Observatory on La Palma, at four different epochs: 27/11/2024, 08/12/2024, 26/02/2025, and 25/09/2025, under proposal 70-404 (PI T. Kami\'nski). We used the E2V 231-42 2k$\times$2k CCD, which has a pixel size of 15.0 \micron, equivalent to a plate scale of 0\farcs214/pixel across a 7\farcm3$\times$7\farcm3 field-of-view (FoV). The detector used no binning with a readout time of 200 pixels/s. As well as these observations, archival data taken with ALFOSC in equivalent, but older, narrowband filters from 21/09/1997 and 22/10/1997 were obtained from the NOT archive. These observations were conducted using the older Loral 2k$\times$2k detector. Detailed filter information for the filters used in 1997 is not currently available (see Table \ref{tab: alfosc obs}). K4-47 was also observed in H$\alpha$ continuum in 1997, with an exposure time of 300 s, but K4-47 was not detected. All ALFOSC imaging data was calibrated using a combination of \texttt{ccdproc} v2.4.2\footnote{https://ccdproc.readthedocs.io/en/2.4.2/} and \texttt{IRAF} routines. We performed astrometric calibration using the astrometry.net\footnote{https://nova.astrometry.net/upload} web service, which identified field stars within the ALFOSC FoV and cross-matched them with star positions in public catalogues to build a new WCS for the images. This provided accurate WCS information with a precision of 0\farcs02--0\farcs05. We used the web service as there were only a few frames to calibrate. The ALFOSC observation details are shown in Table \ref{tab: alfosc obs}.
\begin{table}
    \centering
    \small
    \caption{ALFOSC observations}
    \begin{tabular}{cccccc}\hline
    \multicolumn{6}{c}{Imaging} \\\hline
Date & Filter & Central & FWHM & Exp. & Seeing \\
 & & wavelength & & time & \\
& & (\angstrom) & (\angstrom) & (s) & (\arcsec)\\\hline
21/09/97$^1$ & H$\alpha$ & - & - & 600 & -\\
21/09/97 & [\ion{N}{II}] & - & - & 600 & -\\
22/10/97 & [\ion{S}{II}] & - & - & 600 & -\\
22/10/97 & [\ion{O}{III}] & - & - & 600 & -\\
\\[-5pt]
27/11/24 & H$\alpha$ & 6564 & 33 & 900 & 1.5 \\
27/11/24 & [\ion{N}{II}] & 6583 & 36 & 900 & 1.5 \\
27/11/24 & [\ion{S}{II}] & 6725 & 30 & 900 & 1.5 \\
27/11/24 & [\ion{O}{III}] & 5007 & 60 & 900 & 1.5 \\
08/12/24 & H$\alpha$ & 6564 & 33 & 900 & 0.6\\
08/12/24 & [\ion{N}{II}] & 6583 & 36 & 900 & 0.6\\
08/12/24 & [\ion{S}{II}] & 6725 & 30 & 900 & 0.6\\
08/12/24 & [\ion{O}{III}] & 5007 & 60 & 900 & 0.6\\
26/02/25 & H$\alpha$ & 6564 & 33 & 900 & 0.75 \\
26/02/25 & [\ion{S}{II}] & 6725 & 30 & 900 & 0.75 \\
25/09/25 & [\ion{O}{III}] & 5007 & 60 & 900  & 0.5\\
25/09/25 & H$\alpha$ & 6583 & 36 & 900 & 0.5 \\\hline
\multicolumn{6}{c}{Spectroscopy} \\\hline
Date & \multicolumn{3}{c}{Object} & \multicolumn{2}{c}{Exposure time} \\
& \multicolumn{3}{c}{} & \multicolumn{2}{c}{(s)}\\\hline
08/12/24 & \multicolumn{3}{c}{K4-47} & \multicolumn{2}{c}{90} \\
08/12/24 & \multicolumn{3}{c}{K4-47} & \multicolumn{2}{c}{900} \\
08/12/24 & \multicolumn{3}{c}{Hiltner 600} & \multicolumn{2}{c}{60} \\\hline
    \end{tabular}
    \tablefoot{$^1$Central wavelength and FWHM for the 1997 observations are not known.}
    \label{tab: alfosc obs}
\end{table}

As well as imaging observations, we obtained an ALFOSC spectrum, with differing exposure times of 90 and 900 seconds, on 08/12/2024. The spectroscopic observation details are shown in Table \ref{tab: alfosc obs}. The spectra were taken using grism no. 7\footnote{ALFOSC grism information can be found at https://www.not.iac.es/instruments/alfosc/grisms/} and the horizontal slit with a width of 1\farcs0, giving a wavelength coverage of 3650--7100 \angstrom. The resolving power R=650, giving an instrumental profile width of 461 km s\up{-1}. The spectrum was corrected for bias and flat field using the IRAF package. The spectrum of Hiltner 600 was used to perform flux calibration, as well as a reference for the trace of K4-47. We extracted three separate spectra of K4-47 centred on the core, northern lobe, and southern lobe. 

\subsection{SMA} \label{sect: sma obs}
K4-47 was observed with the Submillimeter Array (SMA), located at Maunakea Observatory, Hawaii on 02/07/2019. The on-source time totalled 4.62 h, using the full range of 8 antennas placed in the sub-compact array with baselines between 4--58 m. We used a gain calibrator, 0359+509, and bandpass calibrator 3C 279. The data were calibrated in the IDL MIR environment\footnote{https://lweb.cfa.harvard.edu/rtdc/SMAdata/process/mir/}. The spectral setup covers four sub-bands within 210.15--280.64 GHz at a spectral resolution of 1.12 MHz. We reduced the data using the \texttt{TCLEAN} algorithm in CASA v6.4.1.2, with natural weighting and binned to a spectral resolution of 5 km s\up{-1}. The spectrum was extracted from a region matching the most extended emission, in this case CO  (2--1). Lines were identified using the the Cologne Database for Molecular Spectroscopy \citep[CDMS;][]{cdmsref} and the Jet Propulsion Laboratory database \citep[JPL;][]{jplref}, available in the CASSIS spectroscopic analysis tool \citep{cassis}; previous millimetre-wave studies \citep[e.g.][]{edwards2014,schmidt2019carbonchem} also assisted in line identification.

Our clean beam size at $\sim$230 GHz (i.e., near CO 2--1) was 1\farcs51$\times$1\farcs12 in natural weighting. This means that our SMA observations are the first to spatially resolve the molecular environment of K4-47 at millimetre wavelengths. We used the CASA routine \texttt{uvcontsub} to separate and subtract the continuum emission from the emission lines.

\subsection{VLA}
K4-47 was observed with the Karl G. Jansky Very Large Array (VLA) on 09/10/1983 at 4.86 GHz in array configuration A (PI: S. Kwok, project code AK94), using 25 antennas with baselines ranging between 25--350 m. Another dataset observed in 1984 as part of this project was published within \citet{aaquist1990}. The clean beam size for the VLA data is 0\farcs66$\times$0\farcs40 at natural weighting, and the total integration time was 250 s. The data was extracted from the NRAO archive\footnote{https://data.nrao.edu/portal/}. The original data was observed using FK4 coordinates, which were converted to FK5, using \texttt{astropy}.

\section{Results}\label{sect: results}
\subsection{Optical imaging} \label{sect: imaging results}
\begin{figure}
    \centering
    \includegraphics[trim=0 0 0 0, scale=0.25]{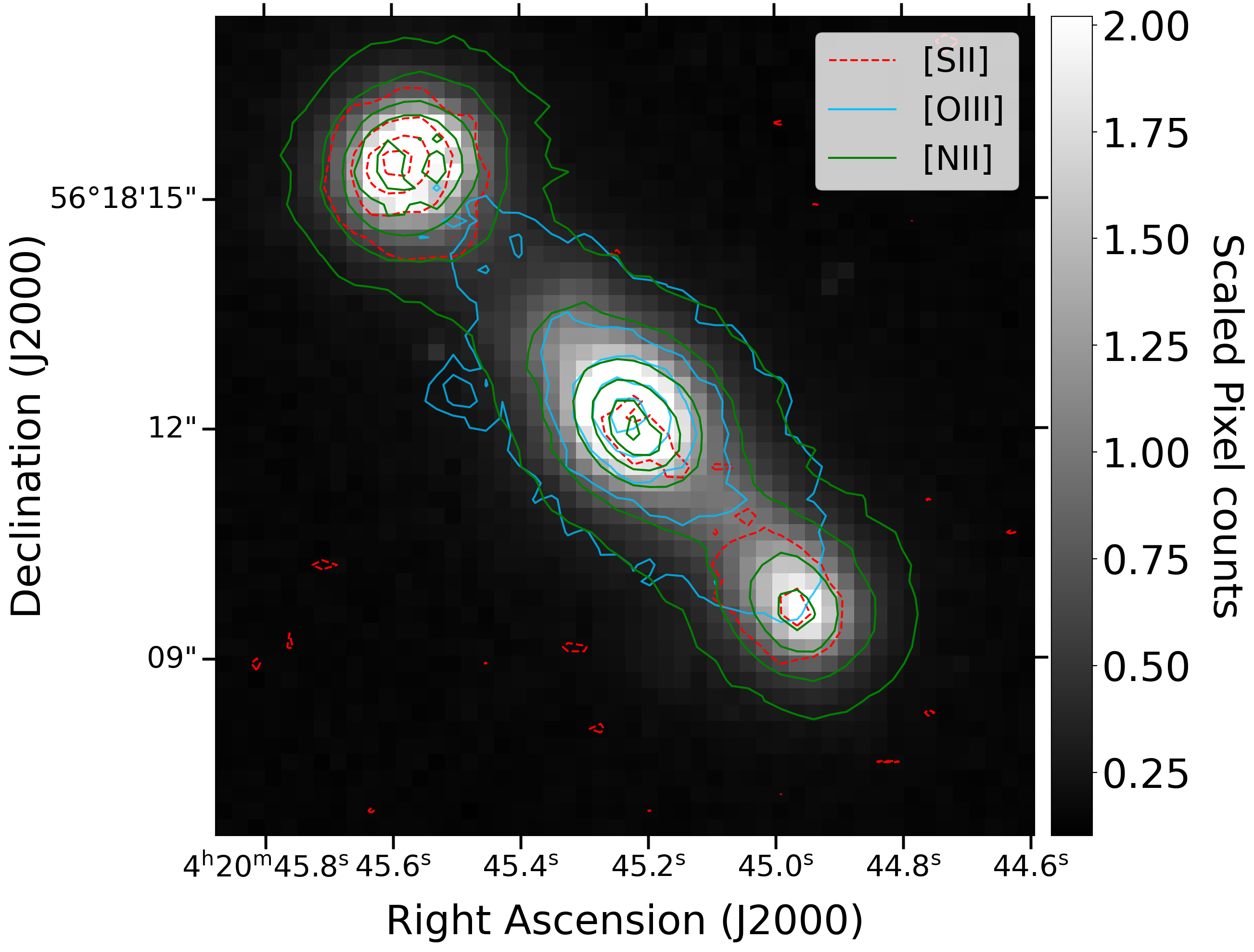}
    \caption{H$\alpha$ image combined over all epochs, overlaid with contours at 20, 40, 60, 80, and 95\% of the peak flux for {[\ion{S}{II}]} (red) and {[\ion{N}{II}]} (green). {[\ion{O}{III}]} contours are overlaid (blue) at 10, 20, 40, 60, 80, and 95\%  of the peak {[\ion{O}{III}]} flux. The contours of the emission lines are taken from the images combined over all epochs.}
    \label{fig: total filter + contours}
\end{figure}

Our images of H$\alpha$, [\ion{N}{II}], and [\ion{S}{II}] (Fig. \ref{fig: total filter + contours}) show the core and bipolar structures as previously seen by \citet{corradi2000} and \citet{goncalves2004}, with the core being only weakly detected in [\ion{S}{II}]. The [\ion{O}{III}] images, in contrast, only show the core emission. We used IRAF routines to combine all available epochs for each narrowband filter to create a single image per filter. This was done by normalising each epoch image to the mean of peak counts measured from five nearby stars seen in all ALFOSC images. Figure \ref{fig: total filter + contours} shows the total H$\alpha$ image, with contours from the total {[\ion{N}{II}]}, {[\ion{S}{II}]}, and {[\ion{O}{III}]} images overlaid. The image shows that H$\alpha$ and {[\ion{N}{II}]} trace each other well spatially. The only slight difference between our ALFOSC images and those of \citet{corradi2000} is that the latter detect clearer connecting emission in [\ion{S}{II}] than seen in our data. This is likely due to observing conditions, rather than an evolutionary difference.

\begin{figure}
    \centering
    \includegraphics[trim=0 0 0 0, scale=0.25]{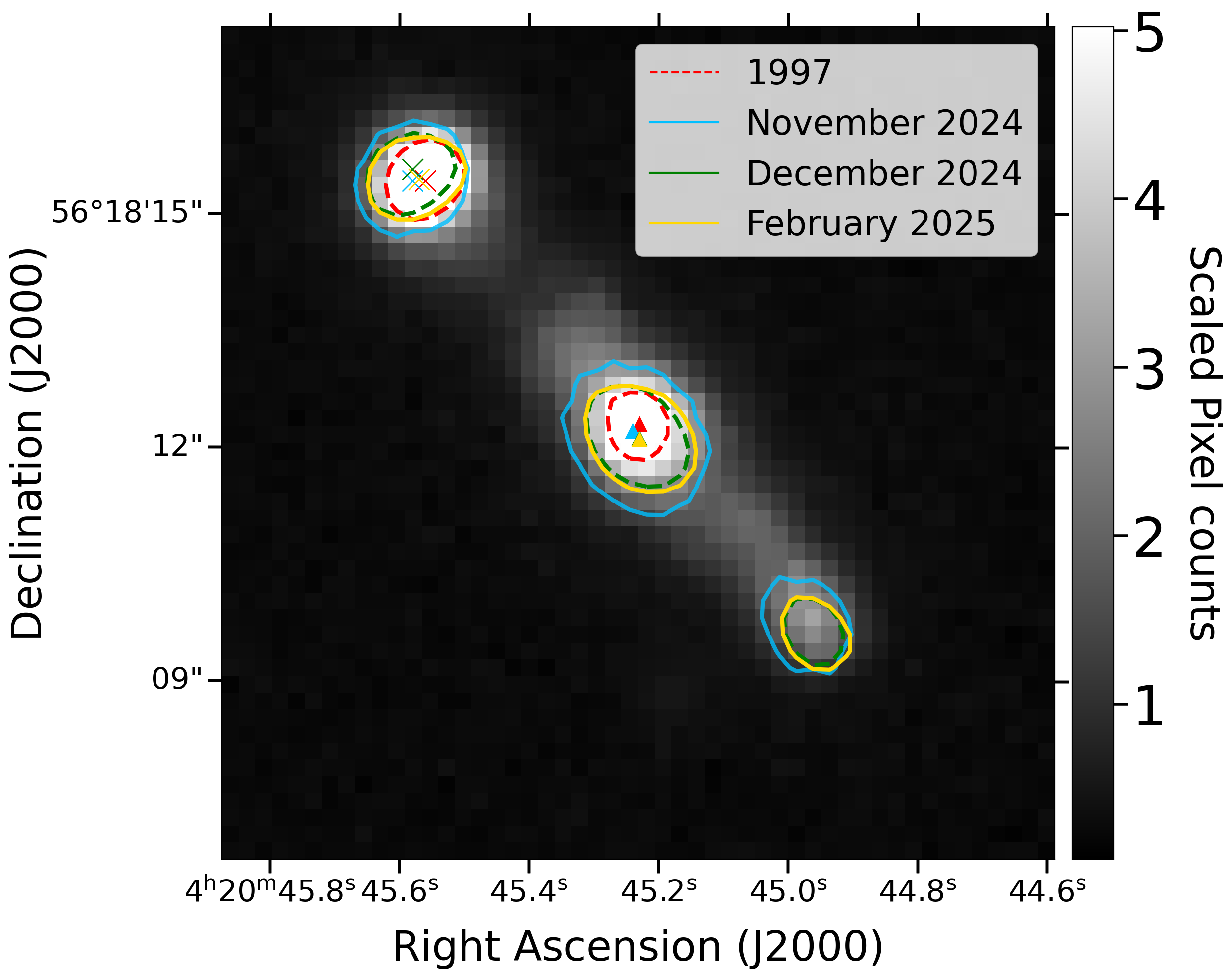}
    \caption{H$\alpha$ image from 1997 with 50\% contours overlaid for H$\alpha$ from each epoch: 1997 (red), November 2024 (blue), December 2024 (green), 2025 (yellow). The light-blue star symbol is located at the coordinates of K4-47, the triangles represent the fitted centre of the corresponding contours for the core, and the crosses show the centres of the corresponding contours for the northern lobe.}
    \label{fig: halpha expansion}
\end{figure}

To determine the expansion/outflow velocity of the lobes that dominate the nebula, we measured the centre of the 50\% level contours of the maximum counts measured for the core and northern lobe in H$\alpha$ for the 1997 and December 2024 epochs (shown in Fig. \ref{fig: halpha expansion}), which had the best seeing out of all available epochs. The 50\% contour level was chosen as, at this level, we have distinct contours of both the northern lobe and the core. Assuming a distance of 5.9 kpc \citep{tajitsu1998}, and an inclination angle to the observer of 67.5\degree\ \citep{corradi2000}, we calculate the deprojected distance between the core and northern lobe for each epoch and calculate the expansion velocity. The measured angular distance change was 0.35\arcsec, from 4\farcs16 in September 1997 to 4\farcs52 in December 2024, which at a distance of 5.9 kpc gives a physical separation change from 2.45\standform{4} AU in September 1997 to 2.67\standform{4} AU in December 2024. The resulting deprojected expansion velocity is 382 $\pm$ 135 km s\up{-1}. This is broadly consistent with the shock velocities of 250--300 km s\up{-1} from \citet{goncalves2004}. We then calculate the kinematical jet age of the optical bipolar outflow as 336 $\pm$ 119 yr, assuming constant velocity. The jet age is independent of distance. This is comparable to the nebula age of 400--900 yr given by \citet{corradi2000}, assuming a distance range of 3--7 kpc. 


\subsection{Optical spectroscopy} \label{sect: spectroscopy results}
\begin{figure}
    \centering
    \includegraphics[trim=0 0 0 0, scale=0.2]{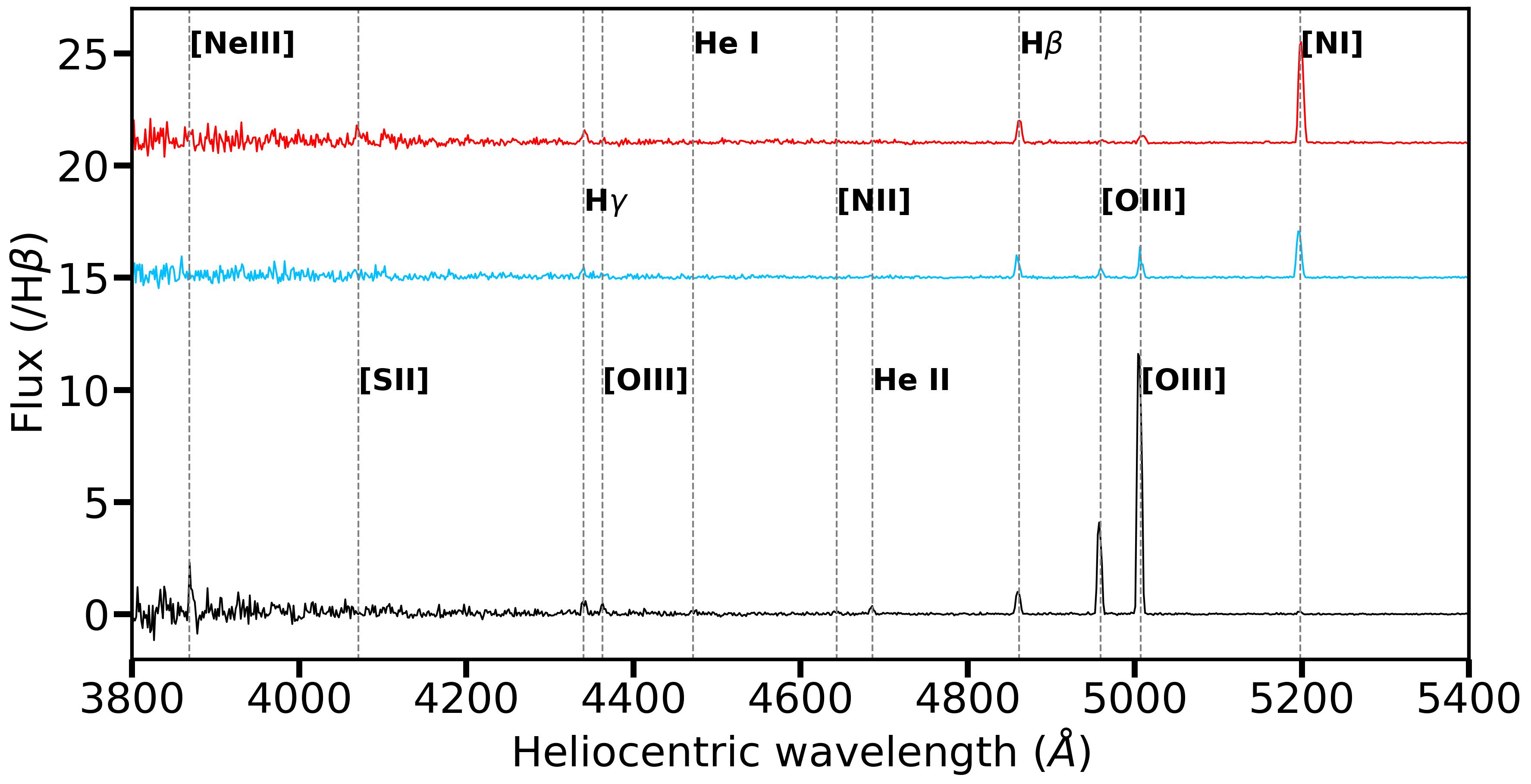}
    \includegraphics[scale=0.2]{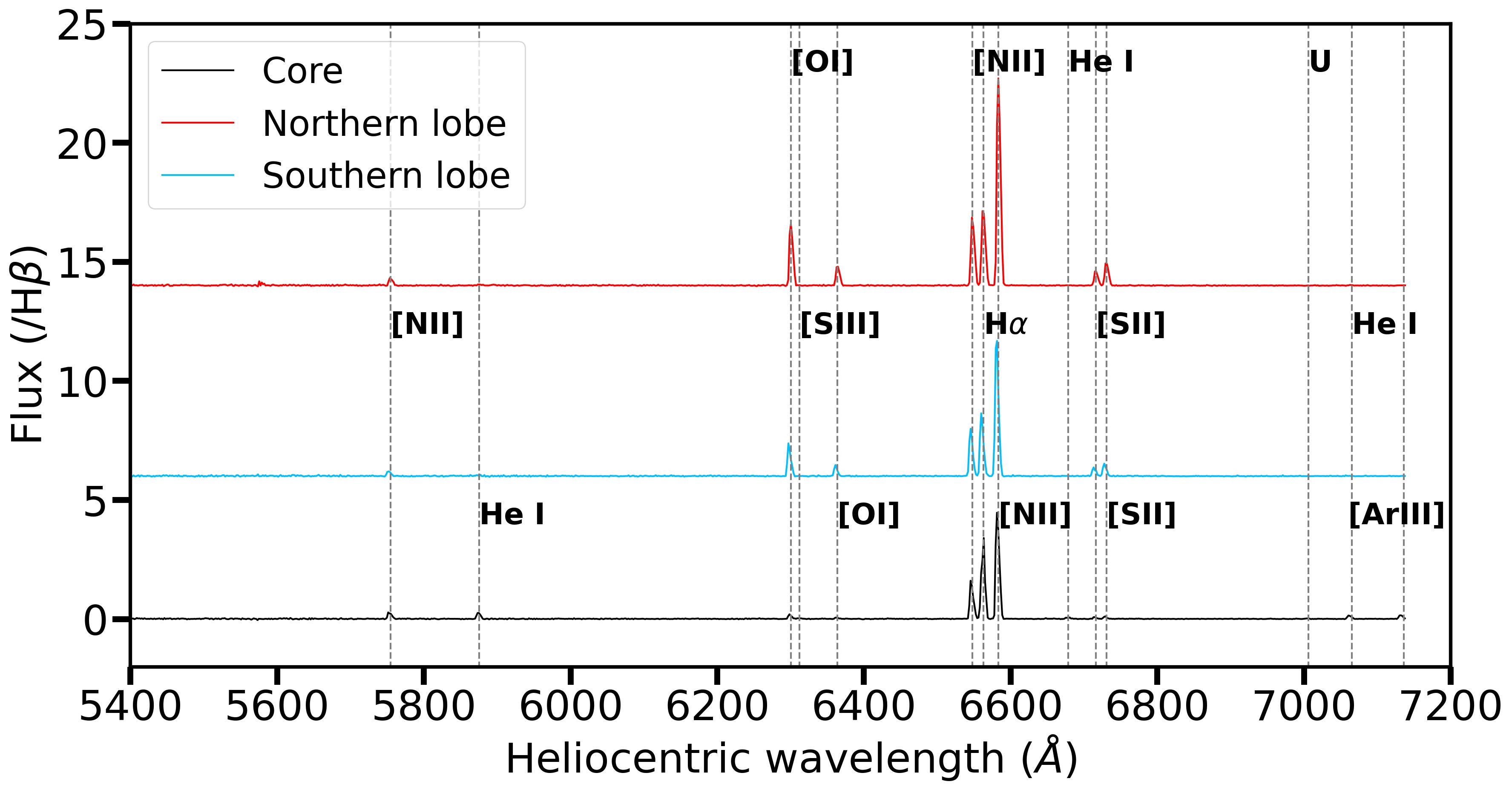}
    \caption{Spectra extracted for each component of K4-47 (see text). Grey dashed lines and annotations indicate rest wavelengths of atomic lines seen in the spectrum.}
    \label{fig: alfosc spec}
\end{figure}

In our optical ALFOSC spectra (Fig. \ref{fig: alfosc spec}), we detect a total of 24 atomic emission lines across the core and bipolar lobes. Many of these lines have been previously detected in K4-47 \citep{corradi2000,goncalves2004}. The strongest lines in the core are the [\ion{O}{III}]$\lambda\lambda$4959,5007 and [\ion{N}{II}]$\lambda\lambda$6548,6584 doublets, as well as the Balmer H$\alpha$ and H$\beta$ lines. For the lobes, the [\ion{N}{II}] doublet and H$\alpha$ are the strongest, followed by [\ion{N}{I}]$\lambda$5198 and [\ion{O}{I}]$\lambda$6300, with the [\ion{O}{III}] lines relatively weak compared to the core. [\ion{S}{II}] is also much weaker in the core compared to the lobes. These differing line strengths are consistent with the distribution of the ionised atomic gas shown in Fig. \ref{fig: total filter + contours}, such as the [\ion{S}{II}] emission found only in the lobes, but is mostly undetected in the core. We also include the tentative identification of the [\ion{Ar}{V}]$\lambda7005$ line, detected at 7004.18 \AA, which was identified by \citet{henry2010}. The tentative nature is marked with a `?' in Table \ref{tab: alfosc line fluxes}, as [\ion{Ar}{V}] is usually expected in PNe with very hot central sources \citep[e.g. NGC 7027, Wray 17-1, K1-2;][]{zhang2005,akras2016}, and the line could be instead [\ion{O}{I}]$\lambda7002$. We favour the [\ion{Ar}{V}] identification due to the previous identification in the literature, as well as the fact that [\ion{Ar}{V}] would have a more consistent line velocity with other lines than [\ion{O}{I}].

The range of measured line widths for the detected emission lines exceeds the instrumental broadening of 461 km s\up{-1}. Therefore, no kinematical information could be extracted from our ALFOSC spectrum. Parameters of all detected lines are listed in Table \ref{tab: alfosc line fluxes}. 

\paragraph{Reddening}
\citet{goncalves2004} identify significant interstellar extinction towards K4-47 across all three components \citep[C\down{H\beta}$\approx$1.1--1.3, equivalent to E(B--V)=$\approx$0.75--0.88 mag, following the extinction law of][hereafter referred to as CCM89]{ccm89}. 
Determining the physical properties of the central source in K4-47 is dependent on obtaining strong constraints of the reddening towards the source. To examine the reddening in our spectrum, we use the Balmer decrement
\begin{equation}\label{eq: balmer decrement}
    {\rm E(B-V)} = \frac{2.5}{\kappa({\rm H}\beta)-\kappa({\rm H}\alpha)}\quad{\rm log}_{10}\left(\frac{({{\rm H}\alpha/{\rm H}\beta})_{\rm obs}}{2.86}\right).
\end{equation}
Here, $\kappa$($\lambda$) refers to the CCM89 extinction curve coefficient. The ratio H$\alpha$/H$\beta$ is the flux ratio of the respective lines (see Table \ref{tab: alfosc line fluxes}). $\kappa({\rm H}\beta)$ and $\kappa({\rm H}\alpha)$ are taken to be 3.61 and 2.53, respectively, which are specific to the CCM89 extinction law. Our measured E(B--V) for the core, northern lobe, and southern lobe, are measured as 1.28 $\pm$ 0.01, 0.97 $\pm$ 0.01, and 0.76 $\pm$ 0.02 mag, respectively, using the measured H$\beta$ and H$\alpha$ fluxes for each component in our ALFOSC spectra (see Table \ref{tab: alfosc line fluxes}).

The extinction for the northern and southern lobes is consistent with the derived logarithmic extinction of \citet{goncalves2004}. The northern lobe shows higher extinction than the southern lobe, which could be explained by the northern lobe being farther away from the observer and therefore may be located behind additional interstellar dust along the line of sight. However, the core shows the highest extinction of the three components. It is not clear if the increased extinction towards the core is only interstellar, or if there is a contribution from circumstellar dust. To test this, we used the GALE{\footnotesize XTIN}\footnote{http://www.galextin.org/} VO service \citep{amores2021} to examine the expected interstellar reddening in the direction of K4-47 based on multiple simulated dust maps. Our reddening analysis is described in Appendix \ref{appendix: reddening}. We extracted interstellar reddening estimates for distances between 3--26.5 kpc \citep[the range of distance estimates for K4-47: ][]{aaquist1990,cahn1992,vandesteene1994,zhang1995,tajitsu1998, corradi2000} for several dust maps \citep{amores2005,sale2014,green2015,green2018,green2019}. Most dust maps we examined predict $\approx$0.55--0.75 mag of interstellar extinction towards K4-47, with model S from \citet{amores2005} predicting a value closer to 1.0 mag. The average range of extracted E(B--V) values from the dust models at the minimum and maximum distances are in the range of 0.62--0.75 mag, which represents the interstellar component of our measured value of E(B--V). The extinction towards the lobes are within 3$\sigma$ of the interstellar extinction range predicted by the various dust maps, confirming the expected interstellar extinction contribution for K4-47. However, the extinction towards the core exceeds the dust maps' estimate. This means $\sim$0.3--0.6 mag of the measured extinction is from circumstellar dust for the core of K4-47.

\paragraph{Temperatures and densities}
Despite the depth of literature on K4-47, there are no strong constraints on the central source properties of K4-47. The effective temperature (T$_{\rm eff}$) of the central ionizing source was previously estimated in the range of 115--130 kK by \citet{goncalves2004}, using the Zanstra method. This method assumes photoionisation only, and neglects the effect of shocks. However, subsequent \texttt{CLOUDY} modelling using this value of T$_{\rm eff}$, as well as the derived abundances lead to an underestimation of certain lines such as [\ion{O}{III}]$\lambda$4363 and [\ion{N}{II}]$\lambda$5755, which are frequently used to estimate the electron temperature. This discrepancy makes the derived abundances and T$_{\rm eff}$ uncertain, but remains the only estimate of the effective temperature for the central source in K4-47 to date. We first estimate T$_{\rm eff}$ of the central source using the relation between T$_{\rm eff}$ and the [\ion{O}{III}]$\lambda$5007/H$\beta$ ratio \citep[see][their Fig. 7]{lumsden2001}. From our dereddened spectrum for the core, we see that [\ion{O}{III}]$\lambda$5007/H$\beta$ = 11.74 $\pm$ 0.14. Using the blackbody models in \citet{lumsden2001}, we get a range of T$_{\rm eff}\sim\ $60--90 kK. The implied value of T$_{\rm eff}$ from the linear fit to the \citet{kaler1991} data in the same figure is $\sim\ $65 kK.

To estimate the electron temperature and density (T\down{e}, n\down{e}), we use the PyNeb emission line tool \citep{luridiana2015}. We utilise the cross-convergence function \texttt{pyneb.Diagnostics.getCrossTemDen} to calculate the best-fitting electron T\down{e} and n\down{e} for emission line diagnostics. We use the forbidden line ratios of [\ion{N}{II}]5755/6548 and 5755/6584, and [\ion{O}{III}]4363/5007, which are all sensitive to electron temperature, and the electron density-sensitive forbidden line ratio [\ion{S}{II}]6731/6716. We find that the diagnostics involving the [\ion{N}{II}] ratios give electron temperatures exceeding 20000 K, which is the upper sensitivity limit for these lines \citep[see Chapter 5 and Fig. 5.1 of][]{osterbrockferland2006}. We therefore only use the results of the diagnostics of the [\ion{O}{III}] and [\ion{S}{II}] ratios, which gives T\down{e}=19900 $\pm$ 1200 K and n\down{e}=2800 $\pm$ 700 cm\up{-3}. Our value of T\down{e} is consistent with the [\ion{O}{III}]-derived T\down{e} (19900 K) of \citet{mari2023}. We then use these values, along with the He II $\lambda$4686/H$\beta$ ratio, to derive the Zanstra temperature. We calculate the blackbody photon flux ratio Q(He\up{+})/Q(H\down{0}) for a given T$_{\rm eff}$, and calculate the observed photon flux ratio for our measured line fluxes, using Case B recombination coefficients of \citet{storey1995}, and find the minimisation between the difference of the theoretical and observed photon flux ratios using Brent's method. This method yields a Zanstra effective temperature for the central source of 81500 $\pm$ 1800 K.

\paragraph{Mass} As shown in Table \ref{tab: alfosc line fluxes}, we detect multiple transitions from ionised atomic gas in all three morphological components. We calculate the ionised gas mass (M\down{ion}) present in the core from
\begin{equation}
    {\rm M_{ion}} = \frac{{\rm m_p L}_{H\alpha}}{{\rm h\nu_{H\alpha}} \alpha^{\rm eff}_{\rm H\alpha} {\rm n_e}},
    \label{eq: ionised gas mass}
\end{equation}
where m\down{\rm p} is the mass of a proton (1.67\standform{-27} kg), L\down{\rm H\alpha} is the luminosity of the H$\alpha$ line for the core, $\nu_{\rm H\alpha}$ is the frequency of the H$\alpha$ line, and $\alpha^{\rm eff}_{\rm H\alpha}$ is the effective recombination coefficient for H$\alpha$, which at T\down{e}=19900 K is 6.31\standform{-14} cm\up{3} s\up{-1} \citep{osterbrockferland2006}. We calculate M\down{ion} for full range of estimated distances from the literature (3--26.5 kpc). This gives a range of values for M\down{ion} between 6.93\standform{-4}--5.21\standform{-2} M\solar\ for the core, 2.32\standform{-4}--1.74\standform{-2} M\solar\ for the northern lobe, and 9.65\standform{-5}--7.24\standform{-3} M\solar\ for the southern lobe. In Sect. \ref{sect: sed fitting}, we provide distance constraints and the corresponding ionised masses.


\subsection{SMA and H$_2$} \label{sect: sma results}
\begin{figure}[t]
    \centering
    \includegraphics[width=0.95\linewidth]{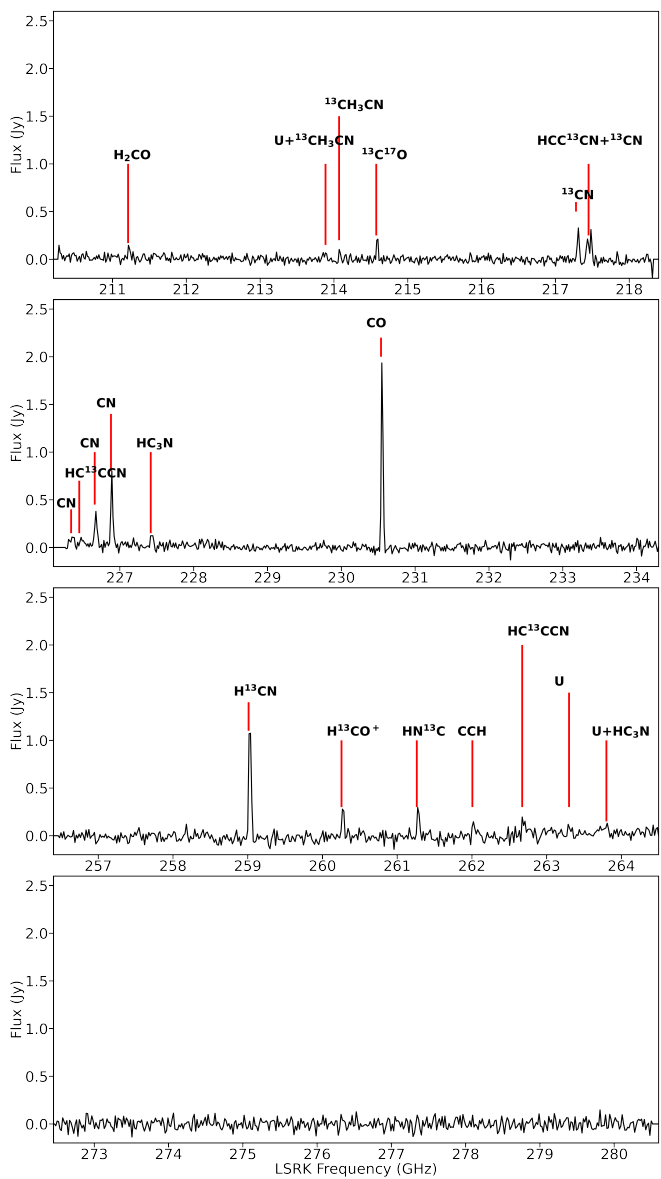}
    \caption{Continuum-subtracted SMA spectrum of K4-47 after rebinning to 20 km s\up{-1} for easier visual inspection. It represents a region encompassing the entire CO (2--1) emission (see Fig. \ref{fig: moment 0 maps 5 sigma}), i.e., the most extended molecular emission observed.}
    \label{fig: sma spec}
\end{figure}

Previous sub-millimetre and millimetre-wave studies of K4-47 \citep{huggins2005,edwards2014,schmidt2016hcnhco,schmidt2017cch,schmidt2017hcnhnc,schmidt2018isotope,schmidt2019carbonchem} have used single dish observations from facilities such as the IRAM 30m telescope, the Arizona Radio Observatory (ARO), and the Sub-Millimeter Telescope (SMT) to detect various molecules in K4-47. In our SMA observations, which spatially resolved the molecular CSM for the first time, we find emission of previously detected molecules such as CO, HC\down{3}N, CCH, CN, and H$_2$CO. We also detect several isotopologue lines of detected molecules, including H\up{13}CN, H\up{13}CO\up{+}, and \up{13}C\up{17}O. All detected lines are listed in Table \ref{tab: SMA line fluxes}. Our sensitivity was insufficient to detect several weak lines (e.g., of SiO) which were previously detected at the covered frequency range within the single-dish surveys. Using the CO (2--1) line, we used the CASA routine \texttt{imfit} to measure a beam-deconvolved source size of (2\farcs36$\pm$0\farcs10) $\times$ (1\farcs12$\pm$0\farcs08).  Our SMA data is not adequate to perform any excitation analysis of the observed molecular emission, as we did not cover multiple transitions of the same molecular species. For source averaged line fluxes, this has been done in \cite{schmidt2019carbonchem} and \cite{edwards2014}, where column densities for main species were derived. They also derived the hydrogen density of roughly 10$^{5}$--10$^6$ cm$^{-3}$ and a temperature of 10--65 K. The SMA spectrum, re-binned to 20 km s\up{-1} for better signal-to-noise ratio (S/N), is shown in Fig. \ref{fig: sma spec}.

To establish the molecular source morphology, we made integrated intensity maps of the detected lines. The exceptions were HC$^{15}$N and HCC$^{13}$CN, which are apparent in the spectrum, but no clear source was visible in the integrated flux maps. For the strongest lines (CO 2--1 and H\up{13}CN 3--2), we also made velocity-field maps (moment-1 maps), where signal below the 3$\sigma$ and 5$\sigma$ noise level was clipped (see Table \ref{fig: moment 0 maps 3 sigma}).

\begin{figure}[t]
    \centering
    \includegraphics[scale=0.19]{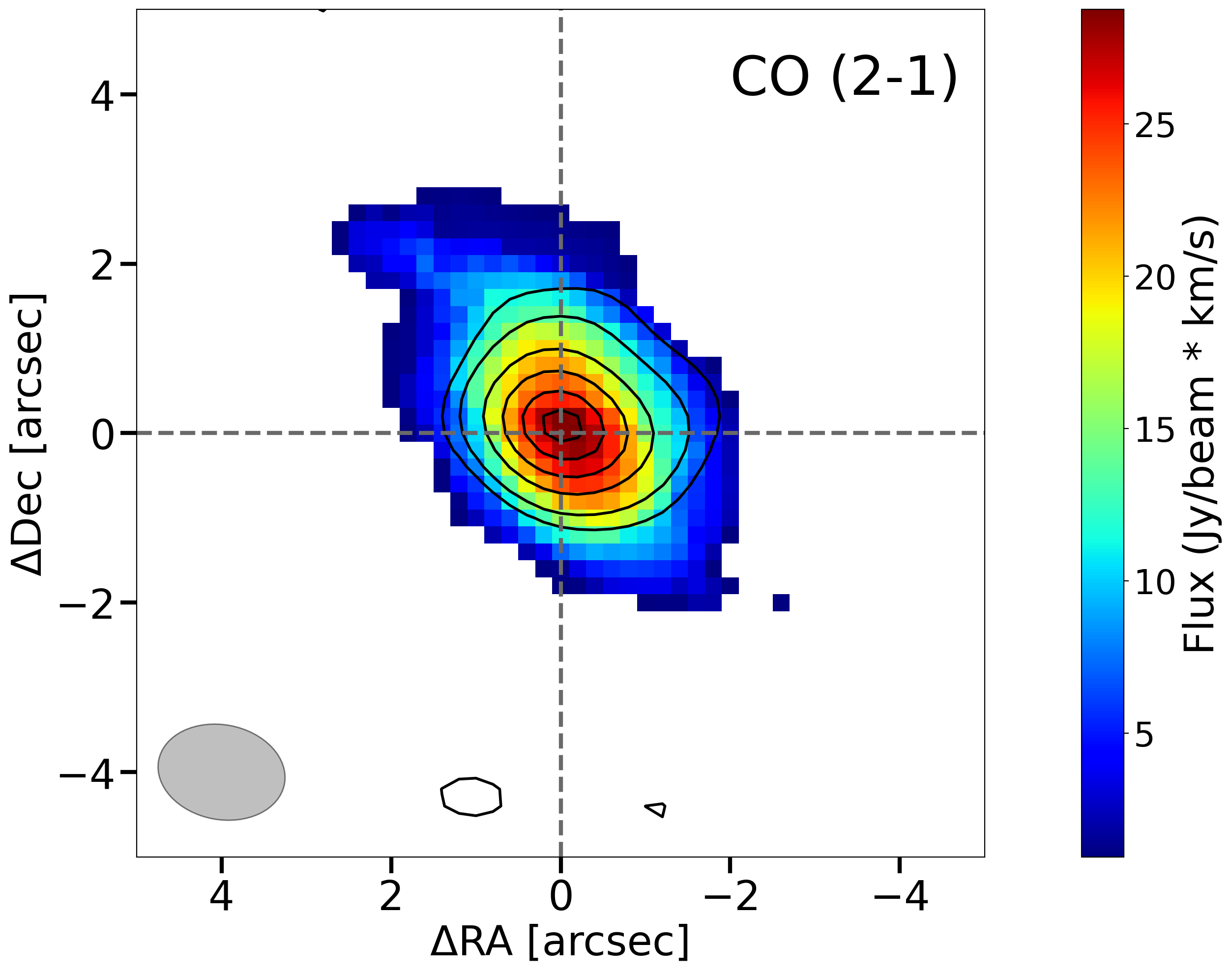}
    \includegraphics[scale=0.19]{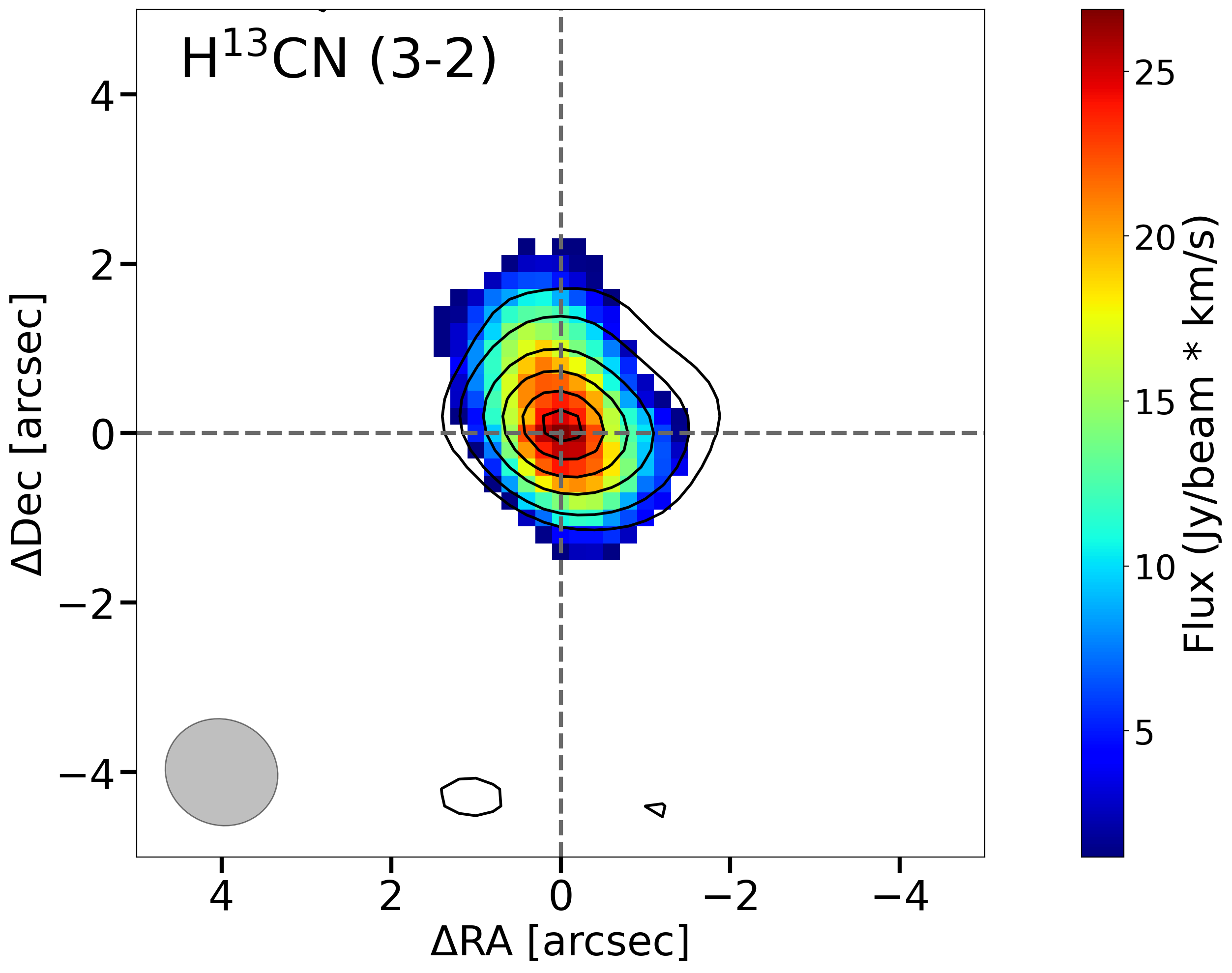}
    \includegraphics[scale=0.19]{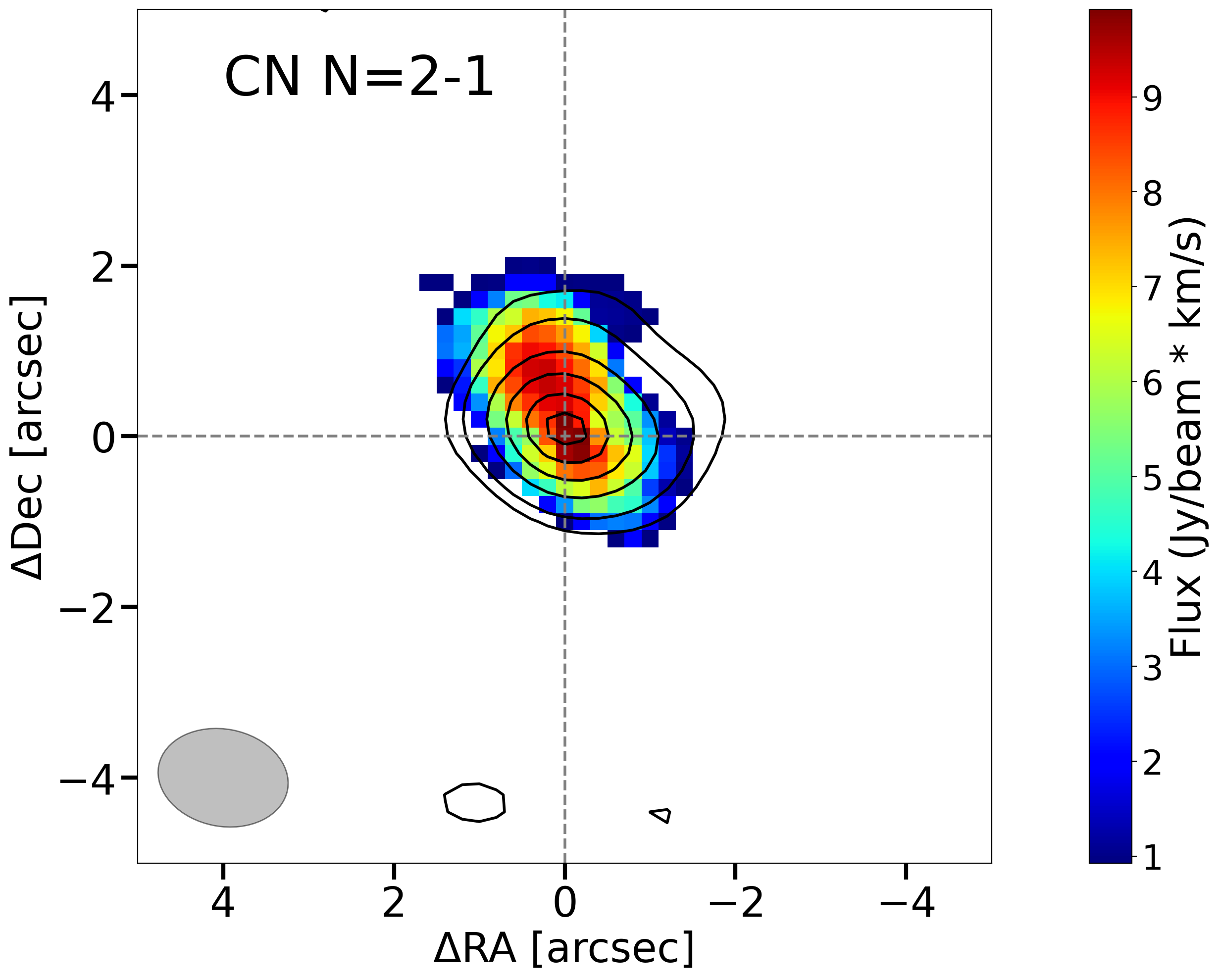}
    \caption{Integrated flux maps of CO $J$=2--1 (left), H$^{13}$CN $J$=3--2 (centre), and CN $N$=2--1 (right). All maps have noise clipped at the 5$\sigma$ level. Black contours are 10, 20, 40, 60, 80, and 95\% of the peak continuum flux. The grey ellipse represents the synthesised beam for the molecular transition.}
    \label{fig: moment 0 maps 5 sigma}
\end{figure}
\begin{figure}[t]
    \centering
    \includegraphics[width= \linewidth]{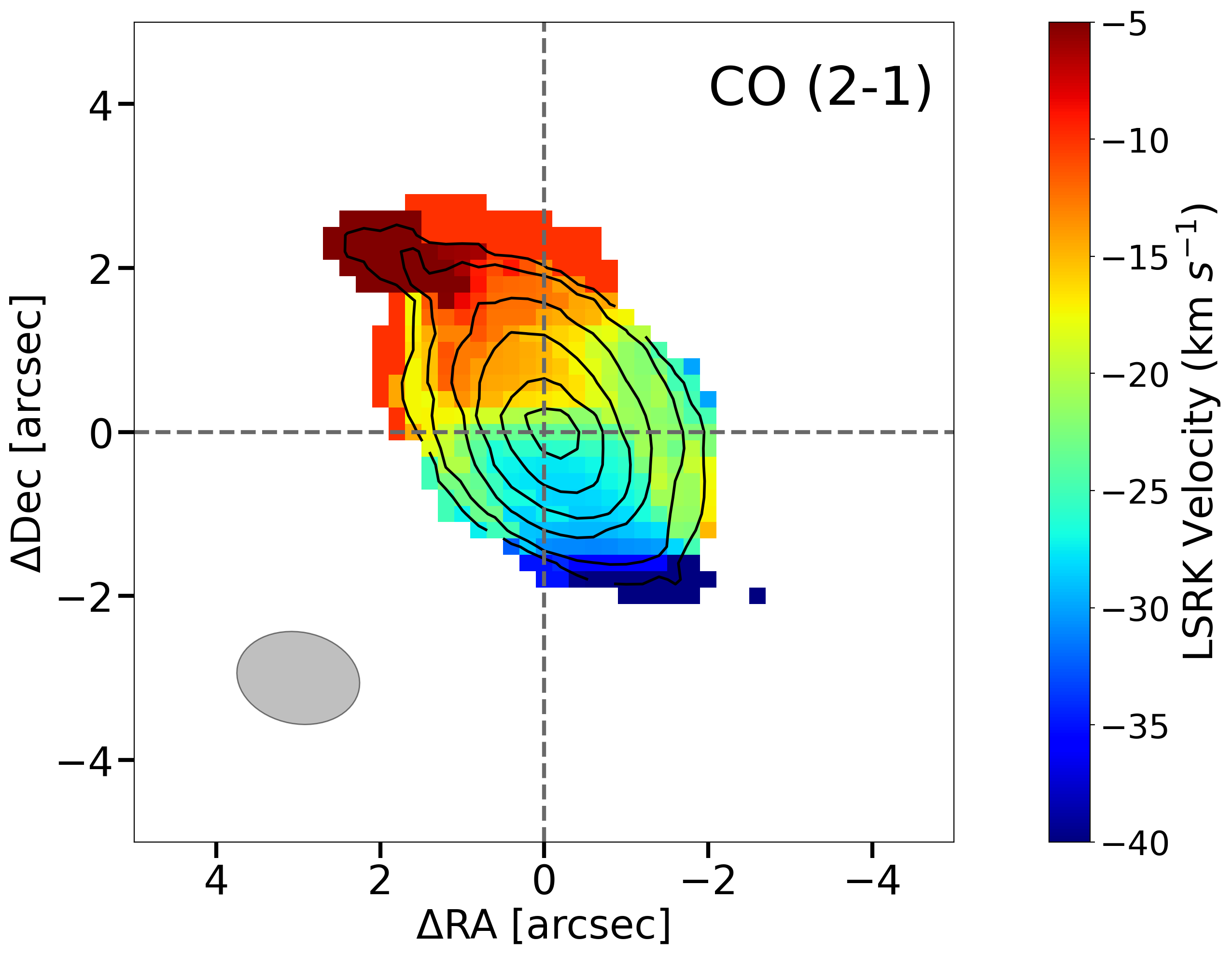}
    \includegraphics[width= \linewidth]{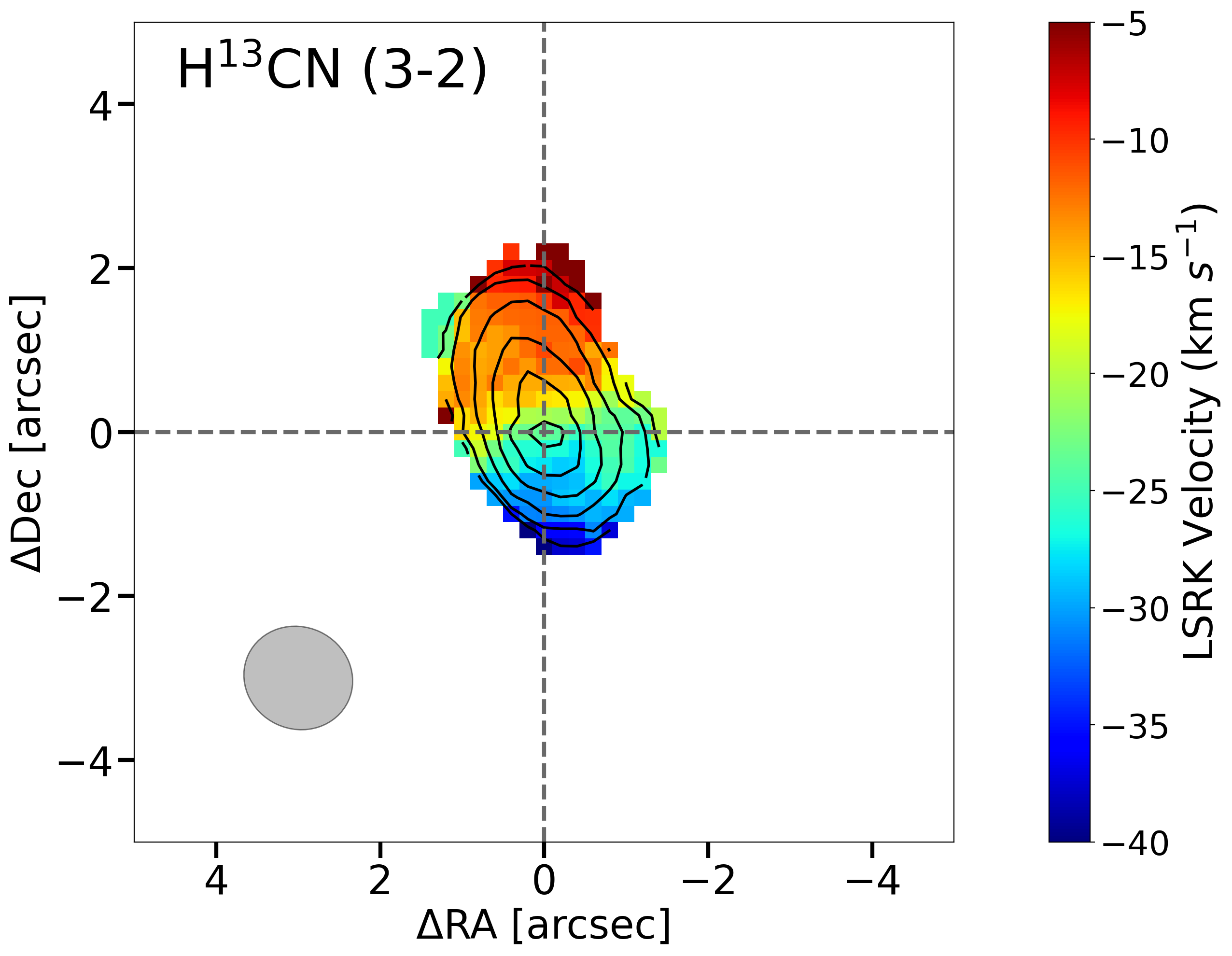}
    \caption{Integrated velocity maps of CO (2--1) and H$^{13}$CN (3--2). Dashed grey lines show the map centre, and grey ellipses show the synthesised beam for each emission line. For the CO map, the velocity range has been set to match that of H\up{13}CN for comparison, even though the maximum redshifted velocity exceeds this range.}
    \label{fig: moment 1 maps}
\end{figure}

We present the integrated flux maps in Figs. \ref{fig: moment 0 maps 5 sigma} and \ref{fig: moment 0 maps 3 sigma} (with sigma clipping at 5$\sigma$ and 3$\sigma$, respectively), and the integrated velocity maps in Fig. \ref{fig: moment 1 maps}. The maps are shown with spatial offsets calculated relative to the peak position of the H\up{13}CN emission, which is not optically thick or contaminated by multiple lines (unlike CO and CN, respectively). The majority of the emission shows an elongated structure compared to the continuum emission, with the exception of H$ ^{13}$CO$^+$ (3--2) and HN$^{13}$C (3--2), which shows a more compact distribution. The elongated structures span from north-east to south-west, at a similar PA to that seen in the optical and near-infrared H$_2$ maps (see Figs. \ref{fig: total filter + contours} and \ref{fig: h2+sma}). The elongation therefore appears to be, at least partially, shaped by the outflows/bullets seen in the optical images (Fig. \ref{fig: total filter + contours}). CO emission is seen to extend farther towards the northern lobe, forming a bridge between the northern lobe and the core that is also seen in the optical images (see Fig. \ref{fig: h2+sma}). The extended part of CO is also seen in the velocity maps of CO (Fig. \ref{fig: moment 1 maps}), where the north-east tip of the CO emission suddenly changes velocity. The extent of the other strong lines (e.g., CN J=$\frac{5}{2}$--$\frac{3}{2}$ and H\up{13}CN 3--2) is less than that for CO, and both extend to approximately half of the angular distance to the northern lobe. No such extension is seen towards the southern lobe. The integrated velocity maps for CO, CN, and H$^{13}$CN show that the northern lobe is redshifted, and the southern lobe is blueshifted. Using the moment-0 map of CO (2--1) shown in Fig. \ref{fig: moment 0 maps 5 sigma}, we measured the CO flux in the core and northern lobe using a 2\farcs0 aperture. We found that the extended CO emission that overlaps partially with the northern lobe constitutes 8.7\% of the core CO flux.

The CO (2--1) emission shows a velocity gradient along a PA of 41\degree. The H\up{13}CN velocity map shows a similar structure at a PA of 13\degree. We made position-velocity (PV) diagrams of the CO (2--1) emission, at PA=41\degree\ \citep[the PA of the atomic outflow; ][]{goncalves2004} and 131\degree, measured from north through east, using the CASA routine \texttt{impv}. These PV diagrams are shown in Fig. \ref{fig: pv diagrams}. The PV diagram along the outflow PA shows a clear outflow structure in both directions, with the outflow on the southern side weaker than the north. The perpendicular PV diagram (PA=131\degree) shows no clear substructure.

Our SMA data use the local standard of rest (LSR) velocity rest frame. The maximum LSR velocity seen in the CO velocity map in the northern direction is 34.6 km s\up{-1}, in the form of a high-velocity tip in the north-east part of the CO (2--1) velocity map (fig. \ref{fig: moment 1 maps}). We estimated the CO `systemic' velocity by taking the average velocity of the central CO region (neglecting the northern lobe and extended emission), and measuring the difference in velocity of the high-velocity tip and the central region. The central region's average velocity is --21.1 km s\up{-1}, giving a molecular outflow velocity of 55.7 km s\up{-1}. We measure the distance between the tip of the northern outflow and the central coordinates (marked by the grey cross in Figs. \ref{fig: moment 0 maps 5 sigma} and \ref{fig: moment 1 maps}) to be 3\farcs41. Using a distance of 5.9 kpc \citep{tajitsu1998} and an inclination angle $i$=67\fdg5 \citep{corradi2000}, we get a deprojected distance of 0.10 pc. Assuming a constant velocity expansion, this gives us a dynamical timescale of 1740 yr for the molecular outflow. It is thus older than the optical outflow component (Sect. \ref{sect: imaging results}). 

\begin{figure}
    \centering
    \includegraphics[width=0.95\linewidth]{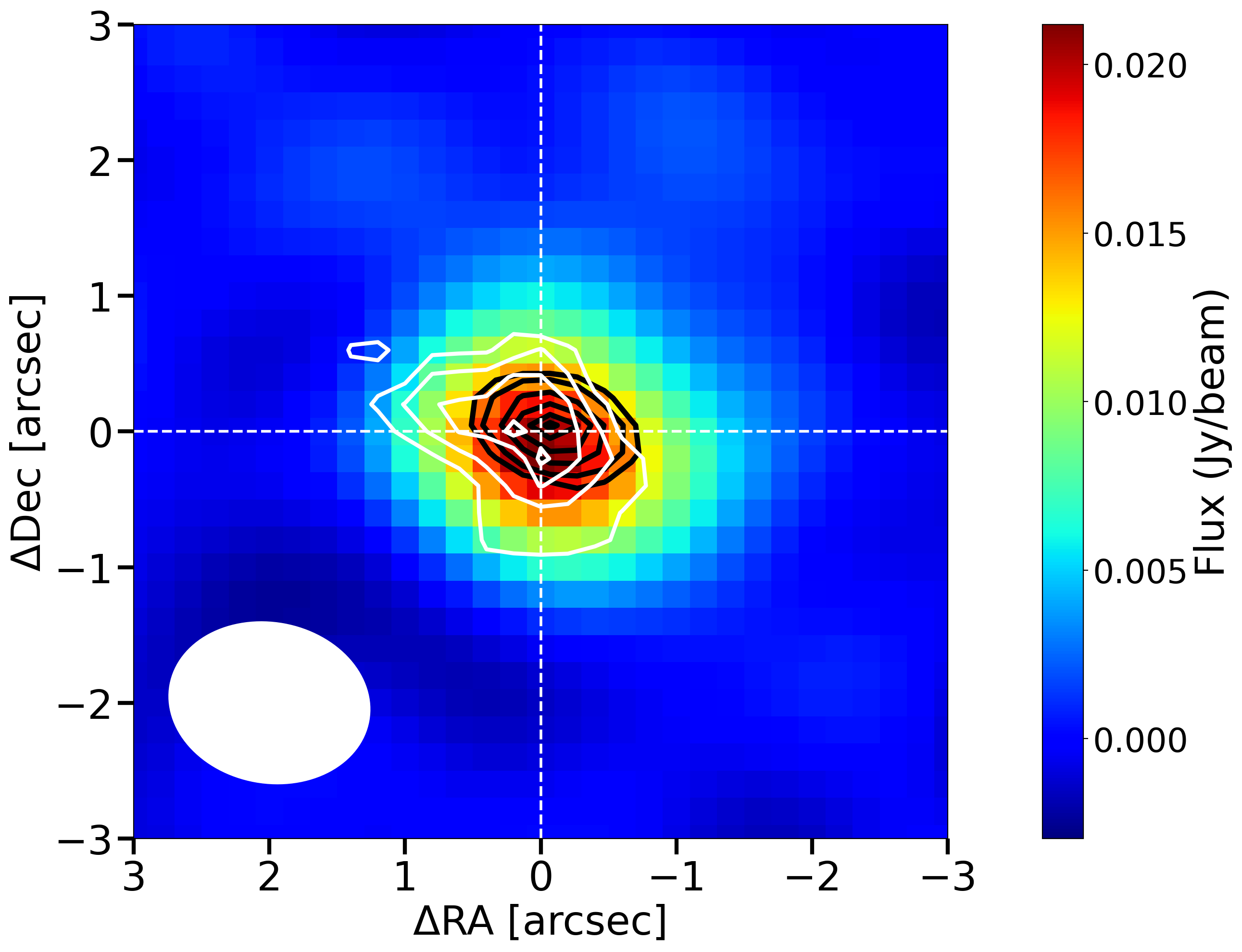}
    \caption{Continuum image of K4-47 observed with SMA near 250 GHz. The black contours are 10, 20, 40, 60, 80, and 95\% of the peak flux of the 1983 VLA observations at 5 GHz. The white contours are 40, 60, 80, and 95\% of the peak of the SMA \up{13}C\up{17}O integrated flux map (see Fig. \ref{fig: moment 0 maps 3 sigma}). The white ellipse indicates the SMA beam size.}
    \label{fig: sma continuum}
\end{figure}

\begin{figure}[h!]
    \centering
    \includegraphics[width=\linewidth]{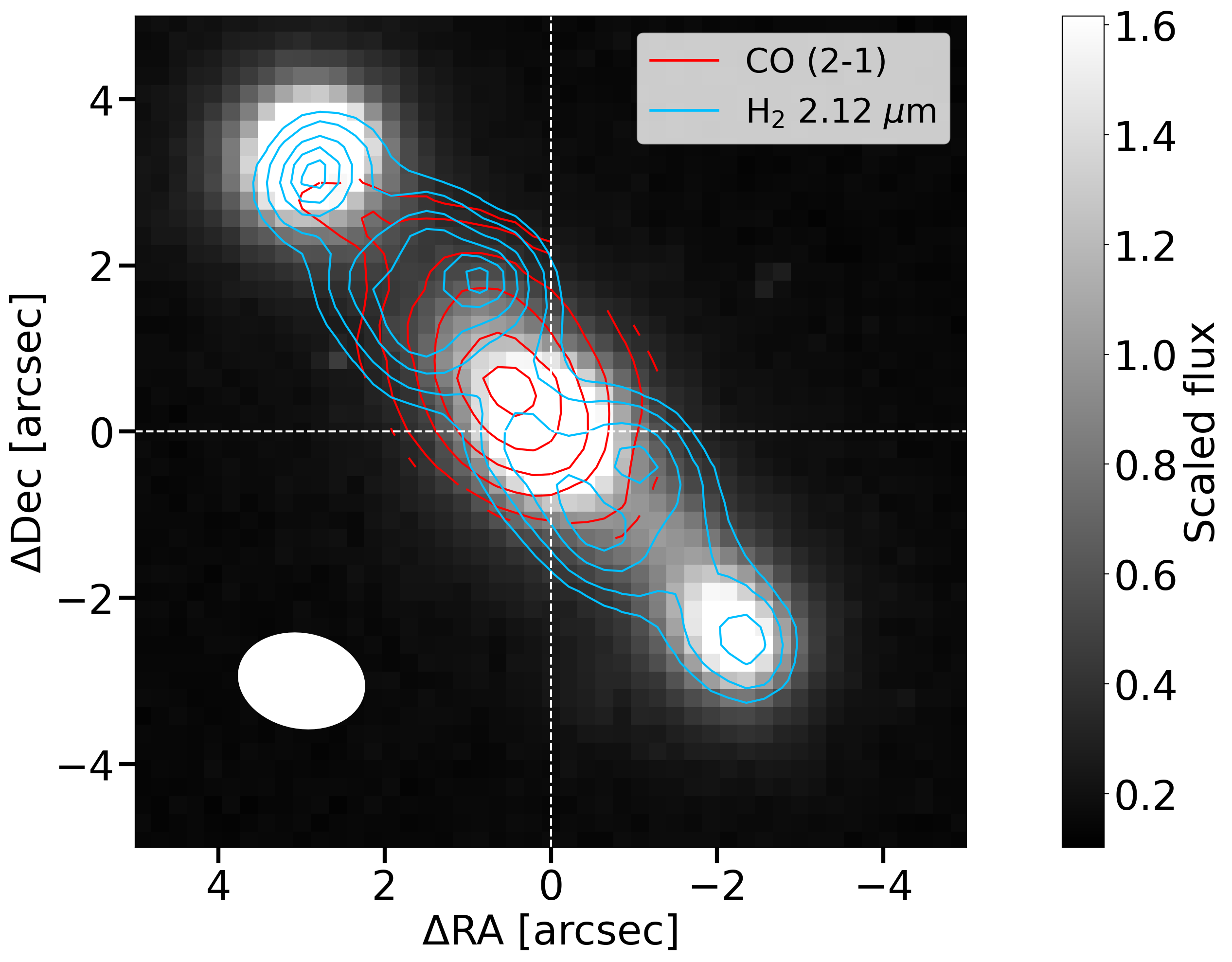}
    \caption{H$\alpha$ image of K4-47, with contours of H$_2$ from \protect{\citet{akras2017}} and CO (2--1) in blue and red, respectively. Contour levels are 10, 20, 40, 60, 80, and 95\% of the peak flux of the corresponding images. The SMA beam is shown in white in the lower right. The H$_2$ contours have been shifted by --1\arcsec\ in declination to account for misalignment in the astrometric solution of SMA and Gemini (see text). The white dashed cross indicates the centre of the H$\alpha$ image, and the white ellipse indicates the synthesised beam size of the CO (2--1) SMA observations.}
    \label{fig: h2+sma}
\end{figure}

We also made a multi-frequency synthesis (mfs) image using \texttt{tclean} to show the continuum emission (Fig. \ref{fig: sma continuum}). The continuum image displays a compact continuum source of comparable size to the radio source detected with VLA \citep[][see also Sect. \ref{sect-vla-results}]{aaquist1990}, and no extended emission. The image was constructed using the entire spectral window between 272.6--280.5 GHz (Fig. \ref{fig: sma spec}, bottom panel), as well as continuum regions in other spectral windows absent of emission lines. The full range of frequencies used were between 211.2--280.5 GHz. The equivalent frequency of the continuum image is 246.46 GHz and the measured flux is 30.6$\pm$2.1 mJy. The central position of the continuum is $\alpha$=04:20:45.2853, $\delta$=+56:18:12.7790. Using \texttt{imfit}, we get a beam-deconvolved continuum source size of (1\farcs02$\pm$0\farcs57) $\times$ (0\farcs78$\pm$0\farcs53).

As well as sub-millimetre observations of molecules with SMA, we examine archival continuum-subtracted H\down{2} 1--0 S(1) ($\lambda$=2.12 \micron) observations from \citet{akras2017}, observed with Gemini North in the infrared. This was the only previously resolved observations of the molecular environment in K4-47. \citet{akras2017} showed that the H\down{2} emission traces the shocked regions surrounding the outflow that gives rise to the lobes seen in optical emission, which represents the walls of the cavity produced by the outflowing material. The provided astrometry for the data from Gemini was inaccurate, resulting in us having to manually shift the astrometric solution by --1\arcsec\ in declination. Typically, the infrared H$_2$ emission requires much higher excitation conditions than the molecules traced at millimetre wavelengths. \cite{lumsden2001nirspec} found excitation temperatures for the H$_2$ gas between 1600--1700 K, whilst \cite{schmidt2019carbonchem} found an excitation temperature of 65 K for pure rotational lines of CO.

In Fig. \ref{fig: h2+sma}, we compare the distribution of the CO (2--1) and H\down{2} emission to the H$\alpha$ image combined over all epochs. The CO (2--1) distribution centre is offset by $\sim$0\farcs5 north-east of the apparent centre of the H$\alpha$ image, as is the H\down{2} distribution from \citet{akras2017}. The two molecular species do not overlap exactly in space, possibly due to different excitation conditions. Emission of molecular hydrogen does not overlap very well with H$\alpha$ emission as well, except for the outer tips of the lobes, which are dominated by shock excitation.

Molecular line velocities and widths are presented in Table \ref{tab: SMA line fluxes}. The central line velocities are in the range of --17 to --35 km s\up{-1}, in contrast to the typical velocities of --27 to --28 km s\up{-1} \citep{edwards2014,schmidt2016hcnhco,schmidt2017hcnhnc,schmidt2017cch,schmidt2019carbonchem}, which were obtained with single dish (and therefore, unresolved) sub-mm observations. \up{13}CH\down{3}CN has a measured central velocity of --6 km s\up{-1}, but this is likely affected by blending or low S/N. The discrepancy between our measurements and the literature is likely due to our modest S/N in the weaker analysed lines. The majority of molecular lines detected by \citet{schmidt2019carbonchem} have similar line widths, although we do not find any lines with the extended pedestal reported for CO (6--5) and HCO$^+$ (3--2) in \cite{edwards2014}, which indicates higher velocity emission. We did not cover any high-excitation lines such as CO (6--5), and while we covered H\up{13}CO\up{+} (3--2), it may have a too low S/N to show the high velocity components. Much higher sensitivity is required to detect the high velocity component, which is expected to be spatially coincident with the lobe tips.


\subsection{VLA} \label{sect-vla-results}
The compact 5 GHz radio source contours detected with VLA are shown, along with \up{13}C\up{17}O (2--1) emission, both overplotted on the sub-millimetre continuum from SMA, in Fig. \ref{fig: sma continuum}. The deconvolved source size is (208 $\pm$ 11) $\times$ (44 $\pm$ 22) mas, with PA=49.9 $\pm$ 4.0\degree, slightly off-axis from the optical bipolar lobes \citep[PA=41\degree;][]{goncalves2004}. The radio emission is centred almost exactly on the SMA continuum peak (Fig. \ref{fig: sma continuum}). The total integrated flux of the source is 5.75 $\pm$ 0.04 mJy, somewhat lower than the \citet{aaquist1990} value of 7.7 mJy from 1984. Our source size is also smaller than that of \citeauthor{aaquist1990}, who obtained an angular size of 0\farcs25. The similar observation epochs mean that the discrepancy in radio flux is likely not a consequence of source evolution.

\begin{figure}
    \centering
    \includegraphics[width=\linewidth]{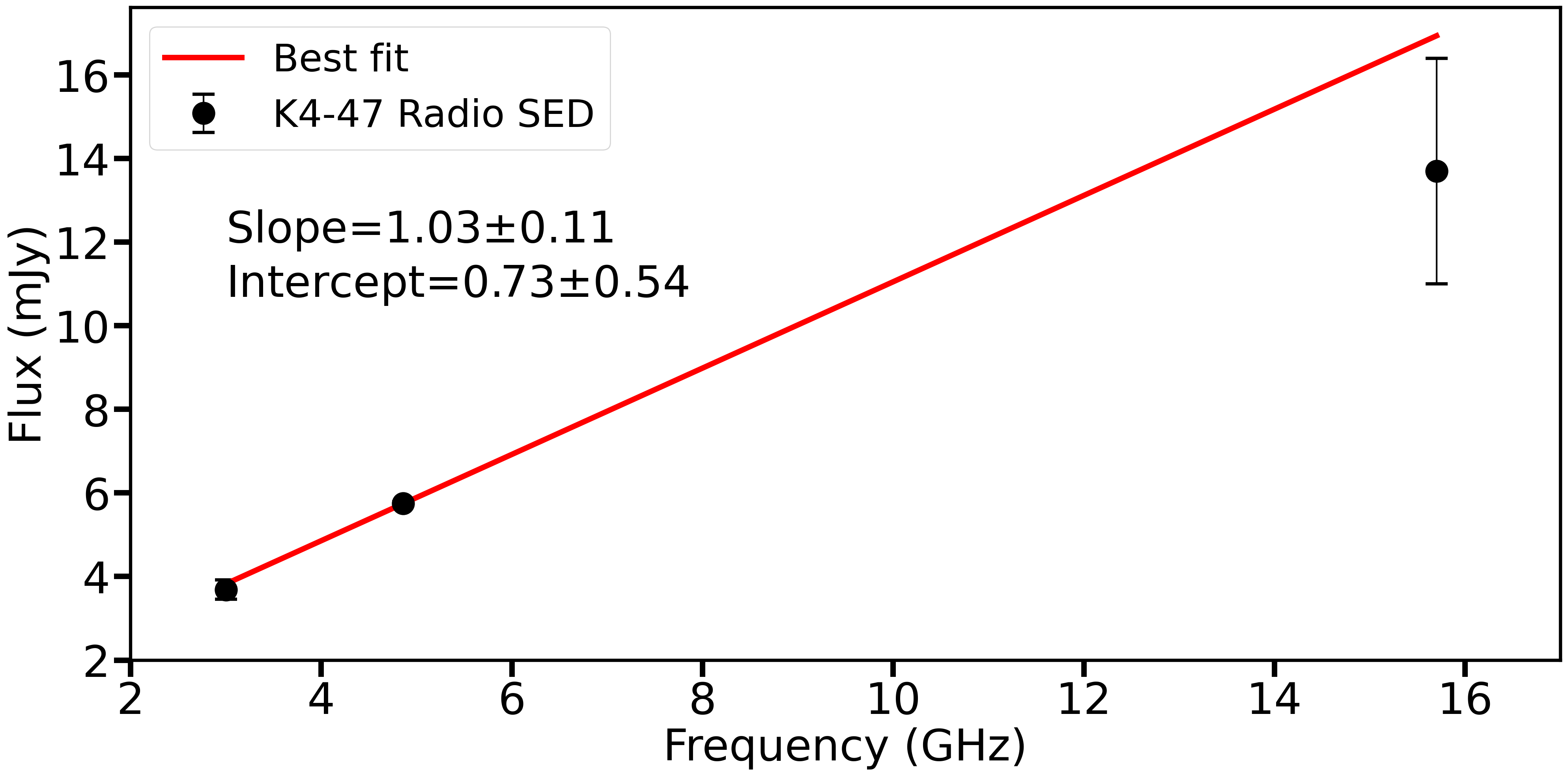}
    \caption{Radio SED of K4-47, featuring our 5 GHz flux from the 1983 observations, as well as the 3 GHz flux of \protect{\citet{gordon2021}} and the 15.7 GHz flux of \protect{\citet{perrott2015}}.}
    \label{fig: radio sed}
\end{figure}

We measure the spectral index of K4-47 in the radio, using the 1983 VLA data as well as VLA-NVSS data from 1994 \citep[< 2.5 mJy;][]{condon1998}, Arcminute Microkelvin Imager (AMI) survey observations \citep[13.7 $\pm$ 2.7 mJy;][]{perrott2015}, and VLASS observations from 2019 \citep[3.69 $\pm$ 0.23 mJy;][]{gordon2021}. The linear fit is parametrised by a slope of $\alpha$ 1.03$\pm$0.11 and an intercept of 0.73$\pm$0.54 (Fig. \ref{fig: radio sed}). Therefore, the fitted spectral index $\alpha\approx$ 1.0, which is consistent with thermal free-free emission in a medium with a fairly large density gradient. This does assume no significant change in the source flux between measurements.

\subsection{SED fitting}\label{sect: sed fitting}
Given the apparent presence of circumstellar dust in the core of K4-47 (see Sect. \ref{sect: spectroscopy results}), the dust temperature is key to derive the dust mass within the core of K4-47. This may also help us to constrain better the distance to the source, and give insight into the progenitor mass. We therefore attempted to derive the dust temperature from modified greybody fitting to the spectral energy distribution (SED) of K4-47.

Using the NASA/IPAC Extragalactic database, we extracted the available photometry for K4-47 in the IR, which was obtained via the 2MASS and WISE surveys, as well as mid-IR photometry from the IRAS satellite. We also obtained optical photometry from the IGAPS survey catalogue for IPHAS, and included our SMA continuum flux, as well as the radio fluxes from the VLA and the literature (see Sect. \ref{sect-vla-results}). Fluxes were dereddened using our measured value for the core of E(B-V)=1.28, assuming R\down{v}=3.1 and using the extinction model of \citet{fitzpatrick2007}. We then attempted to fit modified blackbodies of a certain temperature to the observed SED using in-house routines. The modified blackbodies were parametrised using the equation for dust mass \citep{hildebrand1983}
\begin{equation}\label{eq: dust mass}
    {\rm M_d} = \frac{\rm F_\nu D^2}{\rm \kappa_\nu B_\nu(T_d)},
\end{equation}
where F\down{\nu} is the flux at frequency $\nu$, D is the distance to the source, $\kappa$\down{\nu} is the opacity in cm\up{2} g\up{-1}, and B\down{\nu}(T\down{\rm d}) is the Planck function at an assumed dust temperature $T$\down{d}. Rearranging for F$_{\nu}$ and defining $\kappa_\nu$=$\kappa_0\left(\frac{\nu}{\nu_0}\right)$, we get the equation for the model flux
\begin{equation}\label{eq: modded bb}
    {\rm F_{\nu}=A\ \kappa_0\left(\frac{\nu}{\nu_0}\right)^{\beta}B_{\nu}\left(T_d\right)},
\end{equation}
where A=M\down{\rm d}/D$^2$. As K4-47 is heavily carbon-rich, we assume the dust is dominated by amorphous carbon grains following the dust opacity models of \citet{preibisch1993}, giving $\kappa_0\ $ = 0.65 cm\up{2} g\up{-1} at frequency $\nu_0\ $=246.46 GHz.

\begin{figure}
    \centering
    \includegraphics[width=0.99\linewidth]{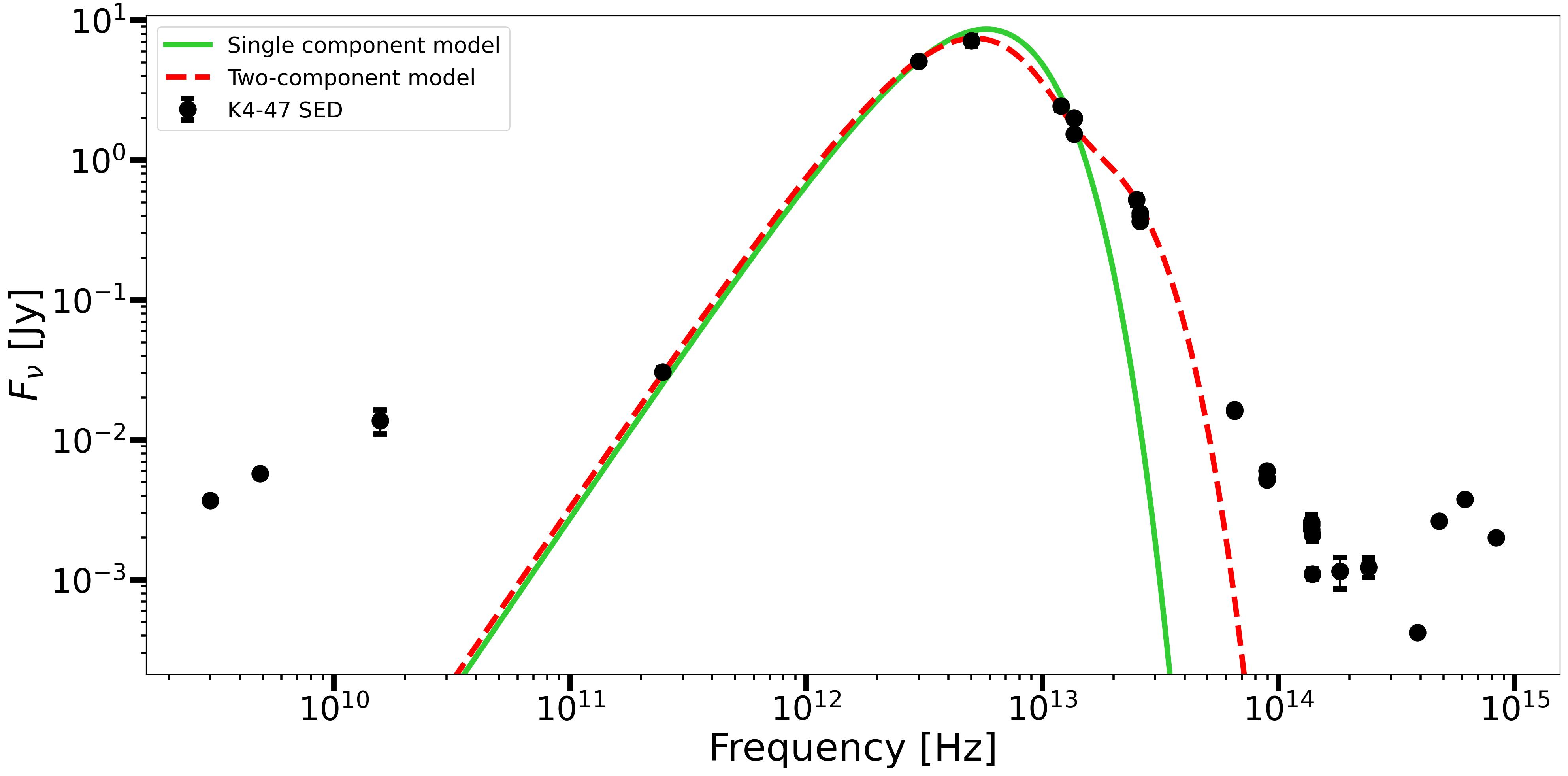}
    \caption{K4-47 SED (black) overlaid with the best-fitting single (green) and double (red) component dust models.}
    \label{fig: sed fitting}
\end{figure}

We do not include our radio data in the fitting procedure, as the radio spectral index is consistent with thermal free-free emission (see Sect. \ref{sect-vla-results}). Initial fitting showed that the mid-IR--mm SED, covered by IRAS and SMA, is best replicated by a two-component modified blackbody. However, this did not fit the NIR-optical part of the SED. We attempted to fit the NIR-optical SED with a third, hot dust component as well as a standard blackbody to resemble the central source. Neither method provided a reasonable fit to the NIR-optical SED. The results of both of these methods are shown in Appendix \ref{appendix: SED}. We found a good fit with both a single dust temperature and double-component model, when only the mid-IR and SMA fluxes were fitted, as shown in Fig. \ref{fig: sed fitting}. When including the NIR 2MASS and WISE fluxes, IRAS fluxes were not well-fitted. This likely indicates that the NIR and optical fluxes are not dominated by dust, and have stellar or emission line or scattered light contributions that cannot be reproduced by our simplified dust models.

Figure \ref{fig: sed fitting} shows that the mid-IR and SMA fluxes are well resembled by a single dust temperature of 82 K. The fitted value of $\log_{10}(A)$=--11.57. This is equivalent to a dust mass range of 1.18\standform{-2}--0.89 M\solar\ across a distance range of 3--26.5 kpc. Assuming a gas/dust ratio of 100, this gives a total released mass range of 1.18--8.9 M\solar\ for the full distance range. Next, we compare our measured dust masses with the predicted initial and final masses to constrain the properties of the progenitor to K4-47, and also constrain the distance.

We show the elemental abundance ratios of \citet{goncalves2004} and \citet{henry2010} in Table \ref{tab: abundances}. We also consider the isotopic ratios of \citet{schmidt2018isotope}. We then compare these ratios to theoretical predictions of the chemistry surrounding post-AGB stars \citep{karakas2016} at solar metallicity. The \up{12}C/\up{13}C=2.2 value implies a lower limit to the progenitor mass of 4 M\solar, as at this initial mass the \up{12}C/\up{13}C ratio drops significantly for the solar and sub-solar metallicity models. The large N/O ratios of \citet{goncalves2004} and \citet{henry2010} (see Table \ref{tab: abundances}), (N/O $\gtrsim$ 2.5 within errors), imply an initial mass of $\sim$6 M\solar, as at initial masses of 6 M\solar, the N/O ratio drops back below 2.5 for the solar metallicity models of \citet{karakas2016}. Therefore we consider a progenitor mass of 4--6 M\solar\ for K4-47.

Recent studies of the WD initial-final mass relation \citep[IFMR;][]{cummings2018,marigo2022} indicate that the resulting WD from 4--6 M\solar\ progenitor AGB stars range between 0.9--1.1 M\solar, implying a total mass loss of 3.1--4.9 M\solar. The combined dust and gas mass calculated using equation \ref{eq: dust mass} exceeds this range at a distance of 6 kpc. We therefore suggest that the distance to K4-47 cannot exceed 6 kpc. This distance should be considered as an upper limit, as it is expected that some of the mass lost during the AGB phase would be contained within faint, extended halos \citep{villaver2002}. Therefore, we would expect the detected mass of the nebula to be below the expected difference between the progenitor mass and the mass of the WD.

A distance upper limit of 6 kpc corresponds to a separation between the core and the northern lobe of 0.13 pc. The outflow velocity at a distance of 6 kpc is 388$\pm$137 km s\up{-1}, or a conservative upper limit of the velocity of $\sim$500 km s\up{-1}. If shocks do contribute to the excitation of the core emission, we would expect the shocks to decelerate the outflow. Therefore, the presence of shocks also infers an upper limit on the velocity. The mass upper limits are 4.7\standform{-2} M\solar\ for dust, and 2.77\standform{-3} M\solar\ for ionised gas, which gives a total detected nebula mass upper limit of 4.74 M\solar for the nebula, assuming a gas/dust ratio of 100. The age of the nebula (336$\pm$119 yr) is independent of distance, although the possible presence of shocks in the core may indicate the age to be an upper limit, due to shock deceleration of the outflows. 




\section{Discussion} \label{sect: discussion}
\subsection{The central source}\label{sect: central source}
Our SMA continuum and VLA 5 GHz observations indicate either a single stellar source, or an unresolved compact binary, within the core of K4-47. No symbiotic or binary properties have yet been discovered in K4-47, making any distinction between a single or binary central system difficult. The derived Zanstra effective temperature of 81 kK is consistent with a WD. In appendix \ref{appendix: galex obs}, we present GALEX survey images of the field of K4-47, showing a non-detection in the near-ultraviolet (NUV,$\lambda_{\rm rest}$=2267 \AA) band. The non-detection can be explained by the significant dust extinction we measure for the core. In Sect. \ref{sect: sed fitting} we show that the progenitor mass can be constrained to $\sim$4--6 M\solar, and that the corresponding expected final WD mass is 0.9--1.1 M\solar. This assumes that the progenitor underwent the final thermal pulse stage.

The collimated outflow in K4-47 could originate from material accreting onto and being ejected by a companion to the central WD \citep[such as for CK Vul, see][]{kaminski2021}. This would mean the central system of K4-47 could potentially be a symbiotic star, which features mass transfer between a cool evolved star and a WD. However, the IR classification scheme developed for symbiotic stars \citep{akras2019systs} show that PNe and post-AGB stars are distinct from symbiotic stars. Likewise, the IR colours for K4-47, $W1$--$W4$=10.33, $K_s$--$W3$=9.39, and $J$--$H$=0.79 mag, satisfy two of the three IR colour classification diagnostics for PNe \citep{akras2019pne}. The high W1-W4 and K-W3 colour indices
are indicative a dusty source, and the IR colours therefore infer the absence of a symbiotic central system. However, we cannot rule out a companion in the centre that is not directly interacting with the WD.

Observed molecules also present some hints on the nature of the central object. HCO\up{+} (including isotopologues) is formed via dissociation of CO and H\down{2}O and subsequent recombination. This can occur via two primary avenues. The first is via UV photo-dissociation \citep{kimura2012,cleeves2017}, and the second via collisional dissociation via shocks \citep{sanchez1997,sanchez2000,sanchez2015, pulliam2011}. The photo-dissociation route is associated with compact HCO\up{+} emission, as seen from the H\up{13}CO\up{+} spatial distribution (Fig. \ref{fig: moment 0 maps 3 sigma}). This is consistent with a hot central source. The lack of H\up{13}CO\up{+} emission in the lobes is inconsistent with the shock-dominated scenario presented by \citet{goncalves2004}.

\begin{figure}[t]
    \centering
    \includegraphics[width=0.99\linewidth]{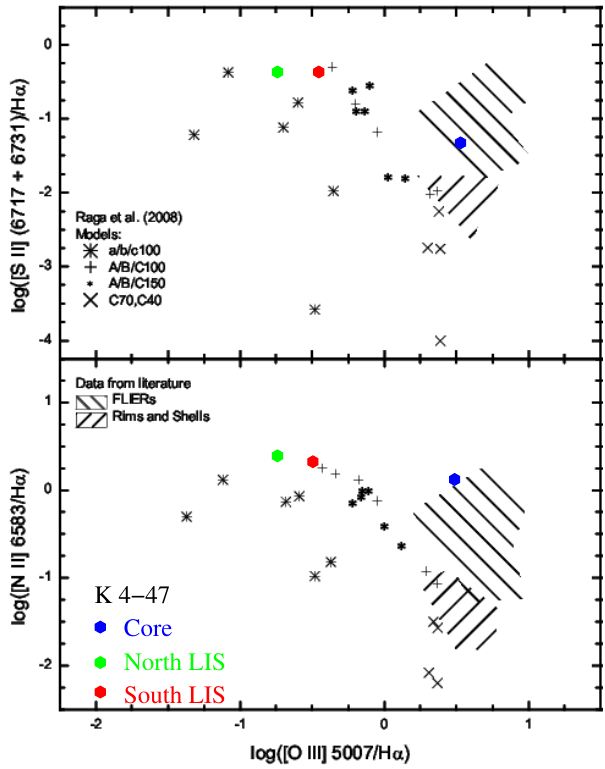}
    \caption{Diagnostic diagrams of [\ion{S}{II}]6717+6730 (top) and [\ion{N}{II}]$\lambda$6583 (bottom) vs [\ion{O}{III}]$\lambda$5007, as shown in \protect{\citet{goncalves2009}}. Black markers denote shock models from \protect{\citet{raga2008}}, and diagonal lines indicate regions occupied by rims/shells and FLIERs. THe numbers in the model names refer to the shock velocity of the models. Blue, green, and red markers show the position of the core, northern lobe, and southern lobe of K4-47, based on the measured emission line fluxes from the ALFOSC spectrum shown in Table \ref{tab: alfosc line fluxes}.}
    \label{fig: k447 raga diags}
\end{figure}

Our discussion of the central source of K4-47 is based mainly on our estimation of the Zanstra temperature of $\sim$80 kK. However, the Zanstra method assumes only photoionisation excites the nebular gas. Comparing the position of the core of K4-47 in the diagnostic diagrams of \citet{mari2023}, in particular that of [\ion{O}{III}] 4363/5007 versus [\ion{N}{II}] 5755/6584 diagnostic ratio, we see that our measured ratios exceed the predicted ratios for photoionisation, implying an additional excitation mechanism. Likewise, we also compare our [\ion{S}{II}]6716+6731, [\ion{N}{II}]$\lambda$6583, and [\ion{O}{III}]$\lambda$5007 line diagnostics with the models of \citet{raga2008}, adjusted from Fig. 6 of \citet{goncalves2009}, in Fig. \ref{fig: k447 raga diags}. We find that the core of K4-47 has similar diagnostic properties to the central core of PN He 1-1 \citep{goncalves2009}. The lobes are consistent with the high-velocity shock models of \citet{raga2008}. The strong [\ion{N}{I}]$\lambda$5200 line (Table \ref{tab: alfosc line fluxes}) and high T\down{e} found at the lobes are also indicative
of fast shocks, in agreement with \citet{goncalves2004}.

To test whether photoionisation dominates the excitation of the atomic gas in the core of K4-47, we examined the photoionisation models in the Mexican Million Models Database \citep[3MdB;][]{3mdb}. We applied filters to examine models that were constructed using input blackbody files with temperatures between 70--90 kK, as well as models with predicted line ratios of [\ion{O}{III}]4959+5007/H$\beta$, [\ion{N}{II}]6548+6584/H$\alpha$, and [\ion{S}{II}]6716+6731/H$\alpha$ within 0.1 dex of the observed line ratios (Table \ref{tab: alfosc line fluxes}). A total of 18 models matched these constraints. We then filtered the remaining models based on the predicted ratios of He II, [\ion{N}{I}]$\lambda$5198, [\ion{O}{I}]$\lambda$6300, and [\ion{O}{III}]$\lambda$4363 relative to H$\beta$. None of the remaining models predict line ratios within 0.1 dex of our observed ratios, although 2 models predicted [\ion{O}{I}]$\lambda$6300/H$\beta$ within 0.2 dex. In particular, the [\ion{O}{III}]$\lambda$4363 line is underpredicted in 3MdB by $\sim$2--3 orders of magnitude. It is apparent, therefore, that several of our line ratios cannot be explained by photoionisation only, and at least one additional excitation mechanism is required. Shock excitation is one such mechanism, which is supported by the unusually high T\down{e} of 20 kK. To rule out photoionisation from a central source hotter than 90 kK, we conduct the same test using an upper blackbody temperature of 150 kK. We find that 3MdB still significantly underpredicts the line ratios of [\ion{N}{I}]$\lambda$5198 and [\ion{O}{III}]$\lambda$4363, meaning that shocks are still necessary to explain the observed line ratios.

An alternative cause of the high T\down{e} is the low metallicity in the core of K4-47 (Table \ref{tab: abundances}). The [\ion{N}{II}]6584/5755 ratio is $\sim$15, 26 and 24 for the core, northern lobe, and southern lobe, respectively. A lower ratio infers a higher T\down{e} in the core than the lobes, despite the lobes being shock-dominated \citep{goncalves2004}. Therefore, shocks alone may not explain the high value of T\down{e} measured for the core of K4-47. However, the range of models explored in 3MdB include those with sub-solar abundances, and lower than those determined for K4-47. These models include those that have input blackbody temperatures of 70--90 K. If the core was dominated by photoionisation, and the high T\down{e} was due to the lower metallicity in K4-47, we would expect such models to be somewhat consistent with the line ratios we use to filter the 3MdB models. However, this is not the case. Therefore, whilst the lower metallicity may contribute to the high T\down{e}, we still find that an additional heating mechanism such as shocks is required to explain the line ratios in K4-47.

The presence of shocks may have implications for the distance to K4-47. Our measured Zanstra temperature for the central WD of 80 kK is too low for a massive ($\sim$1 M\solar) central star of a PN, even at such a young age. Such stars rapidly cross the Hertzprung-Russell (HR) diagram, reaching effective temperatures in excess of 200kK on timescales of just a few hundred years \citep{bloecker1995,bertolami2016}. As such, the much lower Zanstra temperature could be an indication that the central WD is less massive than expected based on the nebular chemistry or that in this case the Zanstra temperature is unreliable either due to shocks or, perhaps more likely, the nebula not being optically thick. On the other hand, if the Zanstra temperature is accurate, this would require a significantly lower central star mass in order to be consistent with the kinematical age of the nebula, which in turn would imply a closer distance (otherwise the previously inferred nebular mass would exceed the AGB envelope mass, although this is heavily dependent on the assumed dust-to-gas ratio).


\subsection{The circumstellar environment} \label{sect: CSE discussion}
The circumstellar environment that makes up the nebula of K4-47 consists of both atomic and molecular material. We see an elongated structure in CO and H\up{13}CN that is consistent with the bipolar structure of the H$\alpha$ emission (Fig. \ref{fig: total filter + contours}). The sub-millimetre core emission traced by CO (2--1) and H\up{13}CN (3--2) shows a clear velocity gradient, as in Fig. \ref{fig: moment 0 maps 5 sigma}. CO (2--1) shows a PA of 41\degree, but the H\up{13}CN emission has a PA of 13\degree. It is not clear if this discrepancy in the PA of different molecular lines is a signature of multi-polar outflows, as seen in other planetary nebulae \citep{akras2016,garcia2021,gomez2023,wen2024}, or a consequence of the smaller source size of K4-47 in H\up{13}CN compared to CO, which means that the emission is less spatially resolved. It is also possible that the much higher optical depth of CO makes direct comparison to HCN inadequate, as it shows only the outer layer of CO gas.

To test if the PA of the molecular and atomic gas phase match, we examined the flux density of the jets in the CO PV diagrams extracted at different PAs. We found that the flux density peaked at a PA of 41\degree $\pm$ 1\degree, indicating that the CO gas matches the PA of the atomic gas \citep{goncalves2004}. 

Our analysis of the expansion of the optical lobes shows an approximate optical outflow velocity of $\sim$few hundreds km s\up{-1}. This is an order of magnitude greater than the CO outflow velocity ($\sim$50--60 km s\up{-1}). Previous optical studies of K4-47 \citep{corradi2000,goncalves2004} have indicated that the lobe emission is powered by shocks. We also see that the tip of the northern CO outflow meets the northern lobe seen in optical and H$_2$ emission. This could be explained by molecular material that has been shaped by the atomic outflow, although this would not explain the narrow widths of the molecular lines. The presence of H\down{2} gas tracing the cavities of the bipolar atomic outflow \citep{akras2017} suggests that the molecular environment formed first, with the atomic outflow triggering at a later time and `punching' through the molecular material, simultaneously shaping the molecular environment and interacting with the molecular material via shocks that produce the H\down{2} emission. The atomic outflow then terminates at the lobes.

In Sect. \ref{sect: spectroscopy results}, we show that the core has significant circumstellar dust extinction of 0.3--0.6 mag from $\sim$10\up{-2} M\solar\ of dust. The dust mass is about 1 order of magnitude higher than measurements of other PNe \citep{koller2001,clayton2013,vandesteene2015,lau2016,otsuka2017,toala2021}. The relatively high dust mass is beyond the upper limit derived in the models of \citet{ventura2025}, and is consistent with the theoretical range of dust surrounding post-AGB stars \citep[10\up{-4}--10\up{-2} M\solar;][]{ferrarotti2006,zhukovska2013,ventura2014,ventura2020}. Therefore, the dust may have formed during the post-AGB phase in conjunction with the compact molecular CSM, before the bipolar outflows were triggered and the source reached the PN phase. 

\subsection{Comparison to CK Vul}\label{sect: ck vul comp}
CK Vul is a Galactic luminous red nova (LRN) produced by a WD-red giant merger \citep{kato2003,kaminski2015} that resulted in a 1670 eruption \citep{shara1982}. In the last few years, it has been extensively observed and characterized in a wide range of wavelengths \citep[e.g.][]{hajduk2007,evans2016,tylenda2019,kaminski2021, tylenda2024}. The source shows a clear bipolar structure in H$\alpha$ that spans 71\arcsec\ \citep{hajduk2007,hajduk2013}, as well as a clumpy core that surrounds the merger remnant. As mentioned, it bears many observational similarities to K4-47, which we critically review below.

Both sources exhibit a bipolar structure, primarily in atomic emission. The tips/lobes at the most extended regions of CK Vul are shock-excited \citep{banerjee2020}, and show molecular emission towards the central region \citep{kaminski2017,kaminski2018alf,kaminski2020}. The age of both sources is similar: our age estimate of K4-47 336$\pm$119 yr is almost exactly 355 yr that past since the 1670 eruption of CK Vul.

In both objects, we see a compact ($\sim$0\farcs1) radio source at 5 GHz \citep[see Fig. \ref{fig: sma continuum} for K4-47, for CK Vul see][]{hajduk2007} that is close to the centre of symmetry in the bipolar structure, but not overlapping with the optical knots. The 5 GHz flux of K4-47 (5.75 mJy, see Sect. \ref{sect-vla-results}), is a factor of $\sim$4 larger than for CK Vul \citep[1.5 mJy][]{hajduk2007}. \citet{hajduk2007} suggested that the radio emission of CK Vul originated from thermal free-free emission in an optically thick medium. The measured spectral index for K4-47 of $\alpha_\nu\approx$1 (see Sect. \ref{sect-vla-results}) indicates the radio emission is consistent with thermal free-free emission in a optically thick medium, but due to the opacity, we cannot rule out shock-induced radio emission, as has been suggested for CK Vul by \citet{kaminski2020}.

Across the various sub-mm/mm studies of K4-47, a plethora of molecules have been detected (see Sect. \ref{sect: sma results}), many of which are seen in CK Vul. CO, CN, SiO, HCO\up{+}, HCN, HNC, CS, CH\down{3}CN, CCH, H\down{2}CO, and N\down{2}H\up{+} (including isotopologues of \up{13}C and \up{15}N) have all been detected in CK Vul. The molecular environments extend out to 3\farcs41 for K4-47 and 6\farcs7 for CK Vul \citep{kaminski2020}. Assuming distances of 5.9 kpc \citep{tajitsu1998} and 3.5 kpc \citep{kaminski2021}, and inclination angles ($i$) of 67.5\degree\ \citep{corradi2000} and 25\degree\ \citep{kaminski2021} for K4-47 and CK Vul, respectively, we calculate the deprojected extent of the most extended molecular emission to be 0.10 pc for K4-47 and 0.12 pc for CK Vul. The similar extent for the molecular environments in both sources, along with the similar nebula ages, indicates that the molecular CSM for both K4-47 and CK Vul may have either been produced or evolved (i.e., shaped by bipolar outflows) via the same mechanism, and across similar timescales. For the extended optical emission, we get deprojected distances of 0.17 pc for K4-47 and 2.87 pc for CK Vul, based on the northern lobe for K4-47 and using the upper limit of $i$=15\degree\ for the H$\alpha$ hourglass structure in CK Vul \citep{kaminski2021}.

Isotopic abundances of \up{13}C, \up{15}N, and \up{17}O were found to be enriched in K4-47 by \citet{schmidt2018isotope}, similar to CK Vul. This led to the idea that both sources have similar origins. In Fig. \ref{fig: isotopic ratios}, we plot the calculated \up{12}C/\up{13}C vs \up{14}N/\up{15}N isotopic ratios for K4-47 \citep{schmidt2018isotope} and CK Vul \citep{kaminski2017}, as well as CRL 618 \citep{wannier1991}, Orion KL \citep{wannier1991}. IRAS 19312+1950 \citep{qiu2023}. CRL 618 is a carbon-rich pre-planetary nebula (PPN) that also shows isotopic enrichment in \up{13}C, \up{15}N, and \up{17}O \citep{schmidt2019carbonchem}. Orion KL is a star-forming region that has been proposed as a possible site of a previous stellar merger \citep{bally2005,zapata2011}, and IRAS 19312+1950 is an embedded star within a giant molecular cloud that shows characteristics of both a YSO \citep{cordiner2016} and an evolved star \citep{nakashima2015}, and has also been proposed as a candidate stellar merger remnant \citep{qiu2023}. Also plotted are the ratios for SiC presolar grains from the Presolar Grains Database \citep[PGD;][]{stephan2024}. The isotopic ratios of both sources are consistent with `nova' grains, and the carbon isotopic ratios are also consistent with A+B grains, which are connected with J-type stars \citep{liu2017}. Interestingly, the origin of J-type stars has been attributed to a He-WD + RGB mergers \citep{liu2017}, the same progenitor systems as hypothesized for CK Vul \citep{tylenda2024}. Therefore, we reiterate the postulate that the progenitor of K4-47 may have been a J-type carbon star, which can explain the exotic carbon chemistry of the molecular environment \citep{schmidt2019carbonchem}.

In Table \ref{tab: abundances}, we show the elemental abundances of K4-47 \citep{goncalves2004,henry2010} and CK Vul \citep{tylenda2019}. The K4-47 abundances of \citet{goncalves2004} and \citet{henry2010}\footnote{For \citet{henry2010}, abundances of K4-47 were extracted from the online data repository on Vizier: https://vizier.cds.unistra.fr/viz-bin/VizieR-3?-source=J/ApJ/724/748/table5} are broadly consistent within percentage errors. The ratios of He/H and O/H are a factor of $\sim$2--3 less in K4-47 compared to CK Vul, S/H is $\sim$3--4 less, and Ne/H is a factor of 7 less. Only the N/O ratio is larger in K4-47, by a factor of 2. The nitrogen abundances (N/H, N/O) are enhanced by more than an order of magnitude compared to solar values \citep{asplund2009}, whilst abundances of O, Ne, and S are all lower than the solar values for both K4-47 and CK Vul. 

CK Vul is known as a nitrogen-rich source, based on sub-mm studies \citep{kaminski2015,kaminski2017}, and made clear by the elemental abundances shown in Table \ref{tab: abundances}. K4-47 exhibits an even higher N/O ratio, implying that H-burning in CNO cycles has processed material into nitrogen \citep[see discussion of][]{tylenda2019}. We therefore classify K4-47 as a nitrogen-rich source. The apparent enhancement of nitrogen-bearing species in envelopes of evolved stars could also come from shock chemistry without the need for CNO processing, as was found for the evolved source OH231.8+4.2 \citep{sanchez2015}.

Despite the similarities outlined above, we see many differences as well. The isotopic abundances of \up{18}O are noticeable, with no \up{18}O-bearing molecules detected in K4-47, unlike CK Vul where it is highly enhanced \citep[ $^{16}$O/$^{18}$O$\approx$36;][]{kaminski2017}. The identification of raised \up{18}O for CK Vul was a key characteristic that led to the hypothesis that partial carbon burning is responsible for some of the isotopic ratios in CK Vul. Another key difference between K4-47 and CK Vul is the lack of detection of \up{26}Al in K4-47, which was detected in the form of \up{26}AlF in CK Vul \citep{kaminski2018alf}.
All these isotopic differences may indicate that both objects went through different nucleosynthesis paths, but by themselves do not exclude a merger scenario for K4-47, as the chemistry of merger remnants is still not well understood.

\begin{table*}[h!]
    \centering
    \small
    \caption{Elemental number abundances of K4-47 and CK Vul. Percentage errors are shown in parentheses. Gon\c{c}alves+ and Henry+ refer to elemental abundances from \protect{\citet{goncalves2004}} and \protect{\citet{henry2010}}, respectively, whilst CK Vul abundances come from \protect{\citet{tylenda2019}}. Solar values from \protect{\citet{asplund2009}} are shown for comparison.}
    \begin{tabular}{ccccc}\hline
    Element & \multicolumn{2}{c}{K4-47} & CK Vul & Solar\\
    & Gon\c{c}alves+ & Henry+ & &\\\hline
He/H & 0.139 (14\%) & 0.09 (13\%) & 0.260 (21\%) & 0.085\\
N/H & 3.74\standform{-4} (40\%) & 1.4\standform{-4} (22\%) & 1.69\standform{-4} (29\%) & 6.76\standform{-5} \\
O/H & 7.37\standform{-5} (32\%) & 4.9\standform{-5} (15\%) & 1.22\standform{-4} (34\%) & 4.90\standform{-4}\\
Ne/H & 1.74\standform{-5} (66\%) & 1.00\standform{-5} (22\%) & 7.45\standform{-5} (29\%) & 8.50\standform{-5}\\
S/H & 1.96\standform{-6} (48\%) & 1.50\standform{-6} (23\%) &  5.70\standform{-6} (9\%) & 1.32\standform{-5}\\
N/O & 5.07 (51\%) & 2.90 (20\%) & 1.39 (45\%) & 0.14 \\\hline
    \end{tabular}
    \label{tab: abundances}
\end{table*}
We also highlight differences in our optical spectrum of K4-47 compared to CK Vul (Fig. \ref{fig: k447+ck vul}). CK Vul is missing some prominent lines seen in K4-47, namely [\ion{O}{III}]$\lambda$4363, HeII $\lambda$4685, NII $\lambda$4643, and [\ion{Ar}{V}]$\lambda$7006. The [\ion{N}{II}]$\lambda\lambda$6548,6583 and [\ion{S}{II}]$\lambda\lambda$6716,6731 doublets are both much stronger in CK Vul than K4-47 (relative to H$\beta$), as is H$\alpha$. The derived physical properties of the atomic gas are also different, with the value n\down{e} derived for CK Vul \citep{tylenda2019} differing somewhat from our values derived for K4-47. \citet{tylenda2019} derive values of 130 $<$ n\down{e} $<$ 600 cm\up{-3} for different kinematic components, an order of magnitude less than for K4-47. This infers a much denser environment for K4-47, despite the similar ages, and so infers a higher total ejecta mass for K4-47, and so a higher overall mass-loss rate.

\begin{figure}
    \centering
    \includegraphics[width=0.49\linewidth]{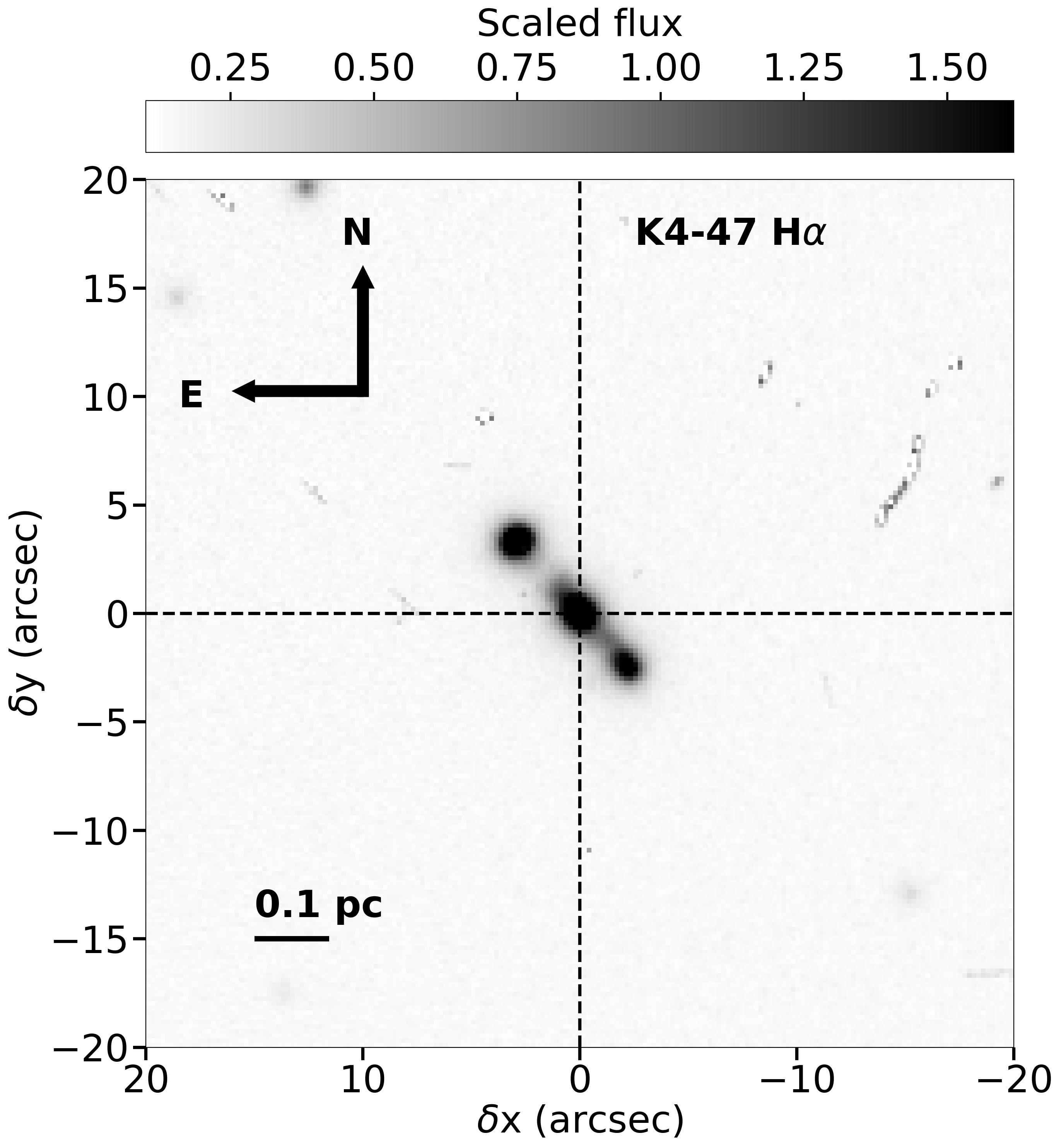}
    \includegraphics[width=0.47\linewidth]{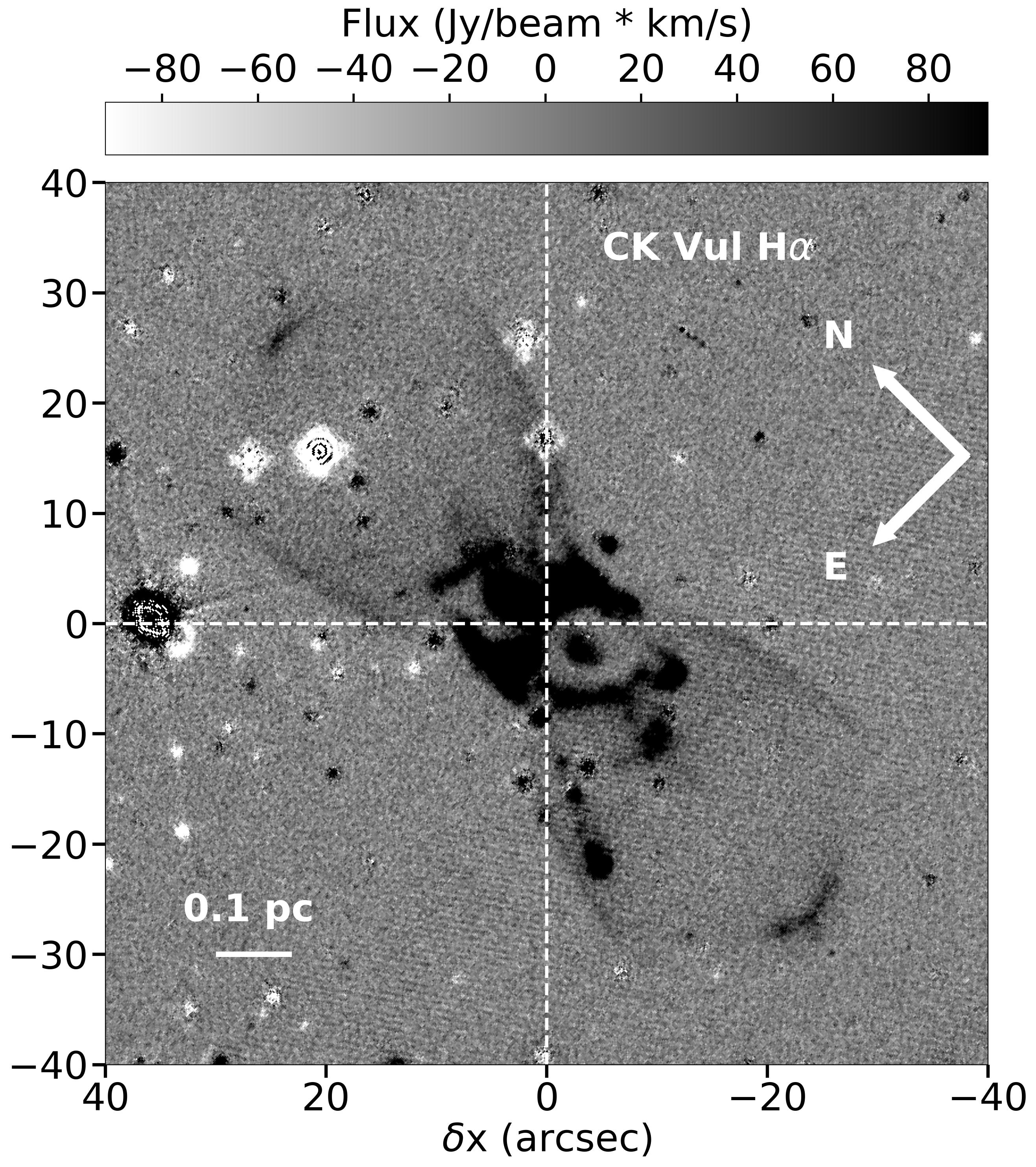}
    \includegraphics[width=0.47\linewidth]{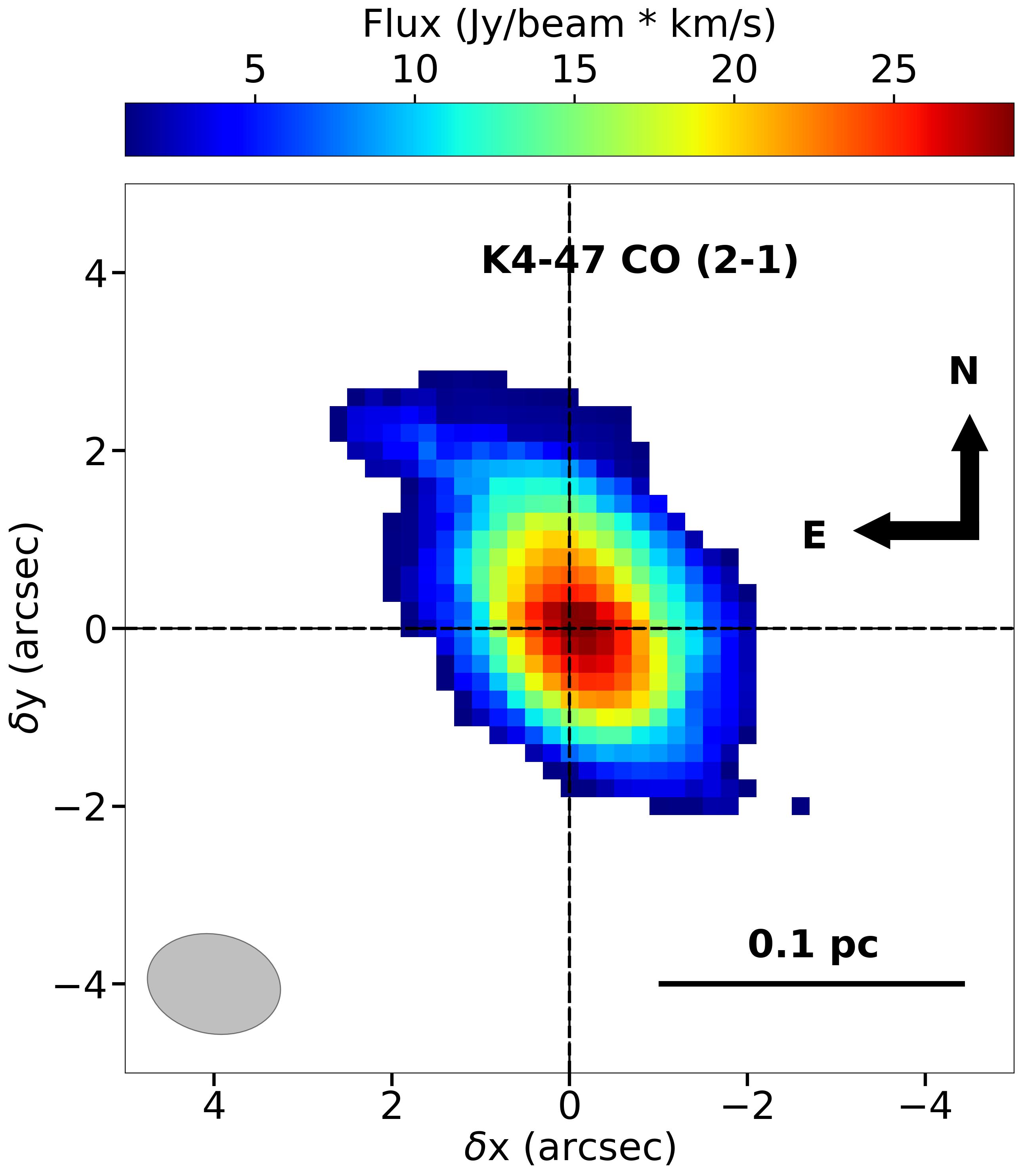}
    \includegraphics[width=0.49\linewidth]{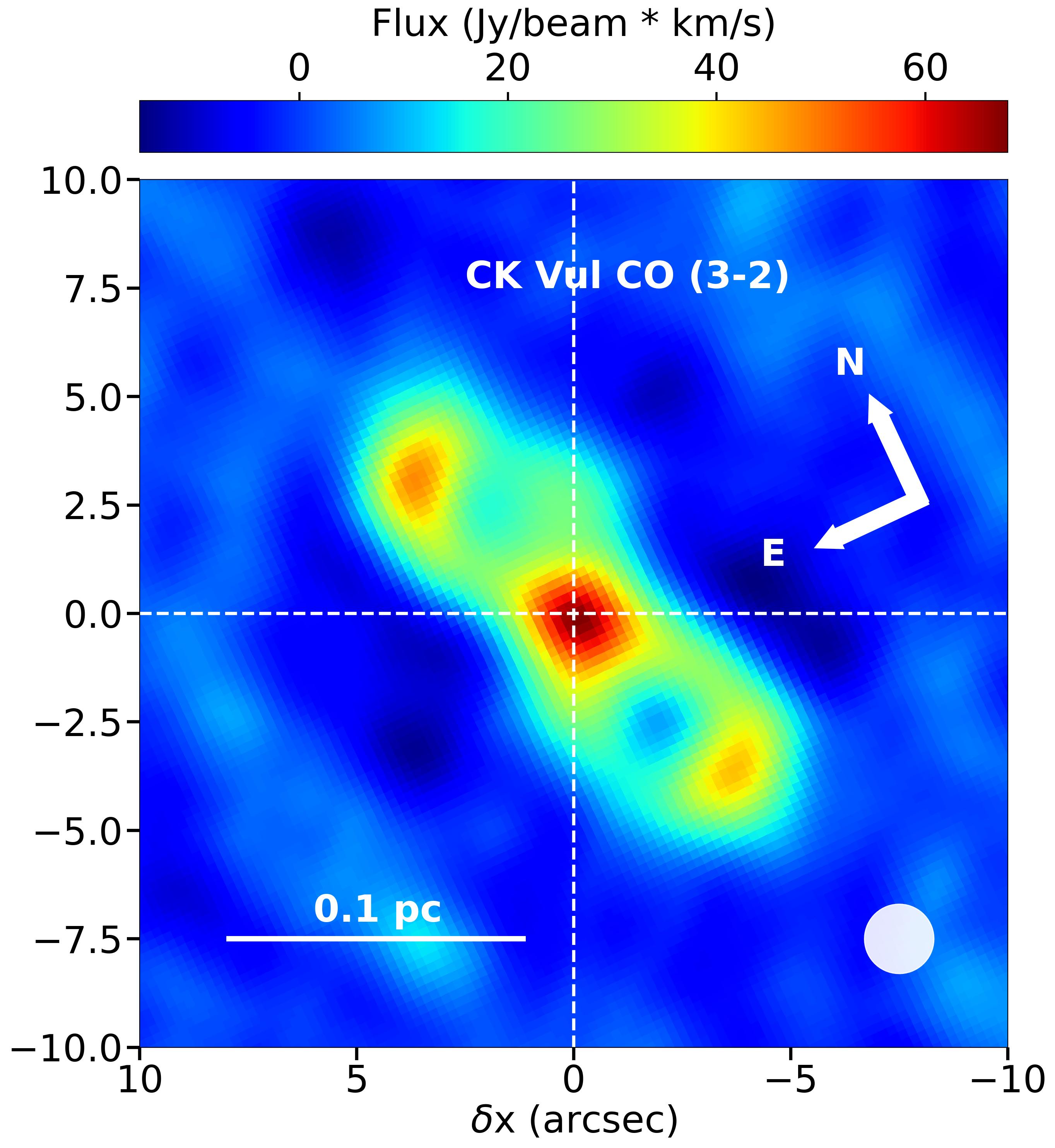}
    \caption{K4-47 (left column) and CK Vul (right column) compared at the same equivalent spatial scale. The top row compares the H$\alpha$ distribution, with the image of CK Vul taken from \protect{\citet{hajduk2007}}. The bottom row compares the CO distribution, with CO (2--1) for K4-47 from our SMA data and CO (3--2) for CK Vul, taken from \protect{\citet{kaminski2020}}. The images of CK Vul were rotated to match the orientation of K4-47.}
    \label{fig: k447 ckvul halpha comparison}
\end{figure}

\begin{figure}[h!]
    \centering
    \includegraphics[scale=0.2]{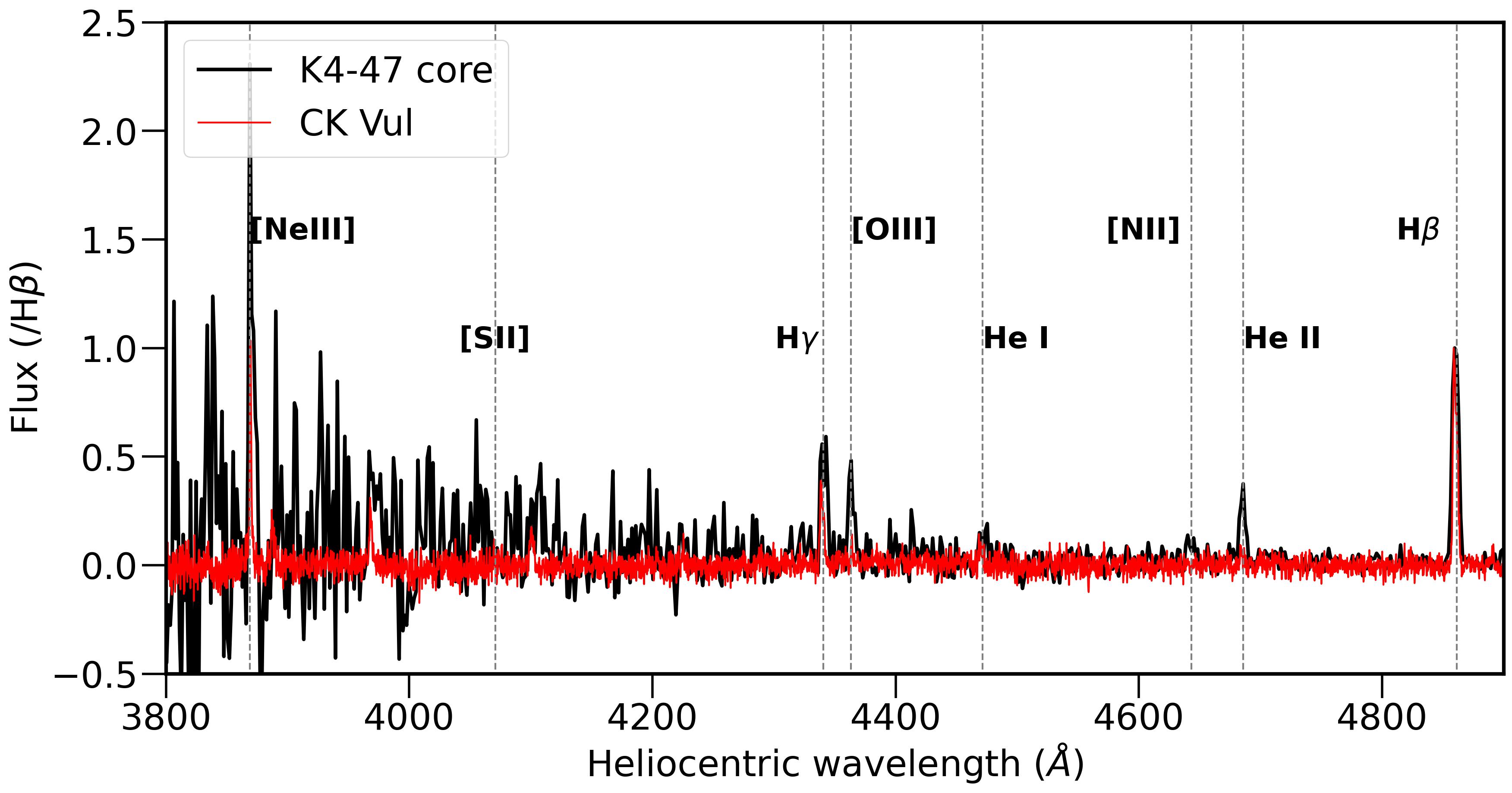}
    \includegraphics[scale=0.208]{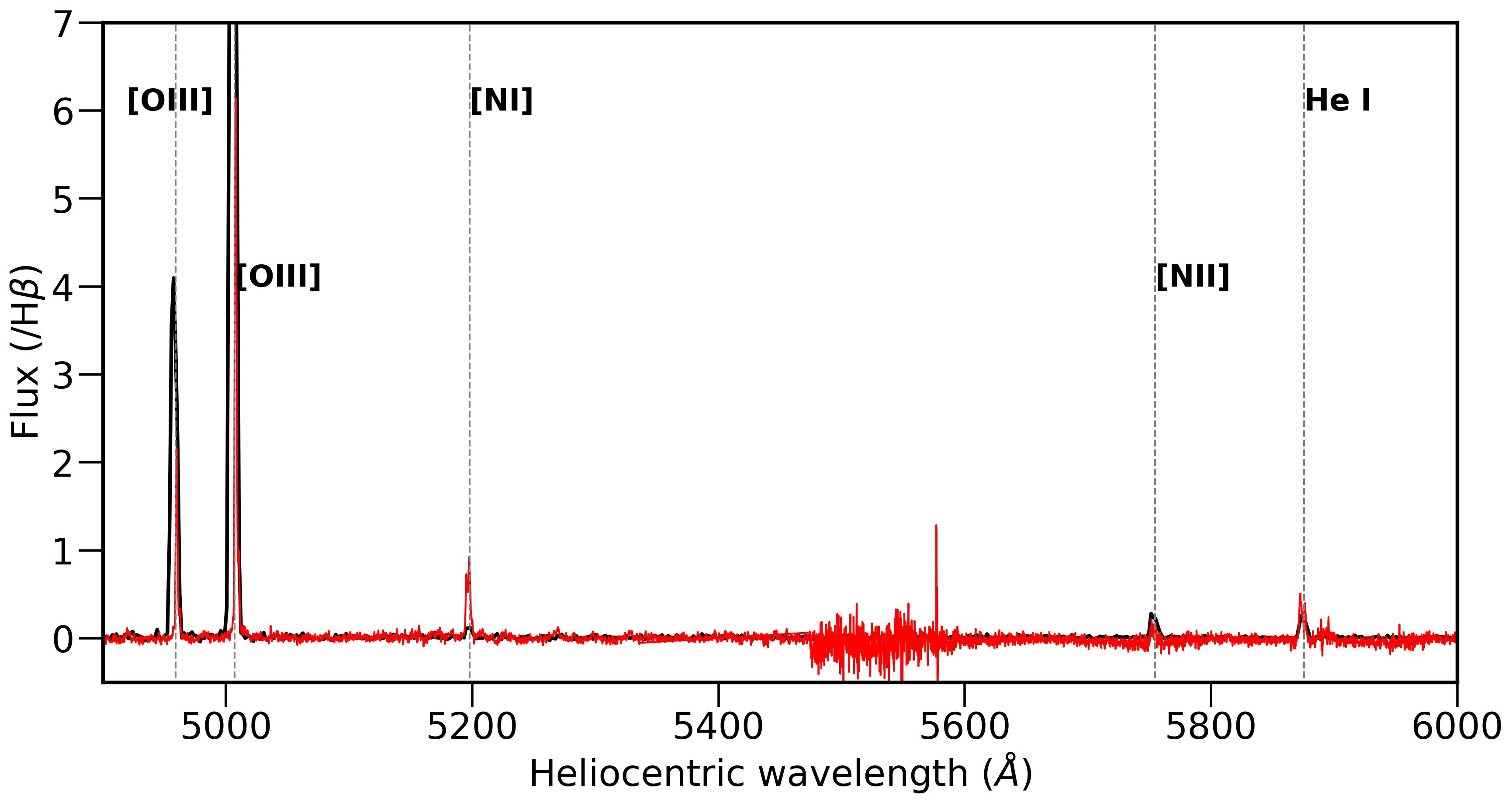}
    \includegraphics[scale=0.205]{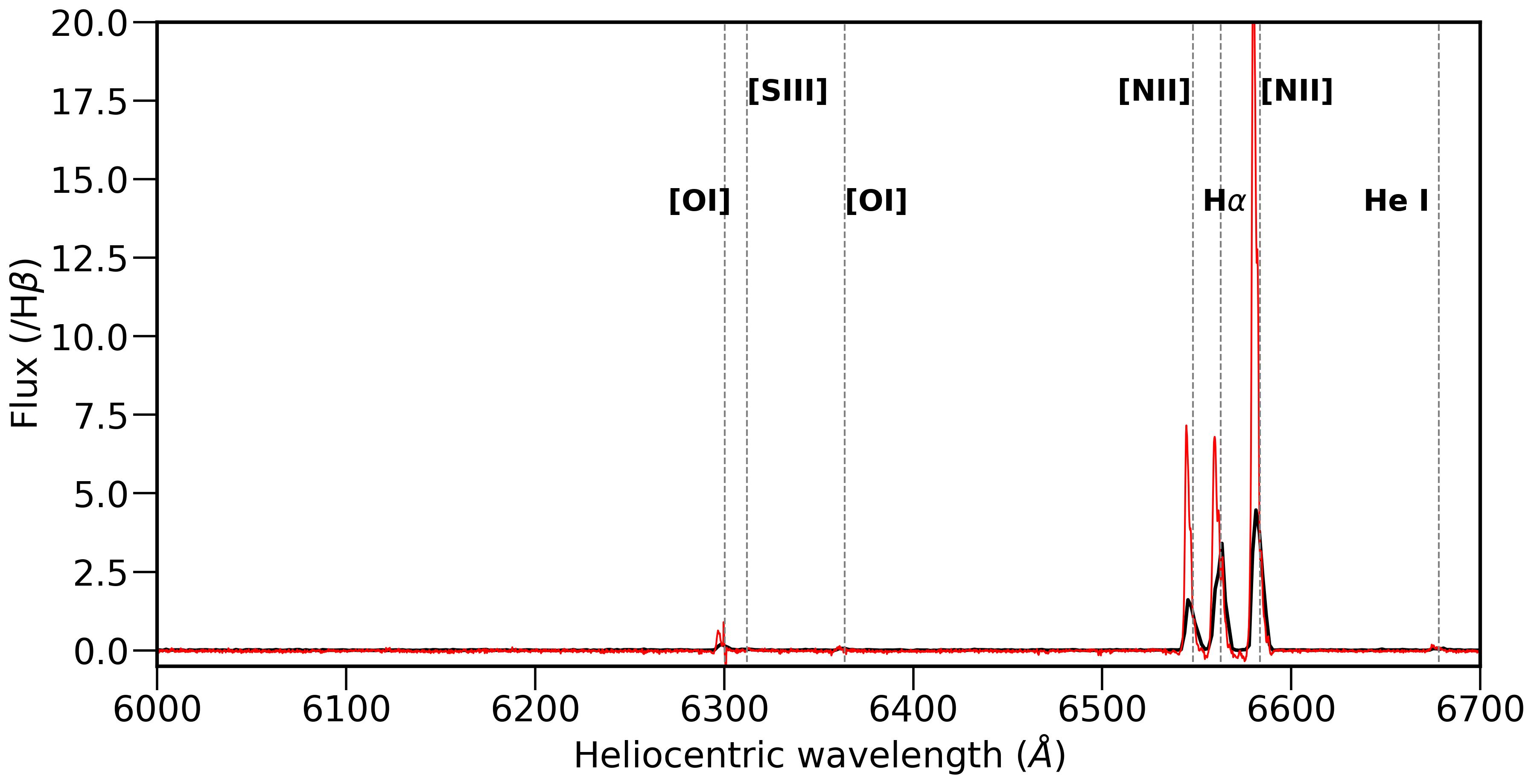}
    \includegraphics[scale=0.20]{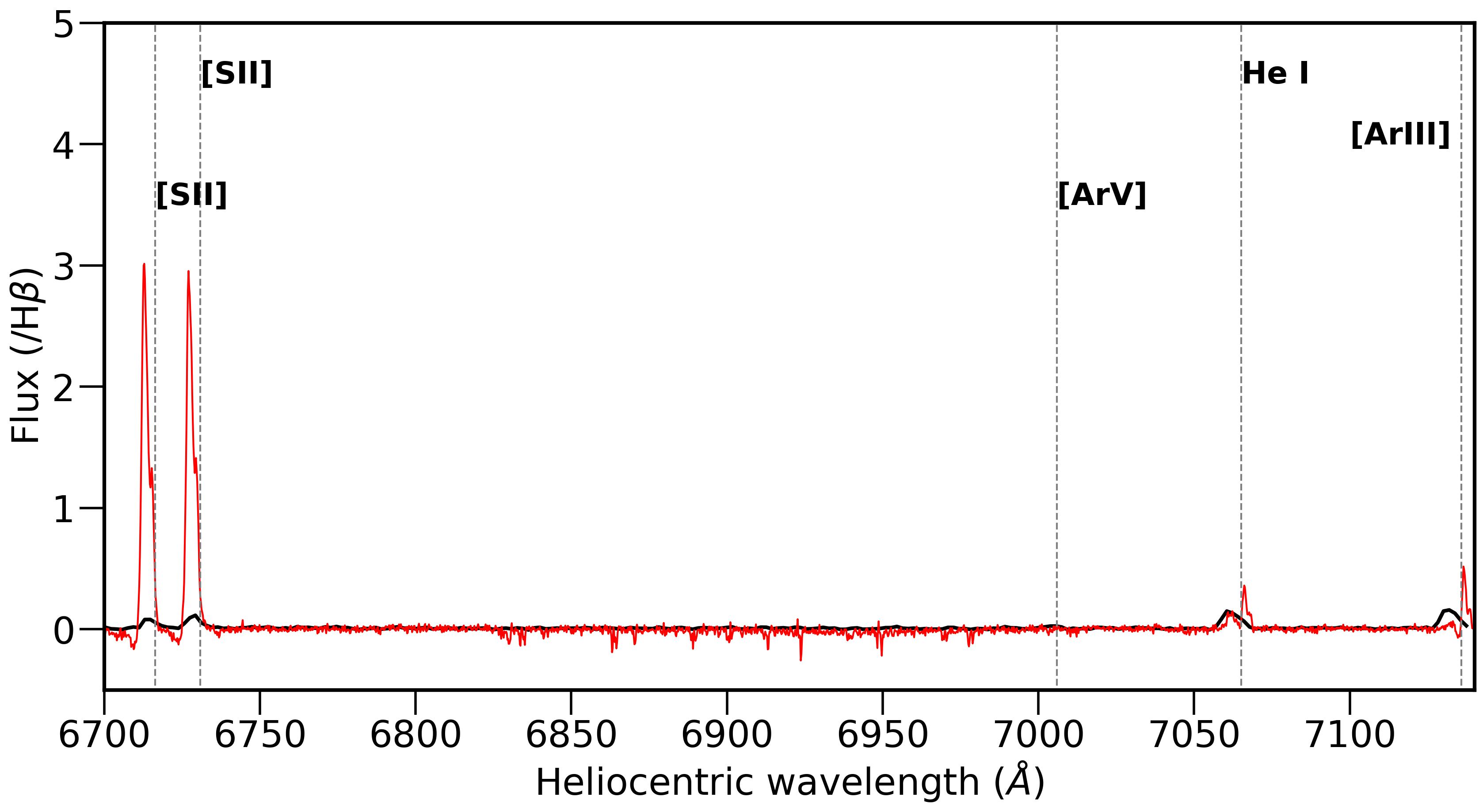}
    \caption{ALFOSC K4-47 total spectrum across all three components compared to the \protect{\citet{tylenda2019}} VLT-XShooter spectrum of CK Vul. Both spectra are normalised to H$\beta$. The y-axes have been truncated to show the weaker lines in more detail. The K4-47 spectrum is in the heliocentric rest frame, and the CK Vul spectrum has been shifted by --150 km s\up{-1} to align the majority of lines from both spectra.}
    \label{fig: k447+ck vul}
\end{figure}

We did not detect an extended optical nebula, as that seen in H$\alpha$ emission for CK Vul \citep{hajduk2007}. In Fig. \ref{fig: k447 ckvul halpha comparison}, we compare the two sources in H$\alpha$ at the same deprojected spatial scale. Here, we assume distances of 5.9 kpc for K4-47 \citep{tajitsu1998} and 3.5 kpc for CK Vul \citep{kaminski2021}. We can see that the optical lobes of K4-47 barely extend beyond the central clumpy region of CK Vul, although the extent of the molecular environments of both sources are comparable. No extended H$\alpha$ structure is seen in K4-47, and assuming that the extent of any such extended structure was present, it would have been covered in the ALFOSC FoV. However, the H\down{2} emission shown in Fig. \ref{fig: h2+sma} traces the cavity walls of the optical outflow as it propagates through the circumstellar medium \citep{akras2017}, which indicates the presence of an extended circumstellar component that is not detected in our observations. This is supported by the fact that the lobes are shock-dominated \citep{goncalves2004}. The mass-loss mechanism for the undetected circumstellar gas is unclear, but may be due to an AGB wind, binary interactions, or even merger ejecta. It is not clear if any extended structure in K4-47 would resemble, morphologically or kinematically, the extended H$\alpha$ structure surrounding CK Vul. We also observe that the dust mass in K4-47 ($\sim$few 10\up{-2} M\solar) is an order of magnitude larger than for CK Vul \citep{banerjee2020}.

The kinematics of the CSM for both sources are also considerably different. Although for both sources we see a fast atomic outflow/expansion velocity, with a much slower (by $\sim$1 order of magnitude compared to the atomic outflow) molecular outflow, the measured velocities are very different. Whilst the expansion velocity of the shocked tips in CK Vul is of the order of $\sim$1700 km s\up{-1}, the atomic outflow in K4-47 is much slower, at 350 km s\up{-1}. As seen in Table \ref{tab: comparison table}, the atomic outflow in K4-47 and the molecular outflow in CK Vul show similar velocities, which may explain why the optical nebula and molecular environment in K4-47 and CK Vul, respectively, extend across similar spatial scales (see Fig. \ref{fig: k447 ckvul halpha comparison}).


There is no consensus on the central source of CK Vul. Previous studies have assumed a central WD, supported by the proposed mechanism of thermal free-free emission to power the 5 GHz radio emission \citep{hajduk2007}, and the presence of the [\ion{O}{IV}]$\lambda$25.89 \micron\ line \citep{evans2016}. However, \citet{kaminski2021} calculated the molecular lifetimes for the lobes in CK Vul when exposed to different levels of radiation shielding by dust. The only combination of stellar radiation and dust shielding that produces molecular lifetimes greater than the nebula age of CK Vul (and so explaining the rich molecular content) is a central WD with T\down{\rm eff}$\approx$20 kK with extinction between the WD and the lobes of A\down{V}$>>$3 mag. This, however, is improbable as such a WD gives a luminosity of $\sim$0.1 L\solar, inconsistent with the derived source luminosity of $\sim$1 L\solar\ \citep{kaminski2015}. Likewise, the extinction value along the lobe-central star axis is unlikely to be as high as A\down{V}=3 mag \citep{tylenda2019}. \citet{kaminski2021} also re-identified the [\ion{O}{IV}]$\lambda$25.89 \micron\ line as the [\ion{Fe}{II}] a\up{6} D 7/2$\rightarrow$9/2 line, which does not require a hot central source. \citet{kaminski2021} subsequently argue that the central star of CK Vul may have the characteristics of a cool, evolved star, similar to other Galactic red novae \citep{tylenda2005,tylenda2011}. \citet{kaminski2021} also argue that the remnant circumstellar structure can be explained by a binary system, making the progenitor of CK Vul a triple system. This can also explain the launching of jets in PNe, and so may be applicable to K4-47.

Contrary to the case of CK Vul, our results indicate that the central source of K4-47 is almost certainly a WD. This is the main reasoning behind excluding the stellar merger scenario for K4-47, as if CK Vul does host a cool, evolved merger remnant like for other Galactic LRNe, this is a key identifying feature for classifying LRNe at later times, even for stellar mergers involving a WD.

The phenomenological comparison of both sources we have presented here cannot be fully conclusive, but we favour the scenario that K4-47 maybe a genuine, albeit very young, planetary nebula. As postulated by \cite{schmidt2018isotope}, the progenitor of the system may be a J-type AGB star, which created the molecular circumstellar environment rich in molecules that are traced at millimetre wavelengths. Dust may have also formed at this time. This cool molecular cocoon was later punched by a pair of slightly asymmetric jets, which we now observe mainly in atomic and H$_2$ emission. Some of the circumstellar wind was swept out by the jets, which is manifested by the CO extensions observed at high velocities and largest distances from the centre. Just as in CK Vul, the jets may be created by a companion to the component that was the source of the circumstellar material.
\subsection{Analogy to other evolved sources}\label{sect: WF comp}
As well as PNe and CK Vul, K4-47 bears some resemblance to a subset of evolved stellar systems known as Water Fountains \citep[WFs; ][]{imai2007review,desmurs2012review}. WFs are often characterised by collimated bipolar jets and high-velocity ($\gtrsim$ 100 km s\up{-1}) H\down{2}O masers. Other components seen are low-velocity ($\lesssim$ 20--40 km s\up{-1}) central toroidal components \citep{gomez2008}, and strong optical extinction \citep{suarez2008,gomez2015}.

On initial inspection, K4-47 possesses multiple morphological similarities with WFs. These include collimated bipolar outflows with velocities exceeding 100 km s\up{-1}, a molecular gas phase, strong apparent obscuration in the optical, as inferred by the GALEX non-detection (Fig. \ref{appendix: galex obs}), and the lack of stellar features from the central WD. \citet{khouri2022} also found that some WFs have high isotopic enrichment of \up{17}O and \up{18}O, which implies initial masses of 1.8--4.0 M\solar. Such stars are expected to become carbon stars eventually \citep{abia2017}, including J-type carbon stars, which we propose as the likely progenitor of K4-47. However, WFs are partly defined by their oxygen-rich chemistry, indicating that the AGB evolution of WFs may have been interrupted. \citet{khouri2022} attribute this to substantial mass loss on short timescales, which they attribute to CEE.

Despite the similarities, there are some glaring differences. Firstly, our progenitor mass estimate for K4-47 is not consistent with the masses of WF stars from \citet{khouri2022}. K4-47 is also dominated by carbon-bearing molecules, as opposed to WF sources, which feature oxygen-rich chemistry \citep{gomez2008,gomez2015,gomez2018}. The fast outflows seen in WFs are seen typically in the molecular phase, whereas our molecular outflow is 56 km s\up{-1}, which is a factor of $\sim$2 slower than the molecular outflows typically seen in WFs.

Another issue is the dynamical age of the molecular outflow. Typically, WF jet ages span $\sim$100--200 yr \citep{tafoya2019}, an order of magnitude less than what we find for the molecular outflow in K4-47. Lastly, the molecular environment shows little to no evidence of a toroidal component (Figs. \ref{fig: moment 1 maps} and \ref{fig: pv diagrams}), due to the extended elongated structure that traces the atomic outflow, and we see no velocity gradient in the PV diagram perpendicular to the outflow (see Fig. \ref{fig: pv diagrams}). Such a structure would be expected to be expanding in a post-CEE source, as well as in LRNe \citep{nandez2014,pejcha2017}. However, we cannot conclude such a structure is not present in K4-47, due to the insufficient spatial resolution of our sub-mm observations, which can resolve the overall molecular structure but not the innermost region. We summarise the overall properties of K4-47, CK Vul, and WFs, in Table \ref{tab: comparison table}.

\begin{table*}[h!]
    \centering
    \caption{Qualitative comparison of various stellar sources/classifications discussed in this work. Certain properties, for instance, the central source temperature of WFs, are not well constrained and represent an approximate range.}
    \begin{tabular}{|cccccccc|}\hline
Source & Central star & Age & Collimated & Extended & \multirow{2}{*}{Shocks} & Isotopic & Expansion velocity \\
&T\down{\rm eff} (kK) & (yr) & jets & nebula & & enrichment & (molecular/atomic, km s\up{-1}) \\\hline
K4-47 & 81$\pm$2 & 400--900 & $\checkmark$ & $\times$ & $\checkmark$ & $\checkmark$ & 56/350 \\
CK Vul & ? & 357 & $\checkmark$ & $\checkmark$ & $\checkmark$ & $\checkmark$ & 312/1700 \\
LRNe & 2--3 & < 40 & $\checkmark^1$ & $\checkmark$ & $\checkmark$ & $\times^2$ & $\sim$200 for both \\
WFs & 3---30\up{3} & 100--200 & $\checkmark$ & Sometimes & $\checkmark$ & $\checkmark$ & 100--200 \\\hline
    \end{tabular}
    \tablefoot{\up{1}Except for OGLE-2002-BLG-360. \\ \up{2}Except for CK Vul. \\\up{3}Heavily dependent on mass and evolutionary stage, uncertain.}
    \label{tab: comparison table}
\end{table*}

Although WFs are typically oxygen-rich sources, this could be a bias based on their characteristic feature: high-velocity water masers. Theoretically, carbon-rich progenitor stars can also undergo CEE and mass-loss. However, it is unclear exactly what a carbon-rich post-CEE system would look like. One such example is PN G054.2-03.4 \citep[The Necklace; ][]{viironen2009}, which is believed to be a post-CE carbon-rich PN \citep{miszalski2012}, but likely features an accretion disk - which K4-47 does not seem to have.


Both the similarities and differences between K4-47 and WFs highlight the difficulty in understanding stellar evolution, particularly post-CEE sources. It is clear that the evolution path for evolved stellar systems hosting carbon-rich stars is still not well understood, but some of the morphological similarities between K4-47 and WF sources might be a clue to understanding these systems. Despite this, the lack of a torus in K4-47 is indicative of no previous CEE phase. Therefore, the similar morphologies mean that the evolution of such sources needs to be investigated further to adequately understand how carbon-rich systems evolve beyond the CE phase, as well as how similar morphologies can be produced by objects of different origins and evolutionary phase.

\section{Summary}\label{sect: summary}
In this paper, we examine the atomic and molecular environment of the unusual planetary nebula PN K4-47, to investigate the similarities between the source and other stellar classifications, and so attempt to better understand the origins of the source. Our data includes the first spatially resolved millimetre molecular maps of the source. We find that:
\begin{itemize}
    \item The source is bipolar, not only in optical emission and shocked H\down{2}, but also in molecular emission. We see a clear velocity gradient approximately along the atomic outflow PA, which infers a maximum molecular outflow velocity of 56 km s\up{-1}.
    \item We determine physical constraints on the circumstellar gas within the core. We estimate from forbidden [\ion{N}{II}] and [\ion{S}{II}] line ratios the electron temperature to be 19900 $\pm$ 1200 K, and the electron density to be 2800 $\pm$ 700 cm\up{-3}.
    \item Using the Zanstra method and the \ion{He}{II} $\lambda$4686/H$\beta$ line ratio for the core, we estimate a stellar effective temperature T\down{\rm eff} = 81 $\pm$ 2 kK, inferring a central WD.
    \item Using greybody fitting to the SED of K4-47, we show that the dust-dominated region of the SED can be fitted by a dust temperature of 82 K. The isotopic and elemental abundances of K4-47 indicate an initial mass of 4--6 M\solar, which implies a current WD mass of $\sim$1 M\solar. The dust and gas mass is consistent with the total released mass of 3--5 M\solar at distances $\le$ 6 kpc. The nebula mass is dominated by neutral atomic gas, with $\sim$few 10\up{-2} M\solar of dust and $\sim$few 10\up{-3} M\solar of ionised gas.
    \item The spatial change in the northern optical lobe, moving away from the centre of symmetry along the bipolar PA, infers a atomic outflow velocity of 350 $\pm$ 88 km s\up{-1} at a distance upper limit of 6 kpc. 
    \item Comparing our line ratios to 3MdB photoionisation models, we find that photoionisation alone cannot explain the excitation of the atomic gas. The high value of T\down{e} supports this statement, although the low metallicity of K4-47 may contribute to the high measured T\down{e}. However, low metallicity alone cannot explain the high value of T\down{e}, ultimately requiring an additional heating mechanism, which we attribute to shocks originating from interaction between the bipolar outflows and the CSE in the core of K4-47.
    \item The lack of an extended nebula, coupled with the clear presence of shocks and H\down{2} emission in the walls of the collimated outflows, infer earlier mass loss long before the atomic outflow was triggered. This is supported by the dynamical age ($\approx$1900 yr) of the molecular outflow. The dynamical age of the atomic outflow is 336 $\pm$ 119 yr. It is possible that the shocks that excite H\down{2} occur as the atomic outflow collides with the older molecular outflow.
    \item Comparisons between the old red nova CK Vul and K4-47 show some clear similarities in the morphology, kinematics, and chemistry. However, the central source of K4-47 likely being a WD is inconsistent with other Galactic merger remnants, including CK Vul.
\end{itemize}
To examine K4-47 further, we would firstly require separate, non-blended spectra of the three components in the optical. Therefore we could investigate the lobes in more detail, addressing their exact nature and therefore understanding better the mass-loss history of the source. Deeper sub-mm observations would also help to examine the spatial distribution of certain molecules, such as HCO\up{+}, and higher-excitation lines which can potentially trace shocks. High-resolution imaging of the core can help to investigate possible sub-structure, including shocked regions, in the core. 
Finally, a dedicated search for a companion star in the central system would be useful, using deeper radio observations with sufficient angular resolution.

This study also highlights the difficulty of identifying LRNe at late times, considering the many similarities that LRNe have with many other classifications of evolved stellar sources. 
\begin{acknowledgements}
TS and TK acknowledge funding from grant no 2018/30/E/ST9/00398 from the Polish National Science Center. DJ acknowledges support from the Agencia Estatal de Investigaci\'on del Ministerio de Ciencia, Innovaci\'on y Universidades (MCIU/AEI) under grant ``Nebulosas planetarias como clave para comprender la evoluci\'on de estrellas binarias'' and the European Regional Development Fund (ERDF) with reference PID2022-136653NA-I00 (DOI:10.13039/501100011033). DJ also acknowledges support from the Agencia Estatal de Investigaci\'on del Ministerio de Ciencia, Innovaci\'on y Universidades (MCIU/AEI) under grant ``Revolucionando el conocimiento de la evoluci\'on de estrellas poco masivas'' and the the European Union NextGenerationEU/PRTR with reference CNS2023-143910 (DOI:10.13039/501100011033). The Submillimeter Array is a joint project between the Smithsonian Astrophysical Observatory and the Academia Sinica Institute of Astronomy and Astrophysics and is funded by the Smithsonian Institution and the Academia Sinica. We recognize that Maunakea is a culturally important site for the indigenous Hawaiian people; we are privileged to study the cosmos from its summit. Based on observations made with the Nordic Optical Telescope, owned in collaboration by the University of Turku and Aarhus University, and operated jointly by Aarhus University, the University of Turku and the University of Oslo, representing Denmark, Finland and Norway, the University of Iceland and Stockholm University at the Observatorio del Roque de los Muchachos, La Palma, Spain, of the Instituto de Astrofisica de Canarias. The NOT data were obtained under program ID P70-404. The National Radio Astronomy Observatory is a facility of the U.S. National
Science Foundation operated under cooperative agreement by Associated Universities, Inc. 
\end{acknowledgements}
\bibliographystyle{aa}
\bibliography{main.bib}
\appendix
\onecolumn
\section{ALFOSC line fluxes}\label{appendix: alfosc fluxes}
\renewcommand{\arraystretch}{1.2} 
\begin{ThreePartTable}
\begin{TableNotes}
\item[a] Flux uncertainties are 1$\sigma$ statistical errors from Gaussian line fitting, including errors of the fitted H$\beta$ line. 
\item[b] Lines marked with `?' denote tentative identifications.
\end{TableNotes}
\begin{longtable}{lcccccccc}
\caption{De-reddened integrated line fluxes from the ALFOSC spectrum taken on 08/12/2024. Fluxes are normalised to the H$\beta$ flux(1.38\standform{-13} erg cm\up{-2} s\up{-1} $\AA^{-1}$). Line velocities are given in the heliocentric frame of reference.}
\label{tab: alfosc line fluxes}\\
\hline
&
& \multicolumn{2}{c}{Core} 
& \multicolumn{2}{c}{North} 
& \multicolumn{2}{c}{South} \\
Species 
& $\lambda_{\rm rest}$
& $\lambda_{\rm obs}$ & Flux/H$\beta$\tnote{a}
& $\lambda_{\rm obs}$ & Flux/H$\beta$\tnote{a}
& $\lambda_{\rm obs}$ & Flux/H$\beta$\tnote{a} \\
& ($\AA$)
& ($\AA$)
&
& ($\AA$)
&
& ($\AA$)
& \\
\hline
\endhead

[\ion{Ne}{III}] & 3868.76
& 3870.01 & $1.46 \pm 0.16$
& -- & --
& -- & -- \\

[\ion{S}{II}] & 4071.00
& -- & --
& 4069.95 & $0.60 \pm 0.06$
& -- & -- \\

H$\gamma$ & 4340.47
& 4340.99 & $0.64 \pm 0.04$
& 4341.59 & $0.55 \pm 0.04$
& 4339.74 & $0.37 \pm 0.04$ \\

[\ion{O}{III}] & 4363.21
& 4363.28 & $0.39 \pm 0.04$
& -- & --
& -- & -- \\

\ion{N}{I} & 4471.49
& 4472.59 & $0.18 \pm 0.04$
& -- & --
& -- & -- \\

[\ion{S}{II}] & 4643.09
& 4642.65 & $0.13 \pm 0.02$
& -- & --
& -- & -- \\

\ion{He}{II} & 4685.91
& 4685.22 & $0.30 \pm 0.02$
& -- & --
& -- & -- \\

[\ion{O}{III}] & 4958.91
& 4957.63 & $3.83 \pm 0.05$
& 4961.11 & $0.13 \pm 0.01$
& 4959.54 & $0.43 \pm 0.02$ \\

[\ion{O}{III}] & 5006.84
& 5005.47 & $11.74 \pm 0.14$
& 5008.90 & $0.44 \pm 0.02$
& 5006.60 & $1.02 \pm 0.02$ \\

[\ion{N}{I}] & 5197.90
& 5197.24 & $0.11 \pm 0.01$
& 5198.83 & $4.64 \pm 0.08$
& 5196.52 & $2.65 \pm 0.05$ \\

{[\ion{N}{II}]} & 5754.59
& 5753.47 & $0.27 \pm 0.01$
& 5754.88 & $0.29 \pm 0.01$
& 5752.31 & $0.25 \pm 0.01$ \\

\ion{He}{I} & 5875.62
& 5874.64 & $0.26 \pm 0.01$
& 5876.17 & $0.05 \pm 0.01$
& -- & -- \\

[\ion{O}{I}] & 6300.30
& 6298.91 & $0.15 \pm 0.01$
& 6300.00 & $2.23 \pm 0.04$
& 6297.97 & $1.37 \pm 0.03$ \\

[\ion{S}{III}] & 6312.06
& 6310.80 & $0.05 \pm 0.01$
& -- & --
& -- & -- \\

[\ion{O}{I}] & 6363.78
& 6362.77 & $0.05 \pm 0.01$
& 6364.39 & $0.74 \pm 0.01$
& 6361.42 & $0.43 \pm 0.01$ \\

[\ion{N}{II}] & 6548.05
& 6546.66 & $1.42 \pm 0.02$
& 6548.34 & $2.45 \pm 0.04$
& 6545.82 & $2.03 \pm 0.04$ \\

H$\alpha$ & 6562.79
& 6562.56 & $2.96 \pm 0.04$
& 6562.98 & $2.91 \pm 0.05$
& 6560.48 & $2.70 \pm 0.05$ \\

[\ion{N}{II}] & 6583.45
& 6582.07 & $4.17 \pm 0.05$
& 6583.63 & $7.57 \pm 0.13$
& 6581.12 & $6.02 \pm 0.11$ \\

\ion{He}{I} & 6678.15
& 6678.35 & $0.06 \pm 0.01$
& -- & --
& -- & -- \\

[\ion{S}{II}] & 6716.43
& 6714.54 & $0.06 \pm 0.01$
& 6716.44 & $0.53 \pm 0.01$
& 6713.68 & $0.36 \pm 0.01$ \\

[\ion{S}{II}] & 6730.80
& 6728.75 & $0.08 \pm 0.01$
& 6730.87 & $0.84 \pm 0.01$
& 6728.06 & $0.54 \pm 0.01$ \\

{[\ion{Ar}{V}]?\tnote{b}} & 7005.87
& 7004.18 & $0.02 \pm 0.01$
& -- & --
& -- & -- \\

\ion{He}{I} & 7065.19
& 7061.86 & $0.15 \pm 0.01$
& -- & --
& -- & -- \\
\hline
\insertTableNotes

\end{longtable}
\end{ThreePartTable}
\onecolumn
\section{SMA line fluxes}\label{appendix: sma fluxes}
    \begin{table}[h!]
    \caption{Identified molecular line properties from the SMA spectrum.}
    \centering
    \begin{tabular}{cccccc}\hline
Line & Transition & Rest frequency & LSR velocity & FWHM & Integrated flux \\
& & (GHz) & (km s\up{-1}) & (km s\up{-1}) & (Jy km s\up{-1}) \\\hline
H\down{2}CO & 3 (1, 3) -- 2 (1,2) & 211.2115 & -20.82 & 51 & 10.20 $\pm$ 4.99 \\
$^{13}$CH$_3$CN & 12 (11) -- 11 (11) & 213.8886 & -6.23 & 30 & 3.83 $\pm$ 3.00 \\
$^{13}$CH$_3$CN & 12 (9) -- 11 (9) & 214.0487 & -34.57 & 32 & 4.75 $\pm$ 3.91 \\
\up{13}C\up{17}O & 2--1 & 214.5732 & -19.40 & 26 & 12.09 $\pm$ 3.54 \\
$^{13}$CN & N=2--1, J=$\frac{5}{2}$--$\frac{3}{2}$, F$_1$=2--1, F=1--1 & 217.2908 & -28.67 & 37 & 16.13 $\pm$ 4.18 \\
HCC\up{13}CN & 24 (23) -- 23 (22) & 217.4196 & -22.65 & 51 & 15.16 $\pm$ 5.19 \\
$^{13}$CN & N=2--1, J=$\frac{5}{2}$--$\frac{3}{2}$, F$_1$=3--2, F=4--3 & 217.4672 & -23.94 & 28 & 12.46 $\pm$ 3.76 \\
CN & N=2--1, J=$\frac{3}{2}$--$\frac{3}{2}$, F=$\frac{1}{2}$--$\frac{1}{2}$ & 226.2874 & -24.21 & 26 & 3.13 $\pm$ 1.81 \\
CN &  N=2--1, J=$\frac{3}{2}$--$\frac{3}{2}$, F=$\frac{5}{2}$--$\frac{3}{2}$ & 226.3419 & -27.50 & 76 & 11.95 $\pm$ 6.63 \\
HC$^{13}$CCN & 25--24 & 226.4542 & -35.04 & 97 & 11.21 $\pm$ 7.11 \\
CN & N=2--1, J=$\frac{3}{2}$--$\frac{1}{2}$, F=$\frac{3}{2}$--$\frac{5}{2}$ & 226.6596 & -20.49 & 54 & 24.51 $\pm$ 4.93 \\
CN &  N=2--1, J=$\frac{5}{2}$--$\frac{3}{2}$, F=$\frac{3}{2}$--$\frac{5}{2}$ & 226.8759 & -18.29 & 37 & 37.21 $\pm$ 4.03 \\
HC\down{3}N & 25--24 & 227.4189 & -19.72 & 53 & 11.43 $\pm$ 4.88 \\
CO & 2--1 & 230.538 & -20.06 & 33 & 87.26 $\pm$ 3.38 \\
H\up{13}CN & 3--2 & 259.0118 & -20.31 & 38 & 66.91 $\pm$ 4.04 \\
H\up{13}CO\up{+} & 3--2 & 260.2553 & -22.85 & 31 & 17.19 $\pm$ 3.86 \\
HN\up{13}C & 3--2 & 261.2633 & -21.40 & 28 & 15.34 $\pm$ 3.65 \\
CCH & N=3--2, J=$\frac{7}{2}$--$\frac{5}{2}$, F=3--2 & 262.0064 & -16.79 & 32 & 8.47 $\pm$ 3.89 \\
HC\up{13}CCN & 29--28 & 262.6733 & -23.88 & 71 & 10.15 $\pm$ 6.13 \\
HC$_3$N & 29--28 & 263.7923 & -17.85 & 41 & 5.00 $\pm$4.44 \\\hline
\end{tabular}
\label{tab: SMA line fluxes}
\end{table}
\onecolumn
\section{SMA moment maps}\label{appendix: SMA moment maps}
\begin{figure}[h!]
    \centering
    \includegraphics[scale=0.2]{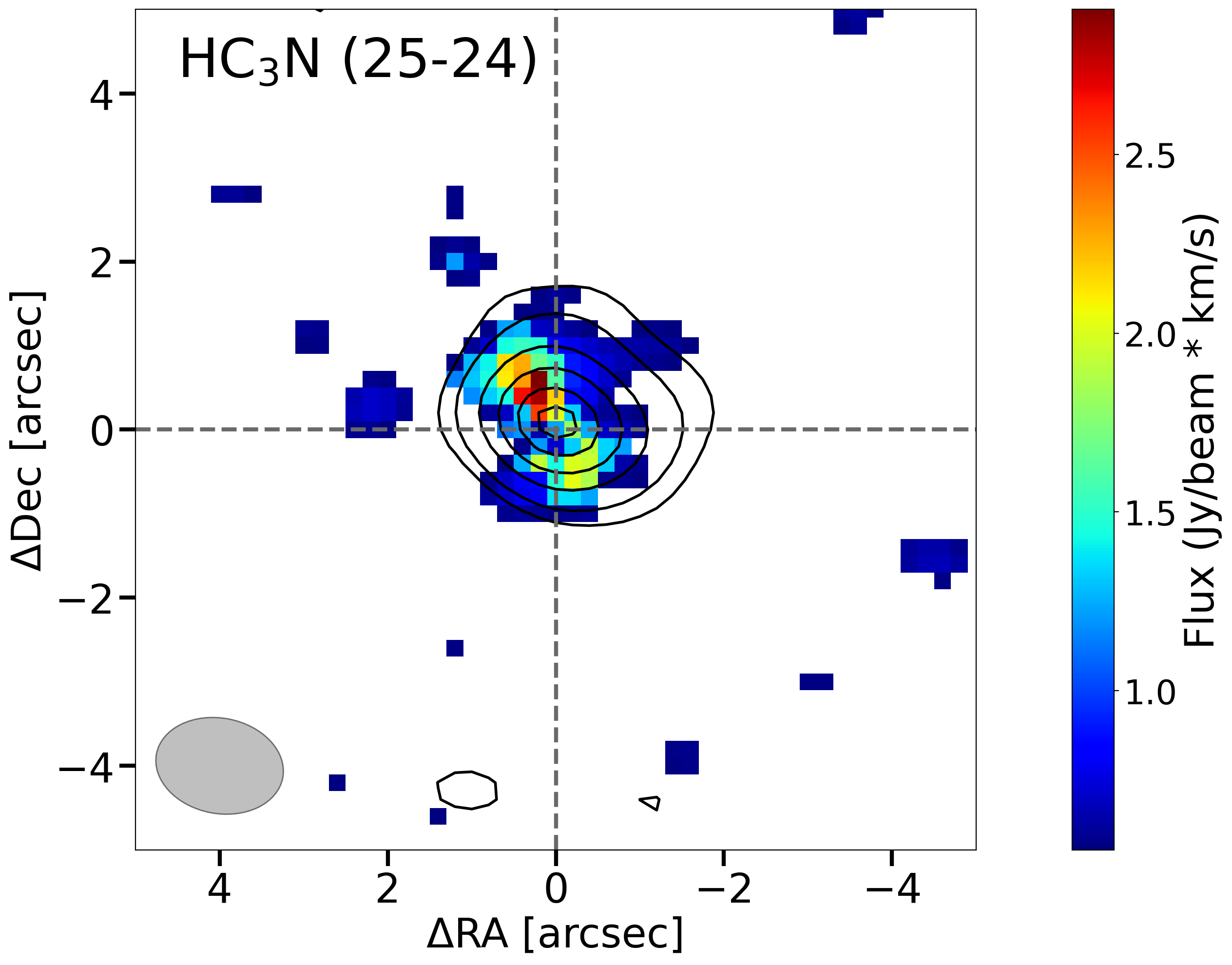}
    \includegraphics[scale=0.2]{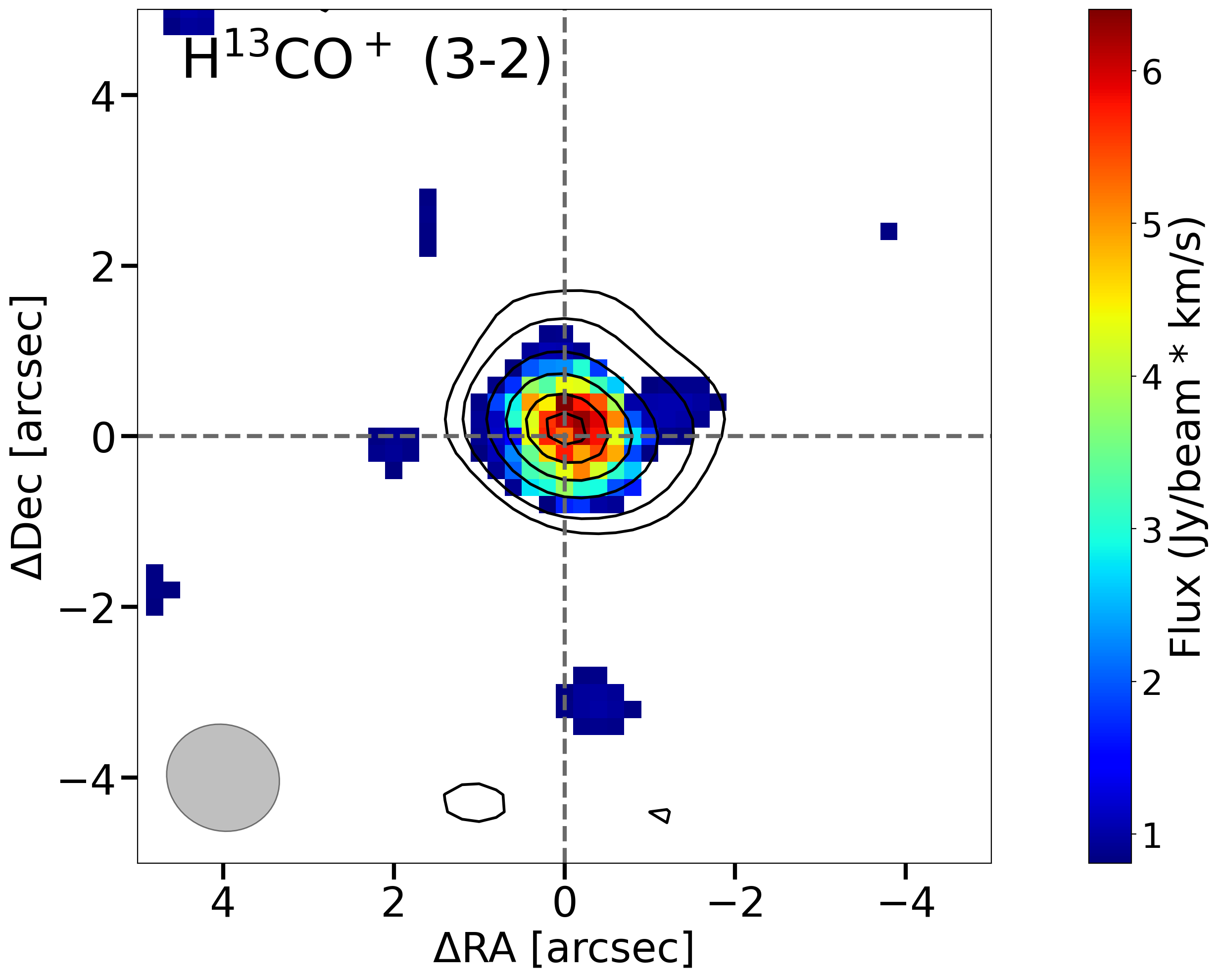}
    \includegraphics[scale=0.2]{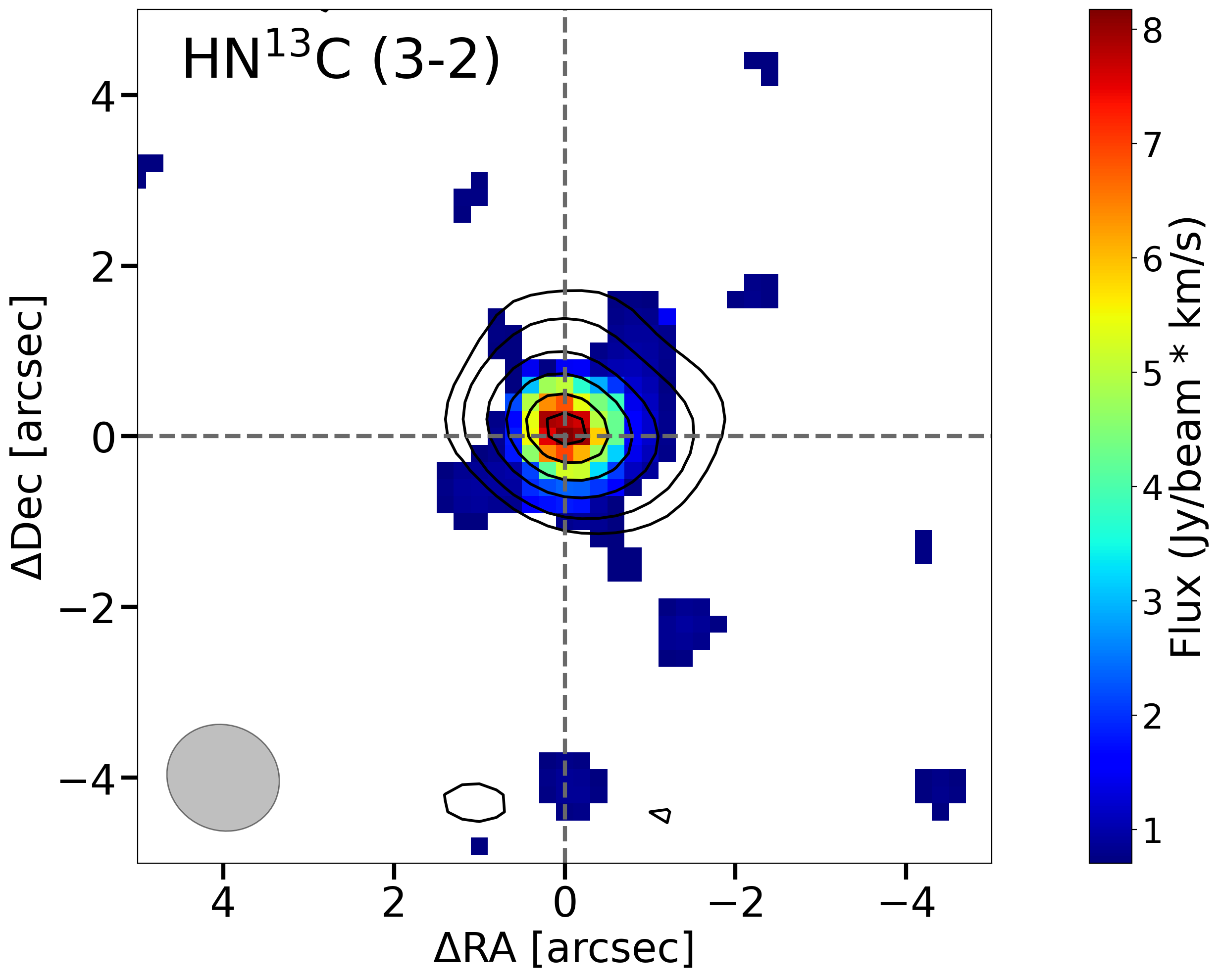}
    \includegraphics[scale=0.2]{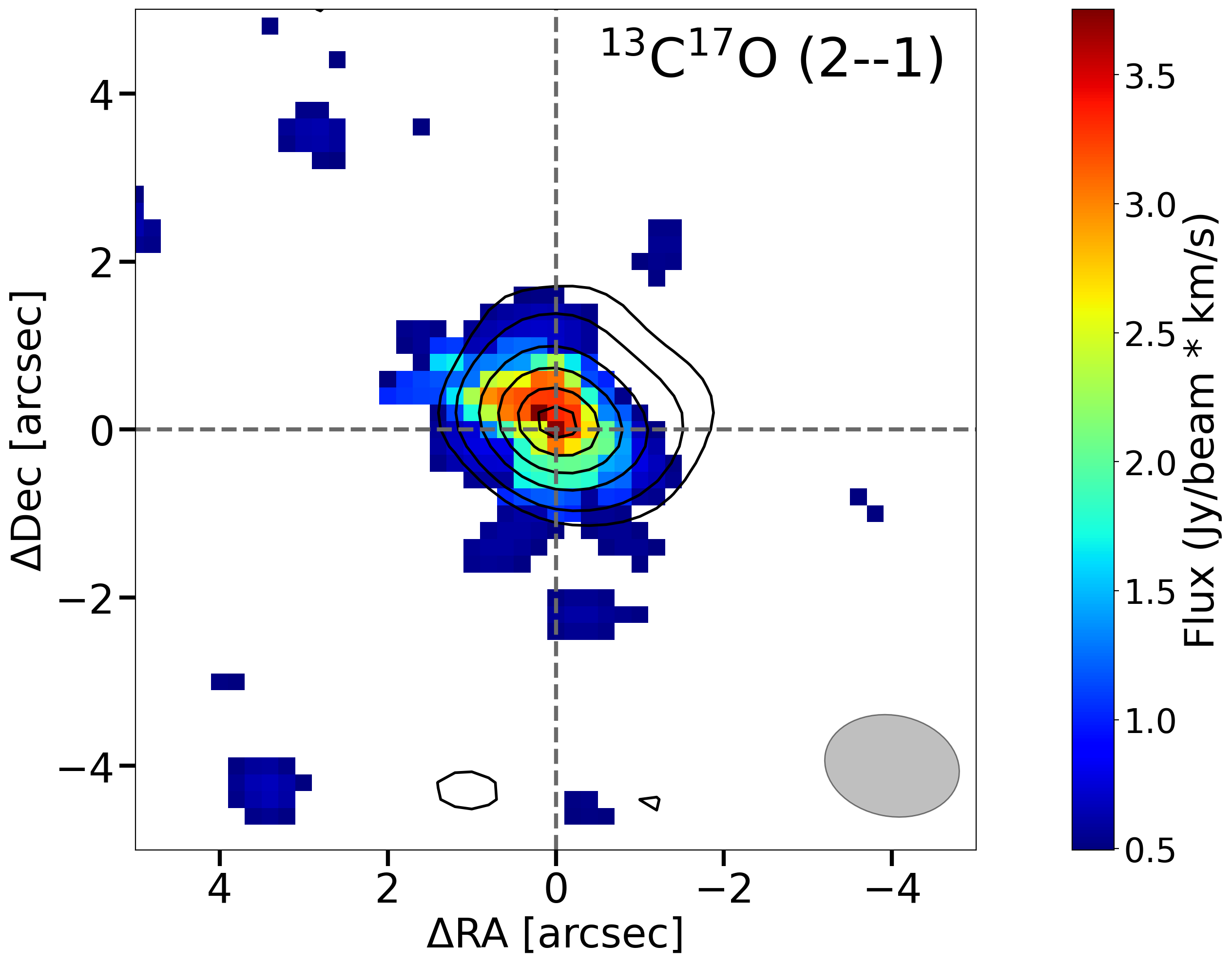}
    \caption{SMA 3$\sigma$ integrated flux maps of HC$_3$N (25--24) (top left), H$^{13}$CO$^+$ (3--2) (top right), HN$^{13}$C (3--2) (bottom left), and CN (bottom right). The contours are the same as for Fig. \ref{fig: moment 0 maps 5 sigma}. The grey ellipse resembles the synthesised beam.}
    \label{fig: moment 0 maps 3 sigma}
\end{figure}
\section{PV diagrams}\label{appendix: PV}

\begin{figure}[h!]
    \centering
    \includegraphics[width=0.49\linewidth]{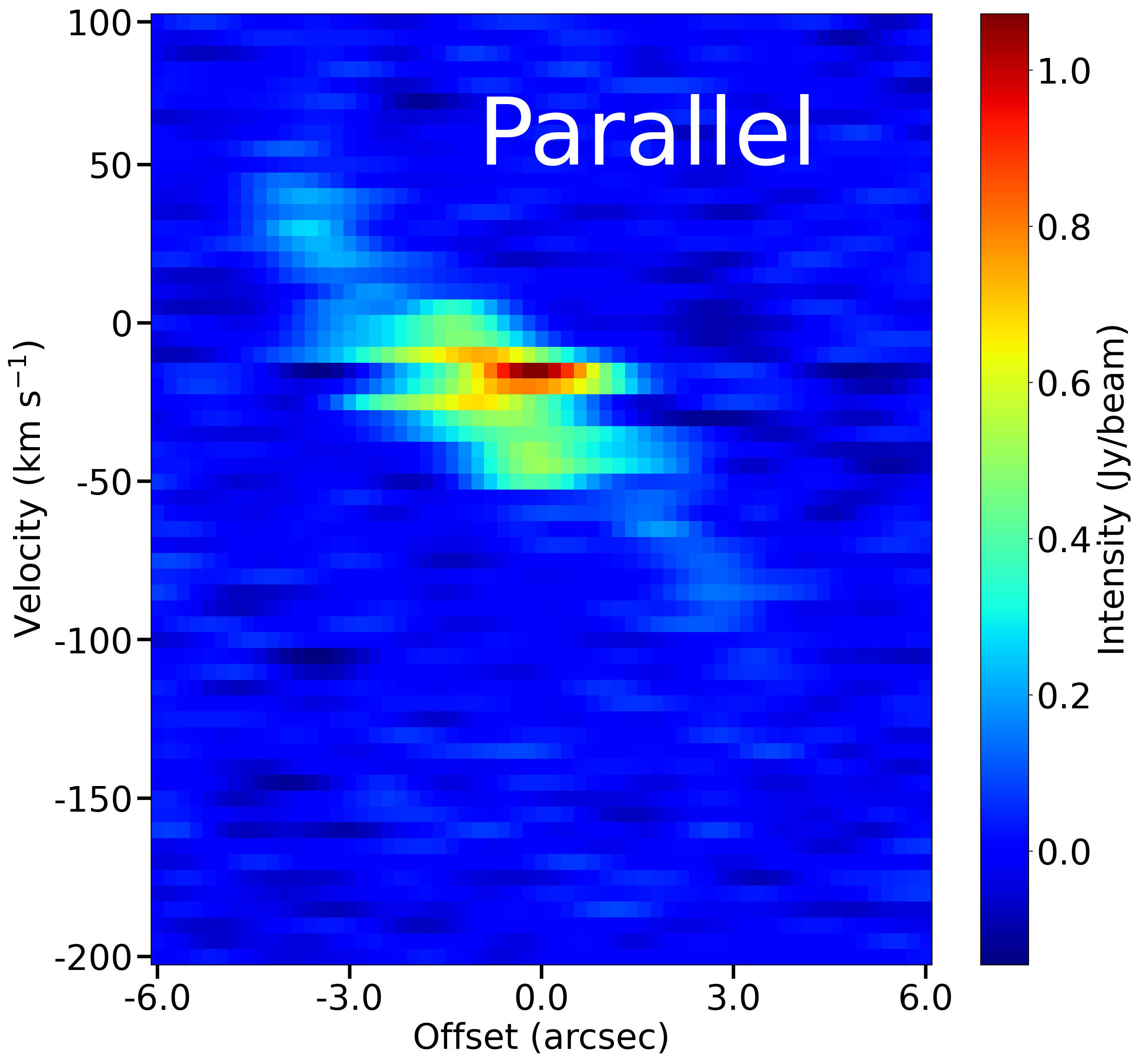}
    \includegraphics[width=0.49\linewidth]{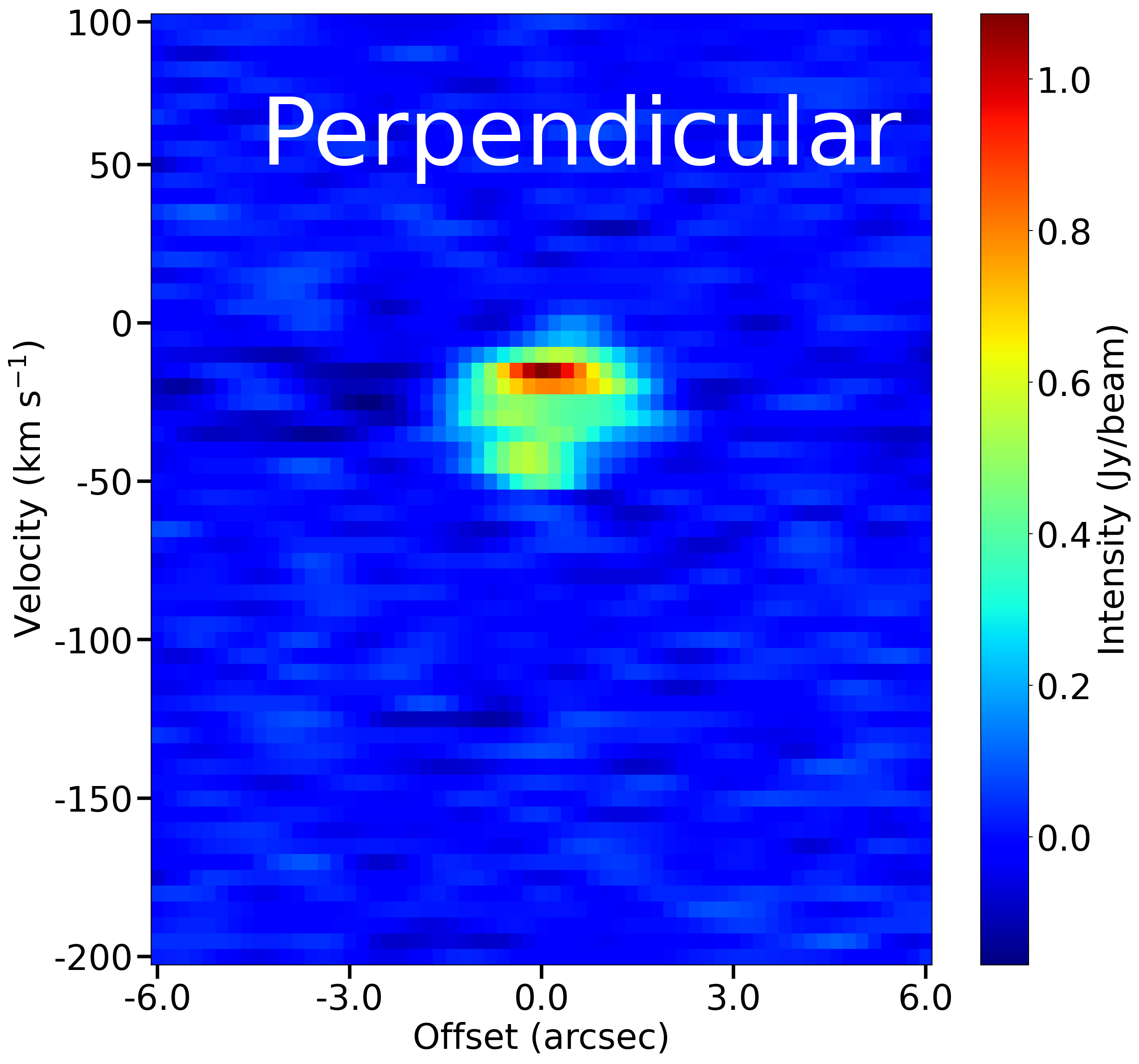}
    \caption{PV diagrams of the CO (2--1) emission, extracted at PA=41\degree\ (top panel) and 131\degree\ (bottom panel). The pixel width was 3 pixels, and the length of the PV extraction slit was 12\arcsec.}
    \label{fig: pv diagrams}
\end{figure}

\section{Interstellar reddening}\label{appendix: reddening}
In order to examine whether the reddening derived for K4-47 has a circumstellar contribution, we used the GALE{\footnotesize XTIN} VO service \citep{amores2021} to extract estimated reddening values in the direction of K4-47 at different distances, using multiple different dust maps of the Galaxy. We utilised both dust models from \citet{amores2005}, as well as the dust maps from \citet{sale2014}, and the Bayestar15,17,19 models \citep{green2015,green2018,green2019}. Each dust map provides estimates of the interstellar extinction at a specified equatorial coordinate and at a specified distance. Using the various distance estimates of K4-47, including 3--7 kpc \citep{corradi2000}, 5.9 kpc \citep{tajitsu1998}, 8.5 kpc \citep{cahn1992}, 20.42 kpc \citep{zhang1995}, and an upper limit of 26.5 kpc \citep{vandesteene1994}, we calculated values of E(B--V) for distance intervals of 0.1 kpc in the range of 3--26.5 kpc. We then convert A\down{V} to E(B--V) using E(B--V)=A\down{V}/R\down{V}, and assuming R\down{V}=3.1.

\begin{figure}[h!]
    \centering
    \includegraphics[width=0.95\linewidth]{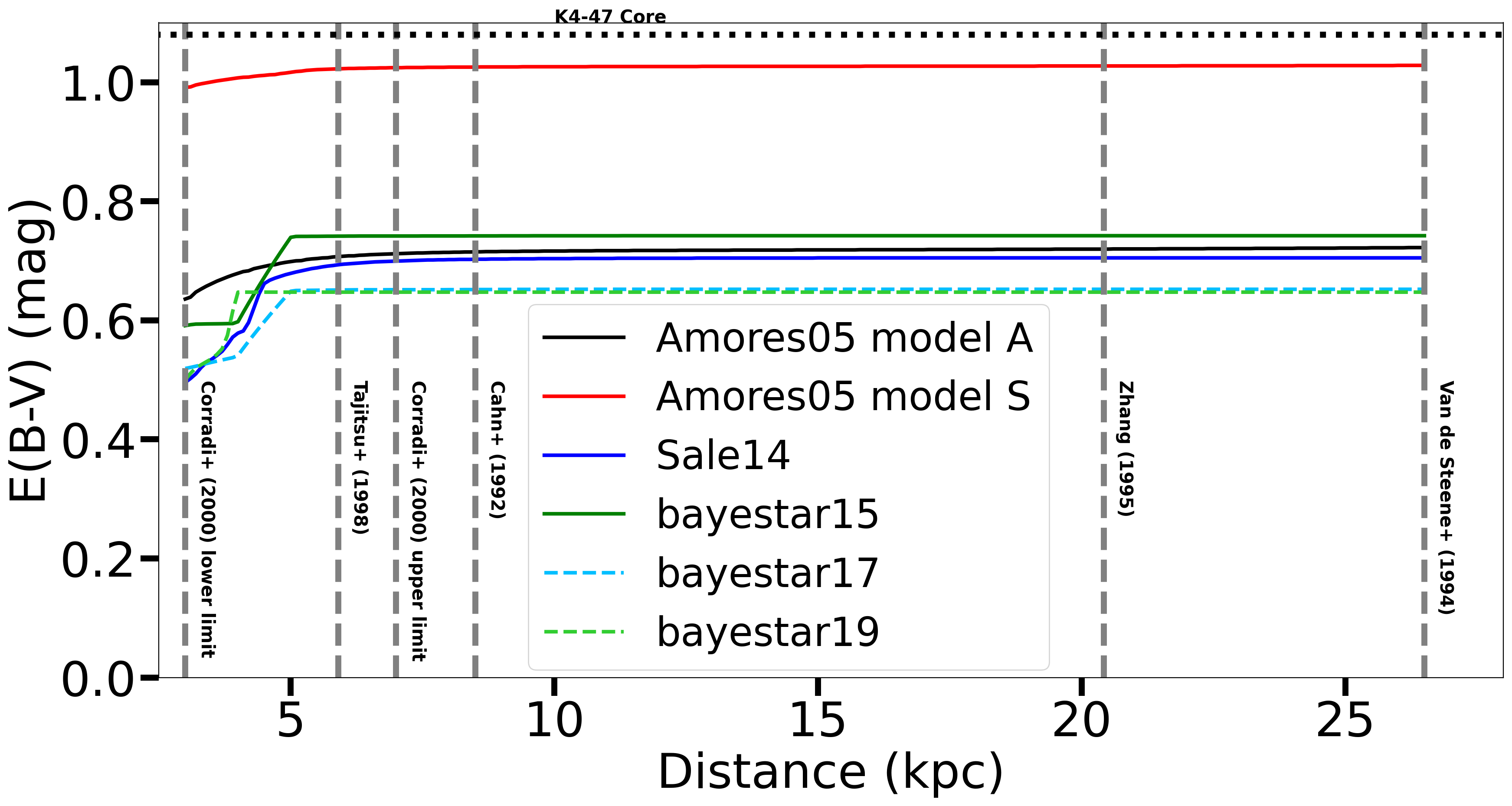}
    \caption{E(B--V) curves from the various models extracted via GALE{\footnotesize XTIN}. The black dotted line indicates the measured E(B--V) value for the core of K4-47, and grey dashed vertical lines indicate the estimated distances to K4-47 in the literature (references shown in the annotations).}
    \label{fig: ebv curves}
\end{figure}

We show the E(B--V) curves vs distance for each dust map in Fig. \ref{fig: ebv curves}. The majority of the outputs for each dust mass did not provide values in units of E(B--V). This was only true for the Bayestar15 model, which provides values for E(B--V) in the same units as \citet{SFD98}. For Bayestar17 and Bayestar19, we assume that these models are not equivalent to the units of \citet{SFD98}, meaning that output values of these models were multiplied by a factor of 0.996 to convert them to the desired units. Details of this conversion are provided via the Argonaut SkyMaps website\footnote{http://argonaut.skymaps.info/usage}. \citet{sale2014} provided extinction values in units of A\down{0}, which is converted to A\down{V} using the relation A\down{V}=A\down{0}/1.003.
\newpage
\section{SED Fitting}\label{appendix: SED}
\begin{figure}[h!]
    \centering
    \includegraphics[width=0.45\linewidth]{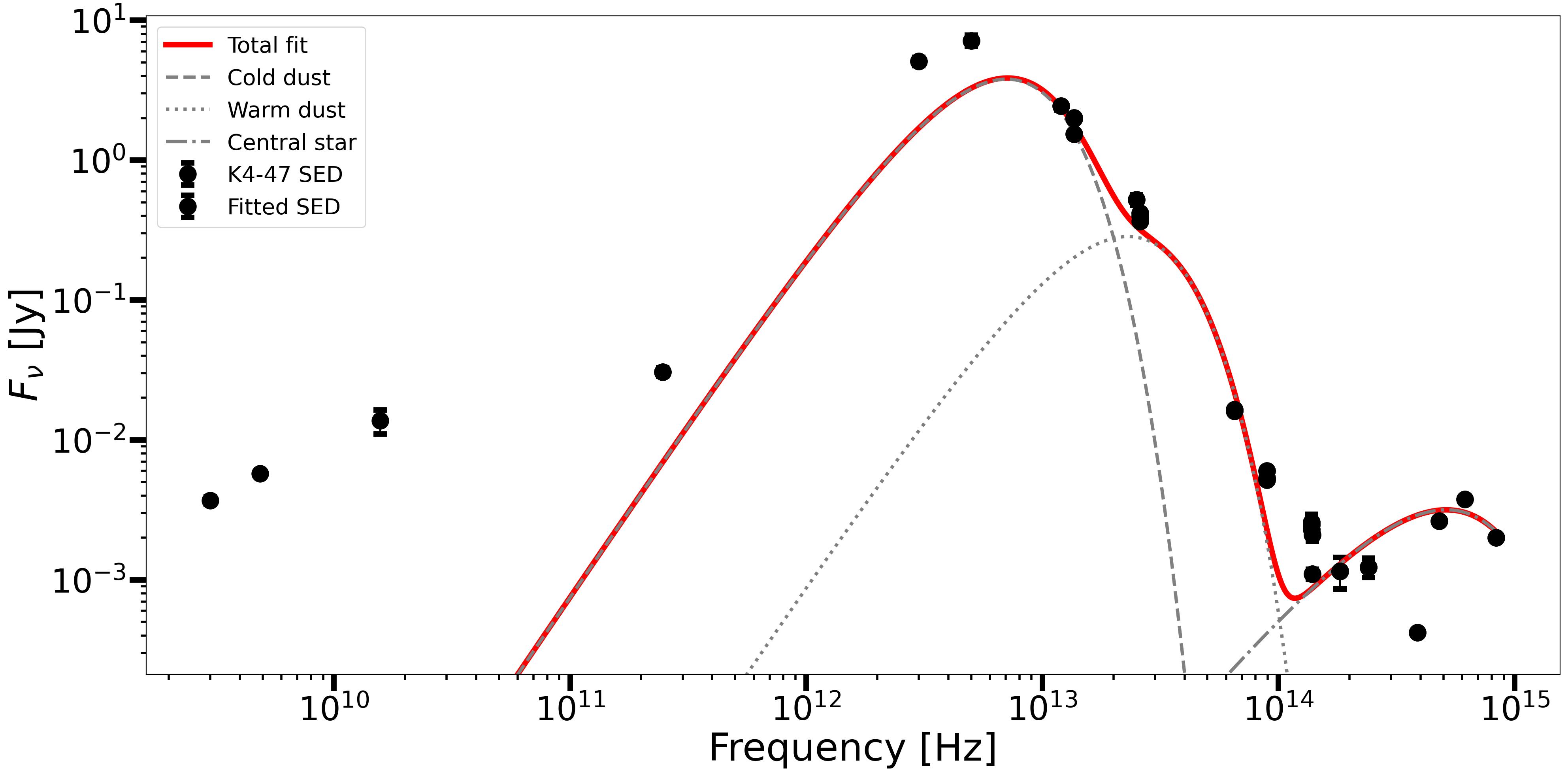}
    \includegraphics[width=0.45\linewidth]{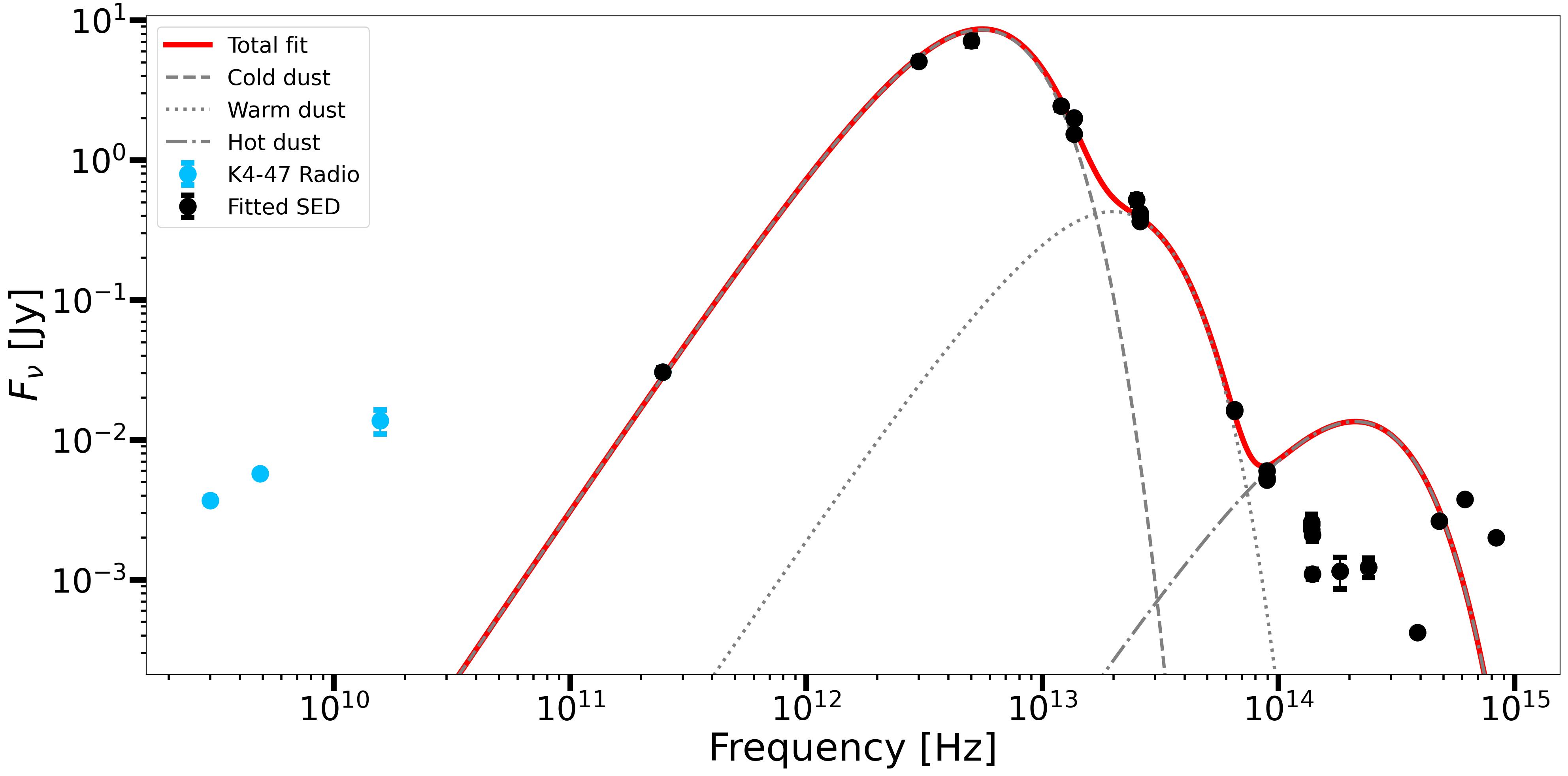}
    \caption{Modified blackbody fits to the full SED of K4-47. The left panel shows the fit to the 2-component dust + stellar blackbody, and the right panel shows the fit to the 3-component dust model. Dashed and dotted curves indicate the contribution from cold and warm dust, respectively, and the dash-dotted curves indicate the stellar flux in the left panel, and the hot dust in the right panel.}
    \label{fig: appendix SED fitting}
\end{figure}
\section{GALEX} \label{appendix: galex obs}
\begin{figure}[h!]
    \centering
    \includegraphics[trim=0 0 0 0, width=0.45\linewidth]{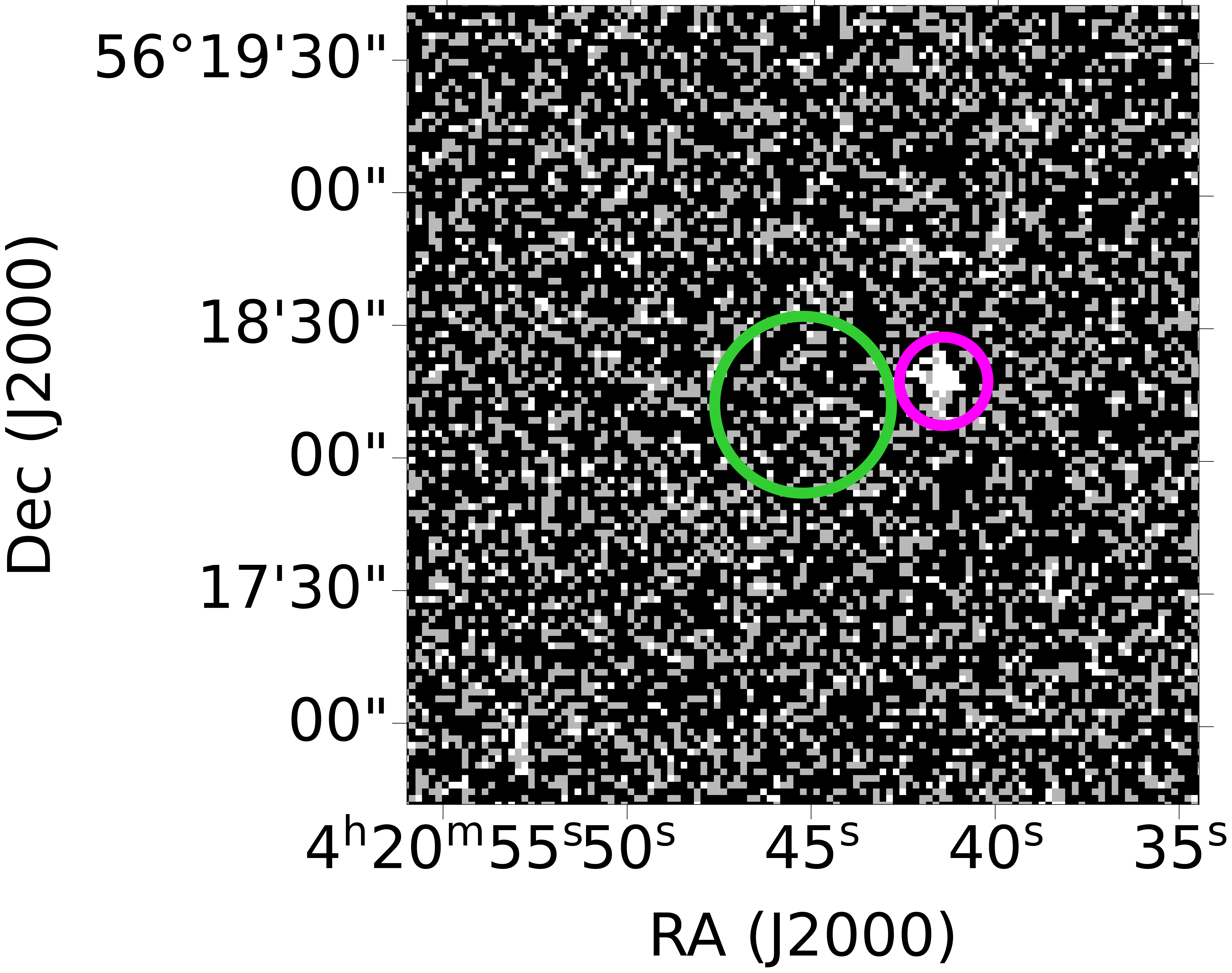}
    \includegraphics[trim=0 0 0 0, width=0.45\linewidth]{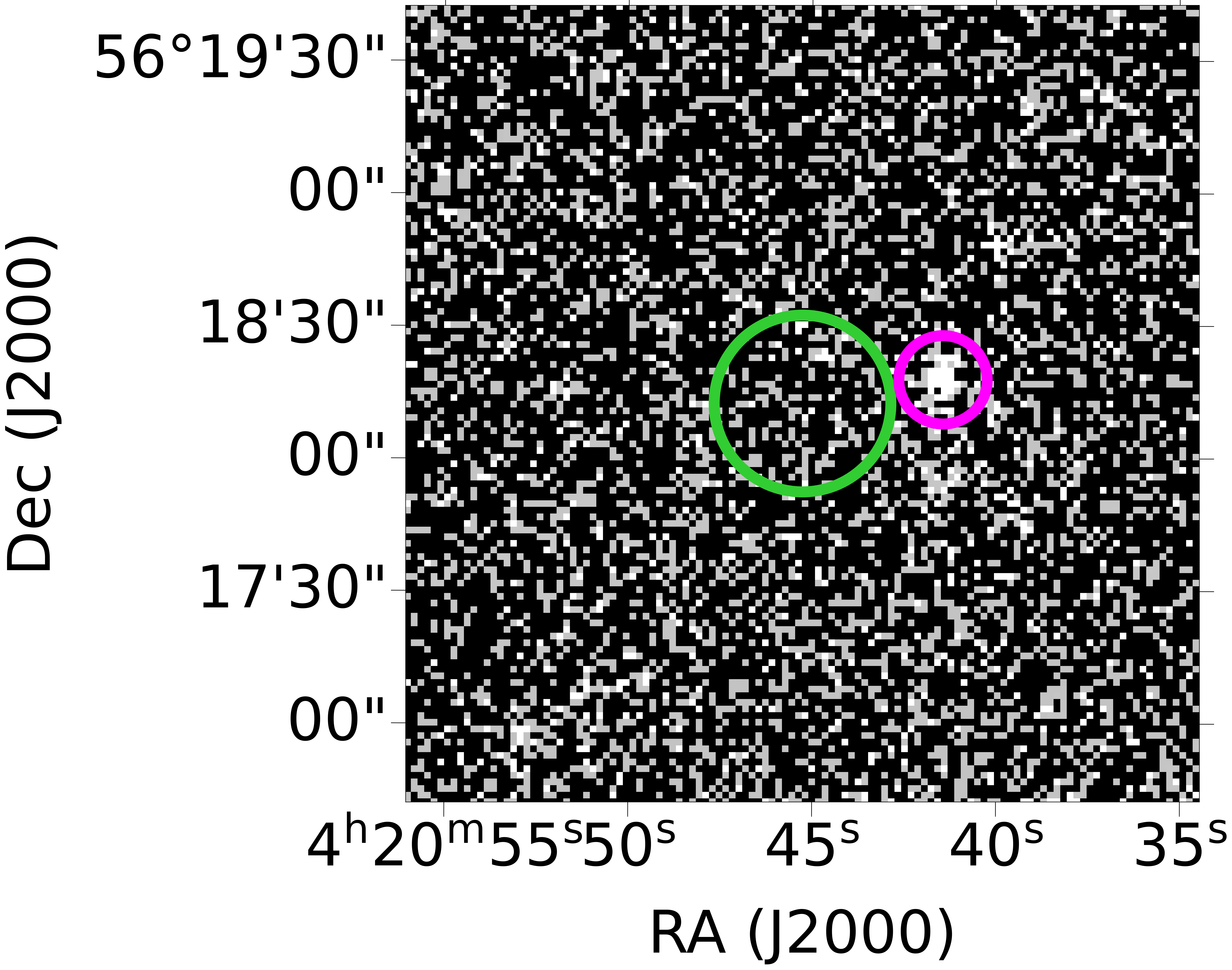}
    \caption{GALEX frames overlaid with a 20\arcsec\ aperture (green) centred on the coordinates of K4-47. The magenta 10\arcsec\ aperture is centred on the location of the nearby Gaia source.}
    \label{fig: galex}
\end{figure}

K4-47 was covered by the GALEX survey on 02/01/2012 in two separate fields: AIS\_53\_0001\_sg79 and AIS\_53\_0001\_sg83. The data was observed with exposure times of 108 and 110 seconds, respectively, in the NUV band ($\lambda_{\rm eff}$=2267 \angstrom, $\lambda_{\rm FWHM}$=616 \angstrom). We found no clear source within a 20\arcsec\ aperture centred on the coordinates of K4-47 ($\alpha$=04:20:45.263, $\delta$=+56:18:12.44; see Fig. \ref{fig: galex}). We calculated 5$\sigma$ upper limits, where $\sigma$ is derived from the root-mean-square (rms) noise within the aperture, for each frame and averaged them. The final averaged upper limit for K4-47 is 37.4 $\mu$Jy at 2267 \AA.

Within our Gaia archival maps, a source can be seen $\sim$30\arcsec\ west of the location of K4-47 (Fig. \ref{fig: galex}). To ensure that the astrometric solution of GALEX is sufficiently precise, so that we can be sure that this source is not in fact K4-47, we checked the Aladin Sky Atlas service\footnote{https://aladin.cds.unistra.fr/\#Overview} to look for nearby sources. A bright source (Gaia EDR3 276843597519980288: $\alpha$=04:20:41.44, $\delta$=+56:18:17.85) catalogued in the Gaia early release DR3 catalogue \citep{gaiaedr3} is seen close to K4-47 and can be identified with the unrelated source found the GALEX maps. We show in Fig. \ref{fig: galex} a 10\arcsec\ aperture centred on the Gaia source. We see that the aperture covers this source exactly, indicating this western source is not K4-47.
\end{document}